\pdfoutput=1

\documentclass[preprint,12pt,5p,twocolumn]{elsarticle}

\def\review{preprint}
\def\rev{rev}

\def\figstyle{color}
\def\bw{bw}

\ifx\review\rev
 \usepackage{lineno,rotating}
\fi

\usepackage{amssymb}

\biboptions{comma,square,numbers}
\journal{Nuclear Instruments and Methods A}

\begin{document}

\begin{frontmatter}



\title{The T2K Fine-Grained Detectors}

\author[triumf]{P.-A. Amaudruz}
\author[regina]{M. Barbi}
\author[triumf]{D. Bishop}
\author[uvic]{N. Braam}
\author[ubc]{D.G. Brook-Roberge}
\author[regina]{S. Giffin}
\author[kyoto]{S. Gomi}
\author[triumf]{P. Gumplinger}
\author[triumf]{K. Hamano}
\author[regina]{N.C. Hastings}
\author[ubc]{S. Hastings}
\author[triumf]{{R.L. Helmer}\corref{cor}}
\ead{helmer@triumf.ca}
\author[triumf]{R. Henderson}
\author[kyoto]{K. Ieki}
\author[ubc]{B. Jamieson}
\author[triumf]{I. Kato}
\author[triumf]{N. Khan}
\author[ubc]{J. Kim}
\author[ubc]{B. Kirby}
\author[alberta]{P. Kitching}
\author[triumf]{A. Konaka}
\author[uvic,triumf]{M. Lenckowski}
\author[regina]{C. Licciardi}
\author[ubc]{T. Lindner}
\author[triumf]{K. Mahn}
\author[regina]{E.L. Mathie}
\author[stfc]{C. Metelko}
\author[triumf]{C.A. Miller}
\author[kyoto]{A. Minamino}
\author[triumf]{K. Mizouchi}
\author[kyoto]{T. Nakaya}
\author[kyoto]{K. Nitta}
\author[triumf]{C. Ohlmann}
\author[triumf]{K. Olchanski}
\author[ubc]{S.M. Oser}
\author[kyoto]{M. Otani}
\author[uvic]{P. Poffenberger}
\author[triumf]{R. Poutissou}
\author[triumf]{J.-M. Poutissou}
\author[stfc]{W. Qian}
\author[triumf]{F. Retiere}
\author[regina]{R. Tacik}
\author[ubc]{H.A. Tanaka}
\author[triumf]{P. Vincent}
\author[triumf]{M. Wilking}
\author[triumf]{S. Yen}
\author[tokyo]{M. Yokoyama}

\cortext[cor]{Corresponding author}

\address[triumf]{TRIUMF, Vancouver, British Columbia, Canada}
\address[ubc]{University of British Columbia, 
Department of Physics and Astronomy,
Vancouver, British Columbia, Canada}
\address[alberta]{University of Alberta,
Centre for Particle Physics, Department of Physics,
Edmonton, Alberta, Canada}
\address[regina]{University of Regina, Physics Department,
Regina, Saskatchewan, Canada}
\address[tokyo]{University of Tokyo, Department of Physics, Tokyo, Japan}
\address[kyoto]{Kyoto University, Department of Physics, Kyoto, Japan}
\address[uvic]{University of Victoria,
Department of Physics and Astronomy,
Victoria, British Columbia, Canada}
\address[stfc]{STFC, Rutherford Appleton Laboratory, Harwell Oxford, United
Kingdom}

\begin{abstract}
\small{
T2K is a long-baseline neutrino oscillation experiment searching for
$\nu_e$ appearance in a $\nu_{\mu}$ beam. The beam is produced at the
J-PARC accelerator complex in Tokai, Japan, and the neutrinos are
detected by the Super-Kamiokande detector located 295~km away in
Kamioka. A suite of near detectors (ND280) located 280~m downstream of
the production target is used to characterize the components of the
beam before they have had a chance to oscillate and to better
understand various neutrino interactions on several nuclei. This paper
describes the design and construction of two massive fine-grained
detectors (FGDs) that serve as active targets in the ND280
tracker. One FGD is composed solely of scintillator bars while the
other is partly scintillator and partly water. Each element of the
FGDs is described, including the wavelength shifting fiber and
Multi-Pixel Photon Counter used to collect the light signals, the
readout electronics, and the calibration system. Initial tests and
{\normalsize\em in situ} results of the FGDs' performance are also
presented.  }
\end{abstract}

\ifx\review\rev
\fi
\end{frontmatter}

\ifx\review\rev
 \linenumbers
\fi

\section{Introduction}
\label{sec:introduction}

The Tokai-to-Kamioka (T2K) experiment~\cite{Itow:2001ee} is studying
neutrino oscillations using a man-made neutrino beam sent from the
Japan Proton Accelerator Research Complex (J-PARC) in Tokai, Japan,
towards the Super-Kamiokande detector~\cite{Fukuda:2002uc}, located
295~km away. T2K is the first experiment to make use of an off-axis
configuration~\cite{Beavis:1995,Mann:1993zk,Helmer:1994ac}, which
provides a relatively narrow band beam, in this case peaked around
700~MeV.  A magnetized near detector called ND280, situated 280~m
downstream from the hadron production target in J-PARC, measures the
neutrino beam's properties before the neutrinos have had a chance to
oscillate.  It can therefore be used to predict the neutrino event
rate and energy spectrum at Super-Kamiokande in the absence of
oscillations. As well, interaction rates or cross sections for several
neutrino reaction channels in the 100~MeV to few GeV energy range will
be measured.

\begin{figure}[htb]
\centering
\ifx\figstyle\bw
 \includegraphics[width=\columnwidth]{jpgFigures/fig01_bw.jpg}
\else
 \includegraphics[width=\columnwidth]{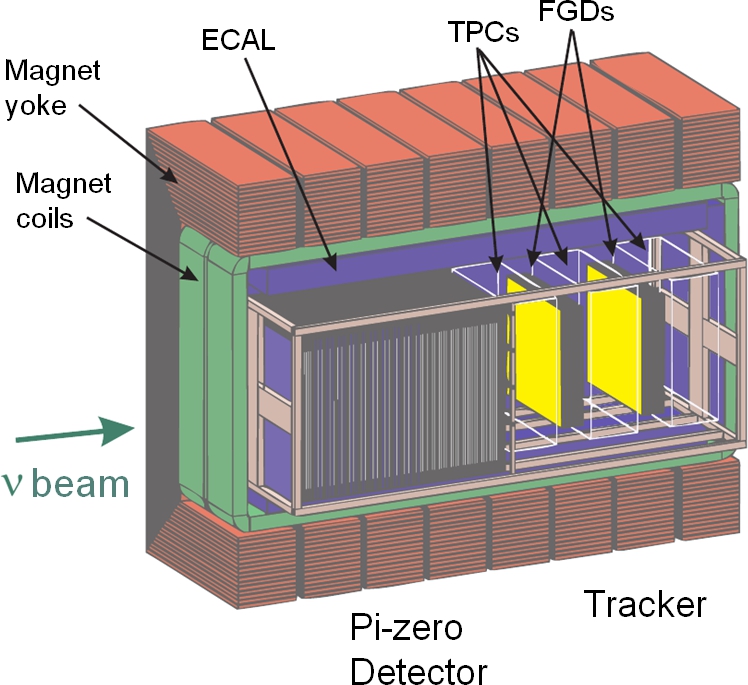}
\fi
\caption{Cutaway view of the ND280 detector.  The two FGDs are located
  between three TPCs.  A Pi-Zero Detector sits upstream of the tracker
  region, and electromagnetic calorimeters (ECAL) surrounds all of the
  central detectors.  The magnetic field is in the horizontal
  direction perpendicular to the beam.}
\label{fig:nd280}
\end{figure}

ND280 contains several subdetectors optimized to measure neutrino
interactions in this energy range~\cite{Abe:2011ks}.  This paper
describes the design and performance of ND280's two massive
Fine-Grained Detectors (FGDs), which provide target mass for neutrino
interactions as well as tracking of charged particles emerging from the
interaction vertex. The FGDs form part of ND280's tracker.  The
tracker, consisting of three large time projection chambers
(TPCs)~\cite{tpc:nim} and the two FGD modules, lies at the heart of
ND280 (see Fig.\ \ref{fig:nd280}).  The primary function of ND280 is
to measure the neutrino beam's flux, energy spectrum, and flavor
composition by observing charged current neutrino interactions.  The
FGDs are thin enough ($\sim 30$~cm) that most of the penetrating
particles produced in neutrino interactions, especially muons, reach
the TPCs where their momenta and charges are measured by their
curvature in the $\sim 0.2$~T magnetic field. While the TPCs provide
excellent 3D tracking and particle identification for forward- and
backward-going charged particles, short-ranged particles such as
recoil protons are primarily measured in the FGDs themselves. The FGDs
therefore have fine granularity so that individual particle tracks can
be resolved and their directions measured.

An especially important reaction to measure is the CCQE interaction
$\nu_\ell + n \to \ell^{-} + p$, which is the most common interaction
at T2K's beam energy.  For these interactions, the energy of the
incident neutrino is calculable from only the energy and direction of
the final lepton, with an accuracy limited by the Fermi momentum of
the neutron in the nucleus.  The CCQE interaction cross section is
relatively simple to model theoretically~\cite{Smith:1972xh}, and is
well constrained by
data~\cite{Ahn:2006zza,Lyubushkin:2008pe,AguilarArevalo:2010zc}.  CCQE
interactions therefore provide an ideal means of measuring the
neutrino beam's energy spectrum and flux in the near detector, which
can then be used to predict the event rate and energy spectrum at the
far detector.

Although the CCQE interaction is the most common interaction mode,
many other processes occur.  An important example is CC single pion
(CC-1$\pi$) production ($\nu_\ell + N \to \ell^{-} + N'+ \pi$).  This
process often proceeds through excitation of a $\Delta$ resonance and
is indistinguishable from a CCQE event in Super-K, where only the
final state charged lepton is above the Cherenkov threshold.  Because
CC-1$\pi$ produces a three-body final state, the initial neutrino's
energy is not a simple function of the charged lepton's direction and
energy.  CC-1$\pi$ events will therefore smear out the energy spectrum
measurement, and every effort is made to exclude them from the energy
spectrum analyses in both the near and far detectors.  At Super-K this
is accomplished by selecting only events with a single charged lepton
in the final state, although CC-1$\pi$ events with the pion below
Cherenkov threshold form an irreducible background.  The ND280 tracker
is used to measure the size of this and other backgrounds to CCQE
interactions at Super-K.

The rates of these CCQE and non-QE interactions from the T2K beam must
be well determined in the tracker so that a satisfactory prediction
can be made of the unoscillated event rates at Super-K.  Because the
tracker can see all charged particles produced in an interaction, it
can identify CCQE events by selecting just those events which contain
a lepton and a recoil proton.  Events containing pions can be rejected
by searching for additional charged tracks near the vertex,
identifying Michel electrons produced by pions stopping in an FGD
through the $\pi \to \mu \to e$ decay chain, or by testing the
consistency of a track's deposited charge, direction, and momentum
with the CCQE hypothesis.

The FGDs must therefore satisfy a variety of design criteria:
\begin{itemize}
\item They must be capable of detecting all charged particles produced
  at the interaction vertex with good efficiency in order to
  determine the type of interaction.
\item They must be thin enough that charged leptons will penetrate
  into the TPCs where their momenta and flavor can be
  determined.
\item The directions of recoil protons must be measured so that CCQE
  events can be selected using kinematic constraints on the recoil
  proton's direction.
\item Particle ID from dE/dx measurements must distinguish
  protons from muons and pions.  
\item The tracker must contain $\sim 1$~tonne of target mass for
  neutrino interactions in order to yield a sufficient statistical
  sample of events.
\item Because the ND280 sits in an off-axis beam, the off-axis angle
  and hence the neutrino energy spectrum varies substantially across
  the face of the tracker.  Nonuniformities in threshold or
  efficiency across the tracker would greatly complicate the
  extraction of the neutrino beam properties and could potentially
  bias the measurement.  Therefore the detector response across
  the tracker needs to be as uniform as practically possible.
\item Because the far detector is a water Cherenkov detector, the
tracker must measure the neutrino interaction rates on water. All of
the relevant neutrino cross sections depend at some level on the
target nucleus through such effects as Pauli blocking, pion
rescattering and absorption inside the nucleus, etc.  These nuclear
effects cannot be reliably corrected for from theory, and therefore
the nuclear interaction rates must be measured on water so that the
rates can be used to predict the rates for these processes in Super-K.
\item The FGD electronics must provide for acceptance of late
  hits such as those due to Michel electrons.
\end{itemize}

\subsection{Overview of the FGD design}
\label{sec:designOverview}

The functional unit of an FGD is a single extruded polystyrene
scintillator bar (described in Section~\ref{sec:scint}) oriented
perpendicular to the beam in either the $x$ or $y$ direction. To
achieve the necessary fine granularity, the bars have a square cross
section 9.6~mm on a side.  They are arranged into modules, each ``XY
module'' consisting of a layer of 192 scintillator bars in the
horizontal direction glued to 192 bars in the vertical direction (see
Section~\ref{sec:xymodule}). FGD1 contains fifteen such modules while
FGD2 contains seven.  Each module has dimensions of 186.4 $\times$
186.4 $\times$ 2.02~cm (not including electronics).  Each scintillator
bar has a reflective coating containing TiO$_2$ and a wavelength
shifting (WLS) fiber (Section~\ref{sec:fiber}) going down an axial
hole. An air gap provides the coupling between the scintillator and
fiber. The fiber extends a few centimeters from one end of the bar to
reach a Multi-Pixel Photon Counter (MPPC)~\cite{Yokoyama:2006nu,
Gomi:2007zz} (see Section~\ref{sec:photosensor}), which senses and
digitizes the light signals. Within an $x$ or $y$ layer alternating
fibers are read out from alternating ends.  To improve light
collection efficiency, the exposed ends of the fibers are mirrored by
vacuum deposition of aluminum.  An LED-based light injection system
that flashes the exposed ends permits {\em in situ} calibration of
photosensor response, saturation, and nonlinearity (see
Section~\ref{sec:fgd_elec_lpb}). The photosensors are mounted along
all four sides of the XY module on busboards screwed directly into the
edges of the module.

In addition to seven XY modules, FGD2 also contains six water target
modules (described in Section~\ref{sec:watermodules}).  These are
sheets of thin-walled hollow corrugated polycarbonate, 2.5~cm thick,
whose ends have been sealed with polyurethane sealant.  The modules
are filled with water to provide a layer of water target.  The water
is maintained under subatmospheric pressure by a vacuum pump system so
that if a leak develops the system will suck air into the modules
rather than spilling water inside the FGD. Comparing the interaction
rates in the two FGDs permits separate determination of cross sections
on carbon and water.

Each FGD has outer dimensions of 230~cm (width) $\times$ 240~cm (height)
$\times$ 36.5~cm (depth in beam direction), and contains 1.1 tonnes of
target material.  They were built with the same geometry, mounting,
and readout for interoperability.

\begin{figure}[htb]
\centering
\includegraphics[angle=-90.,width=\columnwidth]{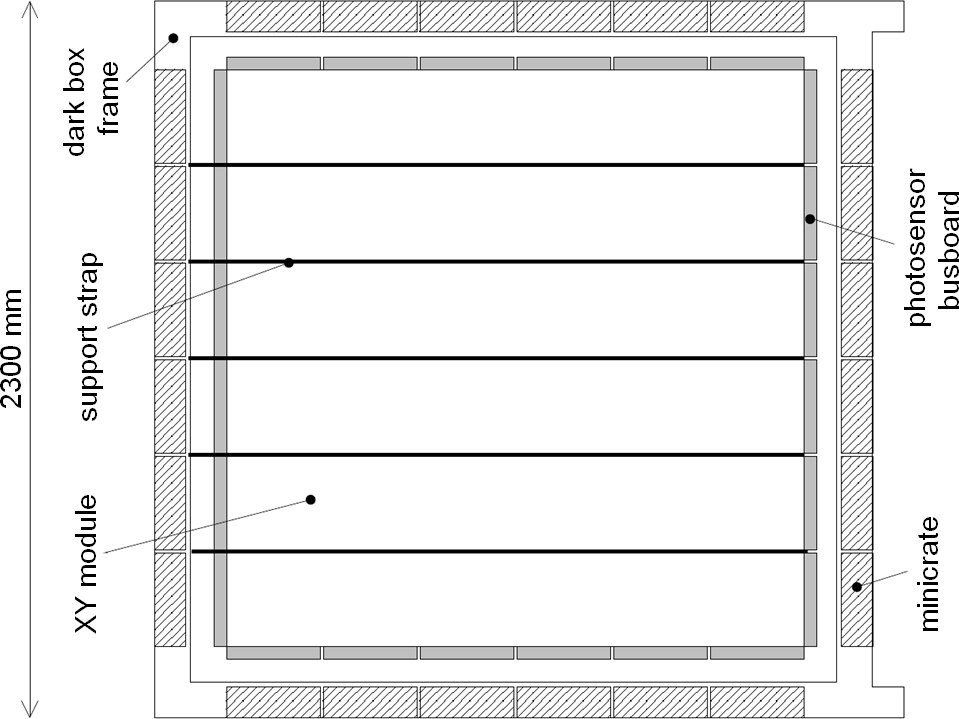}
\caption{Cross-sectional view of an FGD, showing the locations of the
scintillator modules, photosensors, support straps, electronics
minicrates, and dark box.}
\label{fig:fgd_schem}
\end{figure}

The XY modules and target water modules hang inside a lighttight box
called a ``dark box''.  Each module is supported by five stainless
steel straps that loop around the bottom of the module and attach to
anchor points in the top side of the dark box.  The dark box itself is
a sturdy aluminum frame that supports the weight of the FGD modules
and transfers that weight to the detector basket.  The walls of the
dark box are made of thin opaque panels to keep its interior
lighttight. Further details can be found in Section~\ref{sec:darkbox}.

The FGD's front-end electronics resides in 24 minicrates that attach
to the outside of the four sides of the dark box.  Signals from the
photosensors inside the dark box are carried from the photosensor
busboards to the electronics by ribbon cables that attach to the
crates' backplanes, which are mounted over apertures on the four sides
of the box.  The minicrates are cooled by a subatmospheric pressure water
cooling system running along the sides of the frame of the dark box,
and power is carried to the minicrates by a power bus mounted on the
frame.  The electronics is arranged so that all heat-producing
elements are located outside the dark box in the minicrates where they
can be readily cooled by the cooling system, while only elements with
negligible power dissipation (the photosensors themselves) are present
inside the dark box itself. The electronics modules in the minicrates
control the operation of the photosensors and process the signals from
them; details are presented in Section~\ref{sec:electronics}.

A cross-sectional view of an FGD, as viewed by the beam, is given in
Fig.~\ref{fig:fgd_schem}, showing the location of the XY modules,
photosensor busboards, dark box, and minicrates.

Digitized data from each minicrate are read out over optical fiber
links to Data Collector Cards (DCCs) located outside the magnet.  The
DCCs compress the data and pass it to the DAQ system (described in
Section~\ref{sec:daq}).  Slow control systems
(Section~\ref{sec:slowcontrol}) use a separate data and power bus for
redundancy.

This paper will review the design, construction, and testing of each
FGD component, briefly described in the foregoing, and will present
results on the detector performance from calibration efforts
(Section~\ref{sec:calibration}), beam tests
({Section~\ref{sec:m11tests}), and neutrino data-taking periods
(Section~\ref{sec:results}).

\section{Scintillator bars}
\label{sec:scint}

Fabrication of the scintillator bars and associated quality control
measurements are discussed in this section.

\subsection{Geometrical specification}

The geometrical design specifications of each bar are as follows:
\begin{itemize}
\item Length  = $1864.3 \pm 0.3$~mm 
\item Overall width and height = $9.61 \pm 0.2$~mm 
\item TiO$_2$ thickness = $0.25 \pm 0.05$~mm, (co-extruded bar coating)
\item Active dimensions = $9.1 \pm 0.2$~mm 
\item Hole diameter = $1.8 \pm 0.3$~mm   
\end{itemize}
The total number of bars produced, including $\sim$~3500 spares, was 11900.

Section~\ref{sec:xymodule} describes how the scintillator bars are
assembled into XY modules.

\subsection{Materials specification}

The scintillator bars were made of polystyrene doped with PPO and
POPOP, and were co-extruded with a reflective coating consisting of
polystyrene doped with TiO$_2$.  The materials used in extruding the
bars were as follows.  For the basic polystyrene, DOW Styron 663 W-27
general purpose crystal polystyrene in pellet form, without additives,
was used. The fluors were purchased already mixed from Curtiss
Laboratories (curtisslabs.lookchem.com). The primary fluor was 1~\%
PPO (2,5-diphenyloxazole) and the secondary fluor was 0.03~\% POPOP
(1,4-bis(5-phenyloxazol-2-yl) benzene). The TiO$_2$ concentrate was
purchased from Clariant (catalog number WHC-26311-A). This concentrate
has 60~\% TiO$_2$ (rutile form). With 1 part concentrate to 3 parts
plain polystyrene pellets by weight, the concentration of TiO$_2$ in
the coating was 15~\%.

\subsection{Mixing and extrusion procedures}

Preparations for fabricating the bars were carried out at TRIUMF.
The polystyrene pellets were weighed in 100~lb batches and dried for
$\sim$~8 hours at 170~$^\circ$F with dry N$_2$ flowing in the oven. The
pellets were then mixed in a small concrete mixer, which had dry N$_2$
from a liquid N$_2$ dewar flowing through it, with the premixed
PPO/POPOP fluor. This mixture was then stored in 200~lb containers
under dry N$_2$ purge. The TiO$_2$ coating (15~lb polystyrene to
5~lb TiO$_2$ concentrate) was mixed by hand and similarly dried and
stored. The materials, mixing and extrusion procedures were identical
to those utilized in producing extruded scintillator bars for
MINOS~\cite{PlaDalmau:2001en} and K2K SciBar~\cite{Nitta:2004nt}.

The mixed materials were transported to Celco Plastics Ltd, Surrey,
B.C., where they were purged with dry N$_2$, then fed into
hoppers. Separate hoppers were used for the doped plastic scintillator
extrusion and for the TiO$_2$ co-extrusion. The materials were heated
to a molten state and forced through the die and sizer plate, which
determine the bar cross section. The hole in the center of the bar was
formed by passing an air stream through a nozzle in the center of the
die.  The bar was then cooled by pulling it through a water bath and
cut to length by an automated saw which moved with the bar. A laser
feedback system controlled the bar thickness.

During the production run steps were taken to insure the bars met the
following specifications.
\begin{itemize}
\item Each bar was labeled with a bar code (bar code and number) just after
it was cut by the saw.
\item Burrs were removed from the holes and the edges of the bar using a
utility knife.
\item A stainless steel rod was threaded through the bar's hole to check for
blockages.
\item The bar was fed into a jig containing dial gauges to measure the
widths in the $x$ and $y$ directions.
\item The bar code was scanned and then a computer automatically read
out the $x$ and $y$ widths of the bar. This information was stored
with a time stamp.
\item If the bar was out of tolerance ($9.6 \pm 0.1$~mm) the attention of a
Celco operator was raised to investigate the change.

\item Every 1-2 hours a small ($\sim$~5~cm) piece of scintillator was
cut and polished using successively finer graded sandpaper. The piece
was placed into a holding jig so that a photo of the polished end
could be taken. Software to find the TiO$_{2}$ boundaries and hole
boundaries was used to measure the critical dimensions. Fig.\
\ref{fig:scint_fig1} shows a photo of the bar profile with the
measured dimensions for a typical bar.

\begin{figure}[htb]
\centering
\ifx\figstyle\bw
 \includegraphics[width=\columnwidth]{jpgFigures/fig03_bw.jpg}
\else
 \includegraphics[width=\columnwidth]{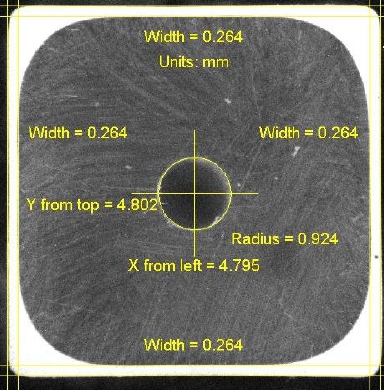} 
\fi
\caption{Photo taken with a CCD camera of a typical scintillator bar
produced at Celco Plastics. A
MATLAB$\textsuperscript{\textregistered}$ edge-finding routine is used
to find the edges of the TiO$_2$ to measure the TiO$_2$ thickness on
all 4 sides, and the edge of the center hole to measure the hole
diameter.}
\label{fig:scint_fig1}
\end{figure}

\item After every shift the bars produced during the shift were taken
back to TRIUMF and 4-8 bars out of every 100 were scanned on the bar
scanner (see section~\ref{sec:barscanner}) to make sure the light
yield was acceptable.
\end{itemize}

\subsection{Bar width results}

This section summarizes the results of the dial gauge measurements on
the bars as they came off the production run. There were four dial
gauges in total: two to measure the width of the bar and two to
measure the height. The resolution of the dial gauges was 0.01~mm. The
distributions are shown in Fig.~\ref{fig:scint_fig2}. The RMS of the
deviations ranges from 0.013~mm for the height to 0.019~mm for the
width.

\begin{figure}[htb]
\centering
\includegraphics[width=\columnwidth]{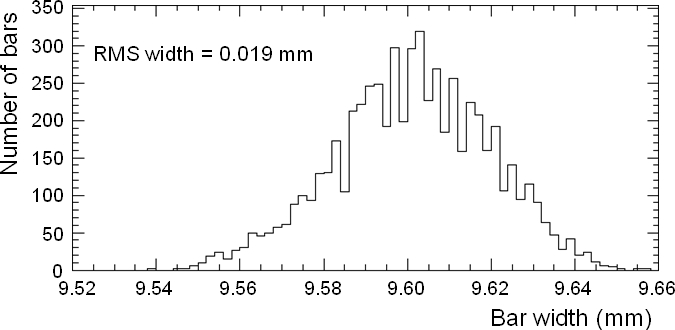}\\
~~\\
\includegraphics[width=\columnwidth]{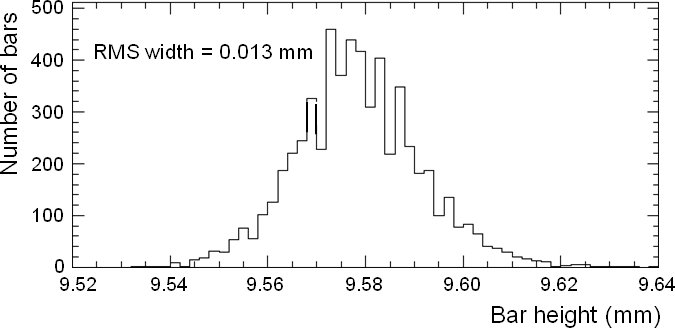}
\caption{Measured distributions of the mean widths and heights of
the  bars. Details of the measurements are given in the text.}
\label{fig:scint_fig2}
\end{figure}

\subsection{Bar scanner measurements}
\label{sec:barscanner}

In order to look for dead spots (areas of the bar which scintillate
less than the rest of the bar) and to compare light yields among the
bars as they came off the production line, the bars were scanned by
moving a $^{106}$Ru beta source along the length of a bar. The light
output from a WLS fiber threaded through the hole in the bar was
measured with an unbiased photodiode whose current was read out by a
Keithley picoammeter. When looking for dead spots, measurements were
made every 2~cm; when comparing light yields between bars they were
made every 50~cm. A special table and computer controlled movable
source holder were constructed so that these measurements could be
carried out reproducibly. The same fiber was used throughout. The
results of a typical scan are shown in Fig.~\ref{fig:scint_fig3}
together with an exponential fit to the data, which yields a
normalized light yield, I$_0$. During production, about 1 in every 4
bars was scanned and the distribution of
light yields was found to be Gaussian with a width of 4.5~\%.

\begin{figure}[htb]
\centering
\includegraphics[width=\columnwidth]{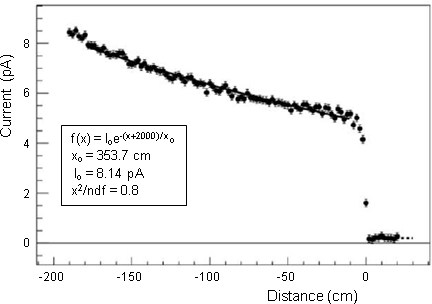}
\caption{Normalized light yield from a typical bar scan. Measurements
were taken with a $^{106}$Ru source placed at various locations along
the bar. See text for details.}
\label{fig:scint_fig3}
\end{figure}

\subsection{Light yield tests of bars}
\label{sec:lightyield}

The M11 secondary beam channel at TRIUMF was used to measure light
attenuation lengths and light yields from prototype scintillator bars
extruded by Celco about 6 months before the main production run. The
channel is described in Section~\ref{sec:m11area}. Electrons, muons
and pions of 120~MeV/c were used and the light yield for each particle
type was measured as a function of distance of the beam from the
photosensor. For these tests a Russian (CPTA) silicon photosensor was
used.  The data are shown in Fig.~\ref{fig:scint_fig5}.

\begin{figure}[htb]
\centering
\ifx\figstyle\bw
 \includegraphics[width=\columnwidth]{jpgFigures/fig06_bw.jpg}
\else
 \includegraphics[width=\columnwidth]{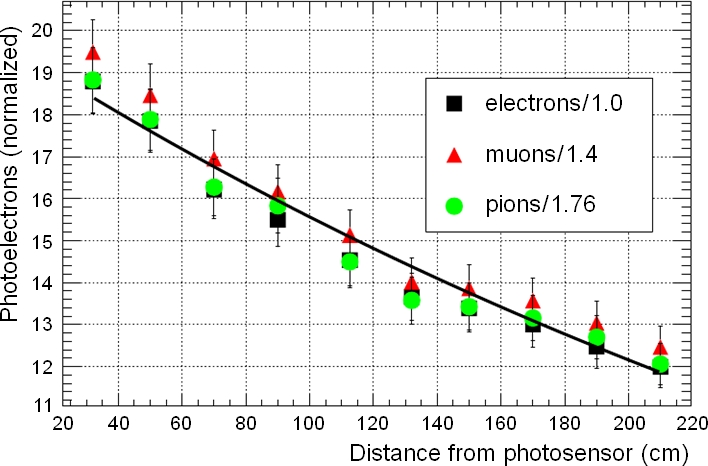}
\fi
\caption{Light yield from a bar for 120~MeV/c particles read out with
1.0~mm Kuraray un-aluminized fiber and a Russian (CPTA) silicon
photomultiplier photosensor, as a function of distance from
photosensor. The data have been normalized by dividing the light
yields by dE/dx for each particle relative to a minimum ionizing
particle.}
\label{fig:scint_fig5}
\end{figure}

Measurements with fibers with reflecting ends showed that the light
yield at the end of the bar furthest from the sensor could be
increased by 30\,-\,40~\% by aluminizing the end of the fiber or by adding
a reflecting cap. Thus aluminizing the far end of a fiber raised the
expected light yield from the far end of the bar up to 16\,-\,18
photoelectrons (pe) for a minimum ionizing particle.  For comparison,
this level of light exceeds that seen in the SciBar detector, which
saw $\sim$~16.5~pe/cm for a minimum ionizing particle incident at the
end \emph{nearest} the photosensor~\cite{Nitta:2004nt}.

\subsection{Optical crosstalk}

Optical crosstalk through the TiO$_2$ coating between bars was
investigated by measuring the amount of light seen in a bar when an
adjacent bar was illuminated with 400~MeV/c protons or with an
LED. The results of these measurements were consistent with each other
and show that, with a TiO$_2$ thickness of 0.25~mm, $0.5 \pm 0.2$~\%
of the light produced in one bar is transmitted through the coating
into an adjacent bar.

Because the FGD uses one photosensor per fiber, there is no crosstalk
between scintillator bars inside the photosensors themselves.

\subsection{Aging tests}

Fermilab estimated that aging of bars produced for MINOS would reduce
the light yield by $\sim$~2~\% per year~\cite{Choudhary:2000ac}. To
check for aging of our bars, we exposed a scintillator bar and
wavelength-shifting fiber to a $^{90}$Sr source and measured the light
yield with a photodiode, comparing our bars to samples of Fermilab
bars with both at room temperature. We also attempted to accelerate
the aging by heating samples of the bars in an oven to 45~$^\circ$C,
60~$^\circ$C and 85~$^\circ$C. The results of these tests may be
summarized as follows:

\begin{enumerate}

\item Scaling the 60 degree and 85 degree aging rates with the
Arrhenius equation predicted a loss of light output of $\sim$~0.2~\% a
day or 75~\% in a year. These results were rejected because the
temperature was too close to the scintillator glass transition
temperature.

\item A light loss of $\sim$~3.93~$\pm$~1.60~\% per year was
obtained directly from the room temperature data in
Fig.\ \ref{fig:scint_fig7} top.

\item A loss of $\sim$~1.93~$\pm$~0.25~\% per year was predicted
from using the Arrhenius scaling rule to the 45 degree data in
Fig.\ \ref{fig:scint_fig7} top.

\item The change in the ratio between the light output of the TRIUMF
bars and the Fermilab bars at room temperature over a 250 day period
is consistent with zero (see Fig.\ \ref{fig:scint_fig7} bottom).

\item The light output from 120 MeV/c muons measured over
about a 1 year period showed no evidence of light loss.

\end {enumerate}

The results of 2), 3) and 4) are all consistent with an aging rate of
about $\sim$~2~\% per year, the same as obtained by
Fermilab~\cite{Choudhary:2000ac}.

\begin{figure}[htb]
\centering
\ifx\figstyle\bw
 \includegraphics[width=\columnwidth]{fig07a_bw.jpg}\\
\else
 \includegraphics[width=\columnwidth]{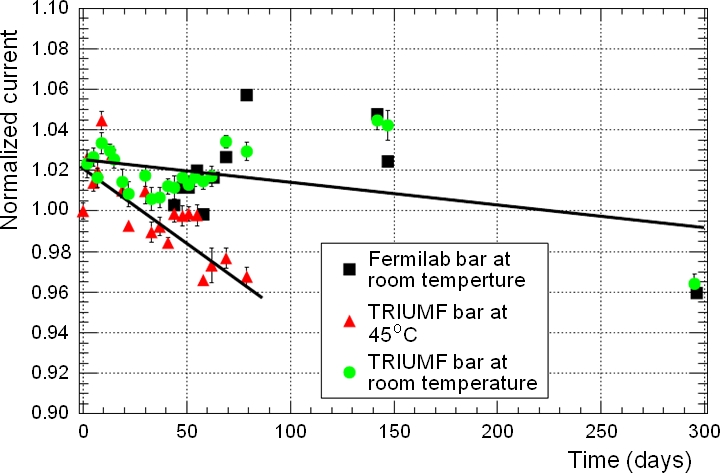}\\
\fi
~~\\
\includegraphics[width=\columnwidth]{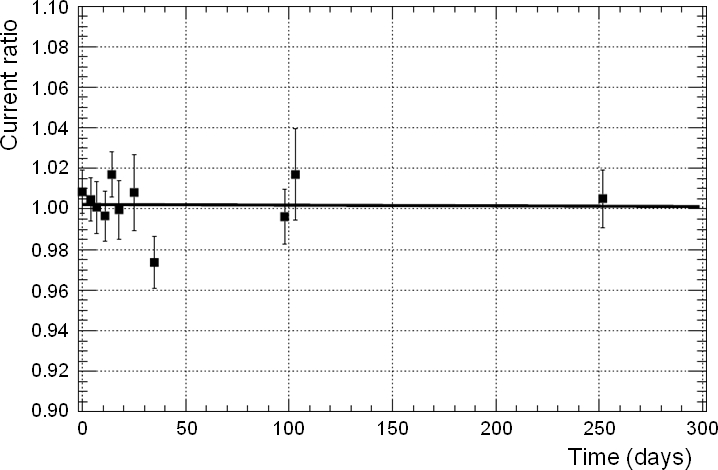}
\caption{Scintillator bar aging study. (Top) Normalized photosensor
current as a function of time from a scintillator bar exposed to a
$^{90}$Sr source. (Bottom) Ratio of current from TRIUMF bar to current
from Fermilab bar as a function of time.}
\label{fig:scint_fig7}
\end{figure}

\section{XY Module assembly}
\label{sec:xymodule}

Each XY module is a glued sandwich consisting of 192 active
scintillator bars in the $x$ direction glued to 192 active bars in the
$y$ direction.  Skins of 0.25~mm thick G10 (obtained from Current
Inc.) are glued to both the upstream and the downstream surface of the
XY module to provide additional mechanical rigidity and to allow
handling of the modules without stressing the glue joint between the
$x$ and $y$ layers.  Each layer of an XY module actually contains 194
bars, but the top edge of the outermost bar on each side is machined
off to provide a surface for mounting the photosensor busboards to the
scintillator layer.  Thus the outermost bar does not contain a WLS
fiber and is only half height.  The outer dimensions of the module are
186.4~cm $\times$ 186.4~cm $\times$ 2.02~cm, not counting the
photosensor busboards.  The twenty-four photosensor busboards screw
onto the four sides of the XY module and support the photosensors and
fibers (see Fig.~\ref{fig:mech1}). Each XY module is supported from
above by five stainless steel straps that wrap under the bottom of the
module, attaching to mounting pads on the bottom of the module and to
the top inner surface of the dark box (see
Section~\ref{sec:modulesupport}).

\subsection{Glue tests}

A large number of tests were carried out to find a suitable adhesive
to bond the scintillator bars together.  The desired adhesive needed
to be able to adhere to both polystyrene and G10. The bonds also had
to be resistant to peeling forces which might arise, for example,
during handling of a glued layer of bars. Thus one test was a simple
snap test in which two pieces of scintillator bar were glued together
along part of their length and it was seen whether the scintillator or
the bond joint failed when a torque was applied to the joint. A
similar test was carried out in which a layer of G10 was glued to a
scintillator bar to see if it was possible to peel the two apart.
Many adhesives failed one or the other of these tests.

The other important characteristic of the adhesive was that it needed
to have a long enough working life that the component materials could
be arranged in position before the glue set. Instant bonding type glues
were not an option for our construction process.

A wide search of different types of adhesive to locate any likely
candidates was carried out. Only two adhesives met all the
requirements - Plexus MA560 and MA590 (obtained from ITW Plexus).
Even for these glues it was found that although they had satisfactory
strength and pot life, they lost strength rapidly from the time they
were prepared if not applied promptly (they are two-part adhesives),
so much so that within 10 to 15 minutes they both failed the snap test
described above. We chose to use MA590 since it had a longer pot life,
but nevertheless it was necessary to carry out the gluing procedure as
quickly as possible.

\subsection{Assembly procedure}

The XY layers were assembled in the largest TRIUMF Detector Facility
clean room. There were three separate bonding steps in which first one
layer was glued, then the second, and finally the two were bonded
together.  The procedure for bonding an individual layer was to align
196 bars in a jig on a granite table, apply adhesive to the top side
of the G10, then flip it over and lay it on top of the bars. Slightly
wider bars were paired with slightly narrower bars to avoid tolerances
from wider or narrower bars from accumulating. Before applying
adhesive, the bar codes were scanned and stored in a database. Thus
the location of each particular bar in the FGDs is known. Also note
that no adhesive was applied between the bars. The G10 sheet was
actually composed of two separate sheets, each covering half of each
bar, for ease of handling. When the G10 sheets were in place, an
aluminum cover plate was placed on top of the layer, four aluminum I
beams were placed over the plate and attached by turnbuckles to a
hydraulic system that was activated to pull down on the beams,
providing pressure while the glue set. The whole procedure took less
than 15 minutes from the moment the glue was mixed. The bond was
allowed to set overnight. See Fig.~\ref{fig:gluingjig} for a photo of
a layer lying on the granite table with the pressure applied.

\begin{figure}[htb]
\centering
\ifx\figstyle\bw
 \includegraphics [width=\columnwidth]{jpgFigures/fig08_bw.jpg}
\else
 \includegraphics [width=\columnwidth]{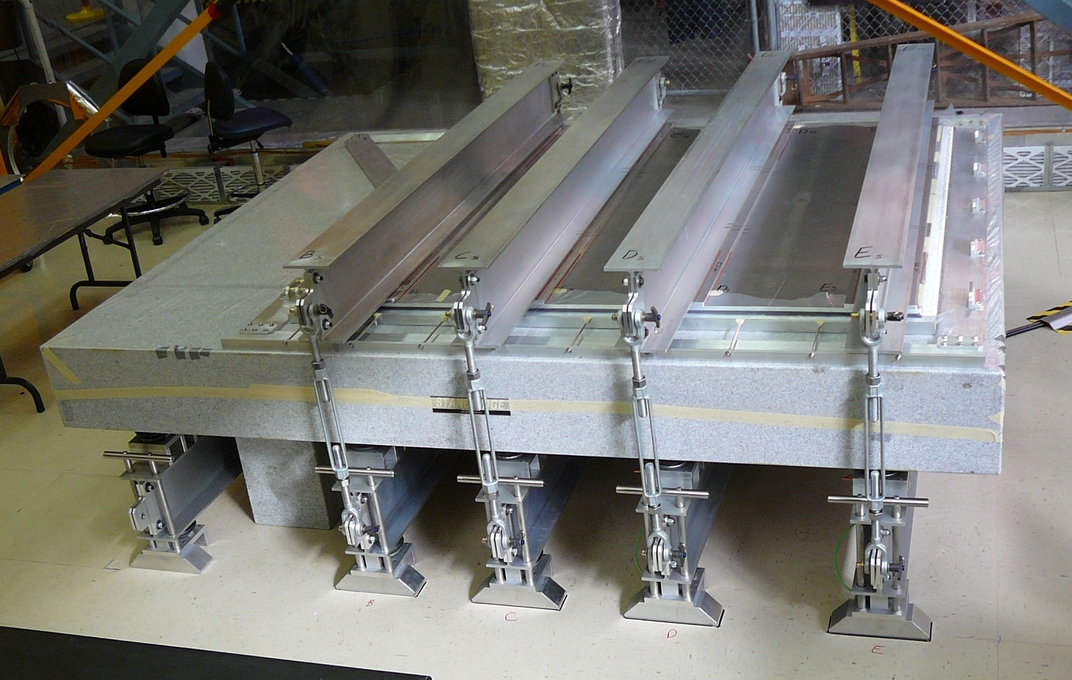}
\fi
\caption{A layer of scintillator bars after the glue has been applied,
the G10 laid over top, the cover plate and I beams put in place and
hydraulic pressure applied.}
\label{fig:gluingjig}
\end{figure}

The procedure for bonding one layer to the other was to set one layer
into the jig, scintillator side up, and apply a layer of adhesive to
the exposed bars. Then the other layer was craned over top,
scintillator side down and with the bars aligned at 90$^\circ$ to the
other layer. The jig ensured the layers remained at right angles as
the top layer was lowered.  Finally the aluminum plate, I beams and
hydraulic system were deployed as before.

\subsection{Final preparation}

Upon completion of the gluing procedure, the module was moved to a
router table for machining to final dimensions. The same router was
used as is described in~\cite{tpc:nim}. First, the outermost bar on
all four edges of the module was machined off.  Then the top half of
the next bar on all four edges was removed to provide a surface for
attachment of the bus boards, and the screw holes to hold the boards
were drilled.  Finally, the module was deburred by hand to remove any
sharp edges.

After the photosensor bus boards and daughter cards were mounted, the
WLS fibers were slid through the holes in the bars and attached to the
MPPCs. A ferrule was pre-glued onto one end of each fiber so that it
was only necessary to slide the ferrule into a coupler in front of the
MPPC (see Fig.~\ref{fig:MPPC-coupler}).  Finally, the stainless steel
straps from which the XY module hangs were attached at five locations
across the bottom of the FGD module.  The XY module was then ready for
installation inside an FGD dark box (see Section~\ref{sec:darkbox}).

\subsection{Elemental composition}

The elemental composition of the modules was determined from the
measured geometry of the scintillator bars and their known
composition, and the measured amounts of MA590 and G10. The
compositions of the latter two materials are not known precisely but
under reasonable assumptions the uncertainties do not contribute
significantly to the overall uncertainty of each element. A
consistency check was made by weighing the modules after machining and
by taking account of the material that had been removed, the areal
density agreed with that calculated from the individual components
within 0.5~\%.

After final assembly it was also necessary to take into account the
contribution from the wavelength shifting fibers. These were assumed
to be composed of 100~\% polystyrene.

The final areal densities of each of the elements contained in the
modules are presented in Table~\ref{tab:water_comp} in
Section~\ref{sec:watermodules}.

\section{Wavelength shifting fibers}
\label{sec:fiber}

The WLS fiber chosen was a 1 mm diameter, double-clad Kuraray Y11
(200) S-35 J-type fiber.  In these fibers, the peak absorption
wavelength is 430~nm and the subsequent emission spectrum is peaked at
476~nm.  The capture and transmission of light in the fiber are
improved with double cladding with the following indices of
refraction: 1.59 (core); 1.49 (inner clad); 1.42 (outer clad).  S-type
fibers are mechanically stronger at some expense to the light
transmission.

Twenty-one kilometers of WLS fibers were obtained from Kuraray in
2.1~m lengths. Random samples of the fibers representing 10~\% of the
total were visually inspected for physical defects upon receipt. The
diameter of the fibers were found to be 1~mm~$\pm$~2~\%, consistent
with the Kuraray specifications. Only 0.80~\% of the inspected
fibers were found to have some problem.

Subsequently, the fibers were sent to the Fermi National Accelerator
Laboratory (FNAL), where one end of each fiber was mirrored for
enhanced light collection.  The mirroring technique consists of
aluminizing the fiber end by vacuum thin-film sputtering deposition,
and then coating with Red Spot UVBT115R5, a one-part epoxy that cures
with UV, for protection. The epoxy extends about 0.15~mm from the
fiber end, while the radial thickness with respect to the surface of
the fiber is less than 0.1~mm. Preliminary measurements with pions,
muons and electrons using the M11 test beam facilities at TRIUMF
demonstrated a substantial gain in the light yield for pulses
originating at the mirror end of the fiber, after mirroring the fibers
(see section~\ref{sec:lightyield}).

Upon their return from FNAL, the mirrored fibers were cut and polished
to a length of 191.5~cm. Subsequently their attenuation properties
were measured with a custom fiber test device (see
section~\ref{sec:fibertester}).  Half the fibers were tested after the
FGD fiber-MPPC coupler (described in section~\ref{sec:fibercoupler})
was glued onto the unmirrored end, and half were tested before the
coupler was glued on.  The two groups of fibers were analyzed
separately; the fibers measured before the connector was glued on are
designated as group 1 fibers and those measured after are designated
as group 2 fibers.

\subsection{Preparation}

A cutting jig was built to shear the fibers slightly longer than the
required 191.5~cm. After shearing, a FiberFin model FF4 diamond fiber
polisher was used to cut and polish to within 1~mm of the specified
length.

A CCD camera looking through a measuring microscope was used to
inspect the fiber end and to verify the length of the finished fiber,
which could easily be measured to within 0.1~mm.

Bicron BC600 glue was chosen to fix the coupler to the fiber.  This
glue was selected because of its characteristics - bond strength,
curing time (workable for about 4 hours at 21~$^\circ$C), slow
aging effect, etc.\ - and also because it is intended to be used with
optical fibers, and therefore is not expected to harm the fibers
chemically.

A gluing jig was developed to ensure the fiber ends were flush with the
coupler end so that the gap between fibers and light sensors is
minimal and reproducible (see Fig.~\ref{fig:glue}). In order to avoid
glue flowing towards the fiber end or out of the coupler, the glue
was allowed to react for approximately two hours before use in order
to increase its viscosity.  An EFD Ultra 2400 Glue Dispensing
workstation was used to provide appropriate small quantities (0.05~ml
of this glue mix) to each fiber, before insertion into the
coupler in the jig.  After 24~h curing time, every glue joint was
tested for strength to withstand 0.5~N force (more than the force
required to remove a coupler from the detector).

\begin{figure}[htb]
\centering
\ifx\figstyle\bw
 \includegraphics[width=\columnwidth]{jpgFigures/fig09_bw.jpg}
\else
 \includegraphics[width=\columnwidth]{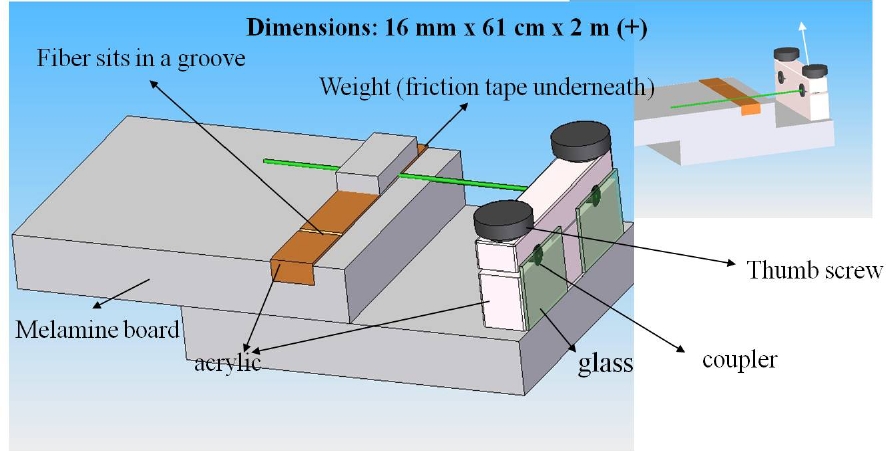}
\fi
\caption{Schematic diagram of the jig used to glue the couplers to
the end of the fibers.}
\label{fig:glue}
\end{figure}

\subsection{Fiber tests}
\label{sec:fibertester}

All fibers used in the FGDs were tested after mirroring.  The fiber
tester is comprised of light sources, light detectors, mechanical
support for the fibers, computer control, multiplexer and a
picoammeter.

The light sources are nineteen 1.5 candela, 5~mm diameter UV
light-emitting diode (LED) assemblies.  In each assembly, the LED
emits light into five clear fibers which fan out to inject light
throughout a piece of FGD scintillator bar whose dimensions are 1~cm
$\times$ 1~cm $\times$ 10~cm. A sixth clear fiber was routed from each
LED to a common large area silicon pin-photodiode to monitor and
correct for LED light intensity variations.  There are ten grooves cut
perpendicular to the long axis of the scintillator bar that serve to
hold the fibers being tested.  The assemblies are located at 10~cm
intervals along a dark box. Darkened aluminum plates fix the spacing
between the bars and support the fiber at the height of the bottom of
the grooves.  The arrangement of the LED assemblies is shown in
Fig.~\ref{fig:tester}.

\begin{figure}[htb]
\centering
\ifx\figstyle\bw
 \includegraphics[width=\columnwidth]{jpgFigures/fig10_bw.jpg}
\else
 \includegraphics[width=\columnwidth]{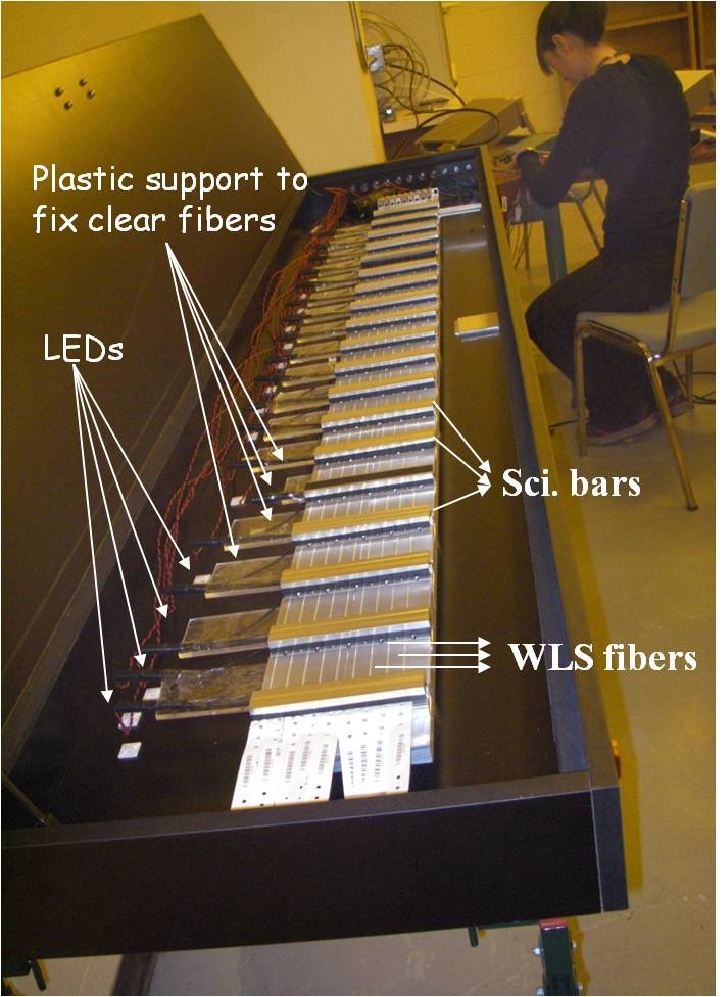}
\fi
\caption{Photo of the fiber tester, showing several light sources and
their corresponding fiber optics and grooved scintillator bars.}
\label{fig:tester}
\end{figure}

An SHE Corporation constant current power supply (CCS) was used to
drive the LEDs, one at a time.  It was found that LED lifetime was
prolonged when the pattern to turn an LED off or on included steps
where the LED was switched while in parallel with a dummy load to
minimize the voltage swing of the CCS.

Ten light-detector elements were positioned to couple to the fibers in
the grooves.  The detector elements were arranged with Ocean Optics
bare-fiber couplers for all group~1 measurements (when testing
preceeded gluing) and with standard FGD fiber couplers for all group~2
measurements (when gluing preceeded testing).  In either case, each
channel was instrumented with a Hamamatsu S1336-18BQ silicon
pin-photodiode, and typically nine of these channels were used, so
that a separate data channel with a constant WLS fiber was available
as reference for normalization purposes at all times.

Calibration of the system was necessary to compensate for variations
in the fiber-to-groove optical coupling at every measuring location.
This was accomplished using a short segment of fiber attached to a
single, movable detector, used at every LED-groove intersection, one
at a time.

Operation of the tester was fully computer controlled using a GPIB
interface and Keithley 7001-S control mainframe with Keithley 7058
scanner cards to switch both LED power and photodetector channel.
Under program control, one LED would be turned on and one fiber
channel read out at a time.  The current from each light detector was
determined with a Keithley 6487 picoammeter, operating with sampling
and filtering. Measurements were repeated and data written to disk for
later analysis. The reproducibility of the measurements was found to
be improved when the LEDs were cycled at the same rate as in data
taking at all times, even between measurements.

\subsection{Attenuation analysis}

A typical attenuation curve for a mirrored fiber after data
calibration and LED light variation correction is shown in
Fig.~\ref{fig:attn}.  The attenuation curve can be empirically
described using equation \ref{eq:attn}:

\begin{equation}
 I  = A\,e^{- x/L} + B\,e^{- x/S}
\label{eq:attn}
\end{equation}
 
\noindent where $L$ and $S$ are referred to as the long and short
light attenuation lengths respectively, and the ratio of $B$ to $A$
indicates the relative contribution of each term. In this expression,
$I$ is the current from the photodetector and $x$ is the distance from
the source to the light detector.

\begin{figure}[htb]
\centering
\includegraphics[width=\columnwidth]{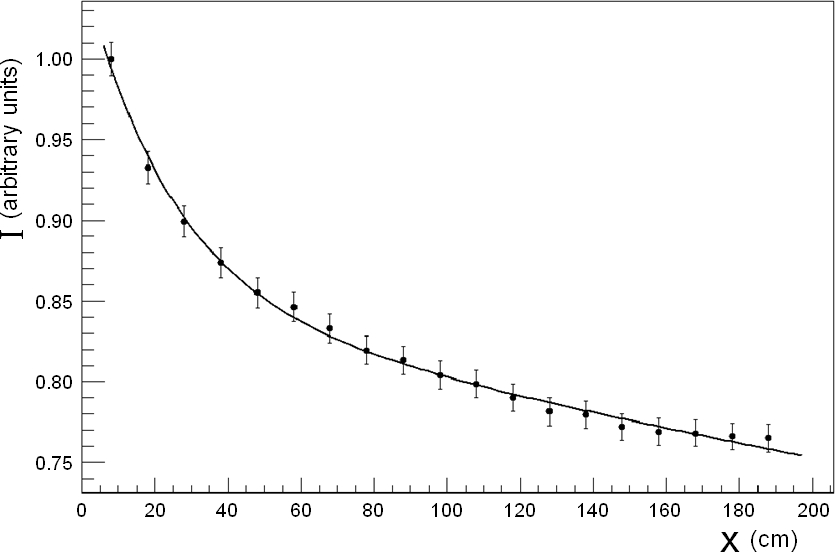}
\caption{A typical attenuation curve for a mirrored fiber. The data
are fit using equation~\ref{eq:attn}.}
\label{fig:attn}
\end{figure}

The attenuation lengths and mean values of $A$ and $B$ are given in 
Table~\ref{tab:fiberTable1}.

\begin{table}[h]
\centering
\caption{Values of the long and short attenuation lengths and their
amplitudes for the two groups of fibers.}
\vspace{2mm}
\footnotesize{
\begin{tabular} {ccccc} \hline
 & $L$ ($\sigma_L$) & $S$ ($\sigma_S$) & $A$($\sigma_A$) & $B$($\sigma_B$) \\ \hline
 &(m) & (m) &  (nA) & (nA) \\
 group 1 & 16.0(3.2) & 0.236 (0.042)&1.47(0.08)& 0.45(0.04) \\
 group 2 & 16.0(3.1) &0.258(0.041)&1.19(0.06)&0.27(0.03) \\
\hline
\end{tabular}
}
\label{tab:fiberTable1}
\end{table}

There were strong correlations among $L$, $S$, $A$ and $B$ such that
they are not robust parameters to test the quality of fibers.  On the
other hand, variables such as the ratio $B/A$ and the comparative light
yield relative to the reference fiber are of particular interest to
the quality control of the fibers as they are expected to be constant
from fiber to fiber.  These two distributions are shown in Figs.\
\ref{fig:quali1} and \ref{fig:quali2}.  The comparative light yield is
the average over all 19 ratios of yield in an individual fiber being
tested to the corresponding yield for the reference fiber.  Numerical
values are given in Table~\ref{tab:fiberTable2}.

\begin{figure}[htb]
\centering
\includegraphics[width=\columnwidth]{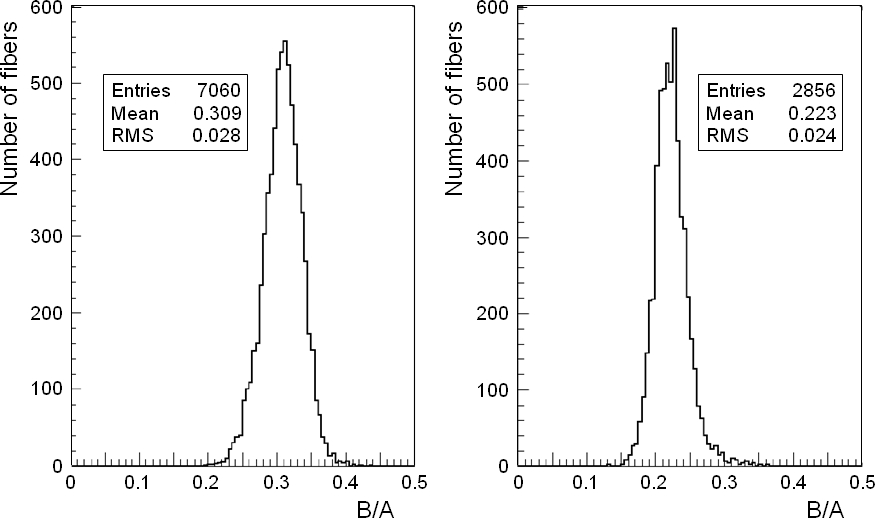} 
\caption{Distribution of the ratio $B/A$ for data group 1 (left) and data 
group 2 (right).}
\label{fig:quali1}
\end{figure}

\begin{figure}[htb]
\centering
\includegraphics[width=\columnwidth]{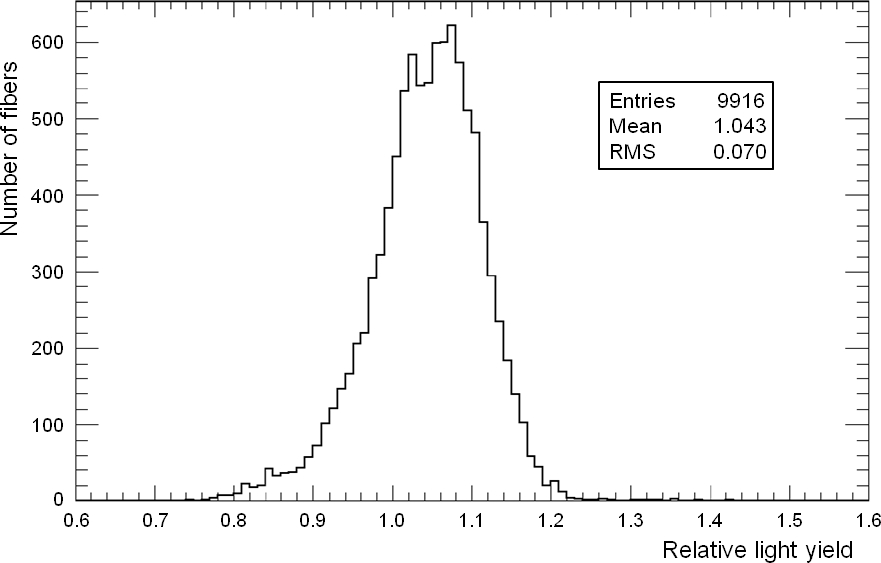}
\caption{Combined distribution (groups 1 and 2) of the relative light yield
between the tested and reference fibers.}
\label{fig:quali2}
\end{figure}
 
\begin{table}[h]
\centering
\caption{Ratios of the amplitudes for the long and short attenuation
lengths for the two groups of fibers and the overall light yield
relative to the reference fiber.}
\vspace{2mm}
\footnotesize{
\begin{tabular}{cccc}
\hline
  & mean  & RMS  & RMS/mean ($\%$)  \\ \hline
 $B/A$ (group 1) & 0.309  & 0.028   & 8.98   \\
 $B/A$ (group 2) & 0.223  & 0.024   & 10.9  \\
{Relative light yield} & 1.043 & 0.070  & 6.72 \\
\hline
\end{tabular}
}
\label{tab:fiberTable2}
\end{table}

Only 46 fibers were finally rejected on the basis of one or more cuts
on the ratio of light output to that of the reference fiber (25
failed), chi-square of the fit to the attenuation curve (15), ratio
$B/A$ (7), and the long and short attentuation lengths (17 and 13,
respectively)~\cite{Licciardi:thesis}.

\section{Photosensor}
\label{sec:photosensor}

The photosensor for the FGDs needs to satisfy tight constraints:
\begin{itemize}
 \item It must be able to count photons down to a few photoelectron level.
 \item It must work in the 0.2~T magnetic field.
 \item It must fit in a very tight space constraint.
\end{itemize}

The Multi-Pixel Photon Counter, or MPPC~\cite{Yokoyama:2006nu,
Gomi:2007zz} , manufactured by Hamamatsu Photonics~\cite{Hamamatsu},
meets all these requirements and was chosen for all the scintillator
detectors in ND280, including the FGDs.

\subsection{MPPC description}

The MPPC is a pixellated avalanche photodiode (APD) that operates in
Geiger mode. A picture of the MPPC used (S10362-13-050C) is shown in
Fig.~\ref{fig:MPPC} and the major specifications are summarized in
Table~\ref{tab:MPPC_spec}.  The outer dimensions of the package are
5~mm $\times$ 6~mm.  The sensitive area of the MPPC is enlarged from
the 1 $\times$ 1~mm$^2$ of those shown in the catalog to 1.3
$\times$ 1.3~mm$^2$ to minimize the light loss at the optical contact.
The size of one APD pixel is 50 $\times$ 50~$\mu$m$^2$ and the number
of APD pixels is 667. A single pair of leads provides for both the
operating voltage and signal readout.

For an MPPC, each APD pixel independently works in limited Geiger mode
with an applied voltage just above (usually $<$ 1 V) the breakdown
voltage ($V_{bd}$).  When a photoelectron is produced, it induces a Geiger
avalanche that is passively quenched by a resistor integrated to each
pixel.  The output charge $Q$ from a single pixel is independent of
the number of produced photoelectrons within the pixel, and can be
written as $ Q = C (V-V_{bd}) \equiv C \delta V, \label{eq:gain-Vdep} $
where $V$ is the applied voltage and $C$ is the capacitance of the
pixel.  The overvoltage, $\delta V \equiv V-V_{bd}$ is the key parameter
that controls the performance of MPPCs.  Combining the output from all
the pixels, the total charge from an MPPC is quantized to multiples of
$Q$ and proportional to the number of pixels that underwent Geiger
discharge.  The number of discharged pixels is proportional to the
number of incident photons if the number of incident photons is small
compared to the total number of pixels.  Thus, the MPPC has an
excellent photon counting capability as long as the number of
photoelectrons does not approach the total number of pixels in the
device.  The MPPC is particularly suitable for the FGDs because of its
excellent photon counting capability with higher quantum efficiency
than photomultipliers for the wavelength distribution produced by the
WLS fibers.

\begin{figure}[htb]
\centering
\includegraphics[width=\columnwidth]{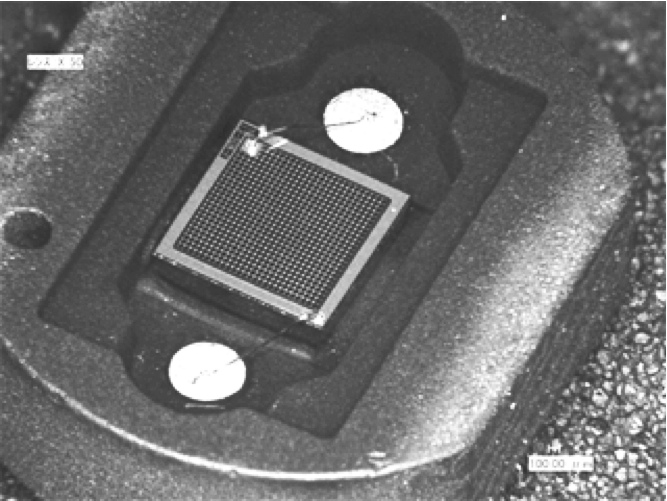}
\caption{Picture of MPPC developed for T2K.}
\label{fig:MPPC}
\end{figure}

\begin{table}[htb]
\centering
\caption{Specifications of T2K-MPPC (S10362-13-050C).}
\vspace{2mm}
\footnotesize{
\begin{tabular}{ccc} \hline
\multicolumn{2}{c}{Item} & Spec. \\ \hline
\multicolumn{2}{c}{Active area} & 1.3$\times$1.3~mm$^2$ \\
\multicolumn{2}{c}{Pixel size} & 50$\times$50~$\mu$m$^2$ \\
\multicolumn{2}{c}{Number of pixels} & 667 \\
\multicolumn{2}{c}{Operation voltage} & 70~V (typ.) \\
\multicolumn{2}{c}{PDE @ 550~nm} & $>$15\% \\
Dark count &($>$0.5~pe) & $<$1.35~Mcps\\
$[$ @ 25 $^\circ$C $]$  & ($>$1.2~pe) & $<$0.135~Mcps \\
\hline
\end{tabular}
}
\label{tab:MPPC_spec}
\end{table}

\subsection{MPPC tests}

All the MPPCs used in the FGDs were tested before integration with the
detectors.  The details of the procedure and the results are described
in \cite{Yokoyama:2010qa}.  In order to characterize a large number of
MPPCs, a system was developed that simultaneously measures 64 MPPCs
with various bias voltages and temperatures.  The gain, dark noise
rate, photodetection efficiency (PDE), crosstalk and after-pulsing
probability of 17,686 MPPCs (including those to be used for other
detectors in ND280 besides the FGDs) were measured as functions of the
operating voltage and temperature.

Table~\ref{tab:MPPC_summary} summarizes the mean value and RMS of
measured performance for these MPPCs at 20~$^\circ$C and
$\Delta V = 1.0 V$.  All the measured MPPCs satisfy the
requirements for the FGDs.

\begin{table*}[!t]
\centering
\caption{Mean value and RMS of gain, dark noise rate, after-pulsing
and cross talk probability, and photodetection efficiency (PDE) for
17686 MPPCs at 20\,$^\circ$C and $\Delta\,V$=\,1.0\,$V$.}
\vspace{2mm}
\footnotesize{
\begin{tabular}{cccc}
\hline
Parameter  & Measured values & RMS & Value/RMS (\%) \\
\hline
Gain    &  $4.85 \times 10^{5}$ & $0.26\times 10^{5}$ & 5.4 \\
Breakdown voltage (V)   & 68.29 &  0.73 & 1.1 \\
Dark noise rate  (Hz)   & $4.47 \times 10^{5}$ & $1.02\times 10^{5}$ & 22.8 \\
After-pulsing and
cross talk  probability                  & 0.070 &  0.036 & 51 \\
Relative PDE    ($\times$PMT)           & 1.53 &  0.33 & 22 \\
\hline
\end{tabular}
}
\label{tab:MPPC_summary}
\end{table*}

\subsection{Coupling to WLS fiber}
\label{sec:fibercoupler}

In order to achieve good optical coupling between a WLS fiber and an
MPPC, a custom connector was developed~\cite{Kawamuko:2006zz}.
As shown in Fig.~\ref{fig:MPPC-coupler}, the connector consists
of two parts: one part holds the MPPC at the bottom of the funnel
(right), and the other part is glued to the fiber (left). The two
parts can be latched together without glue or screws but just by a mechanical
force so that the connection is simple while well-aligned and robust.
The outer dimensions of the coupler are about 8~mm in diameter and
17.5~mm in length when the two parts are connected, fulfilling the
space constraint from the mechanical design.

\begin{figure}[htb]
\centering
\includegraphics[width=\columnwidth]{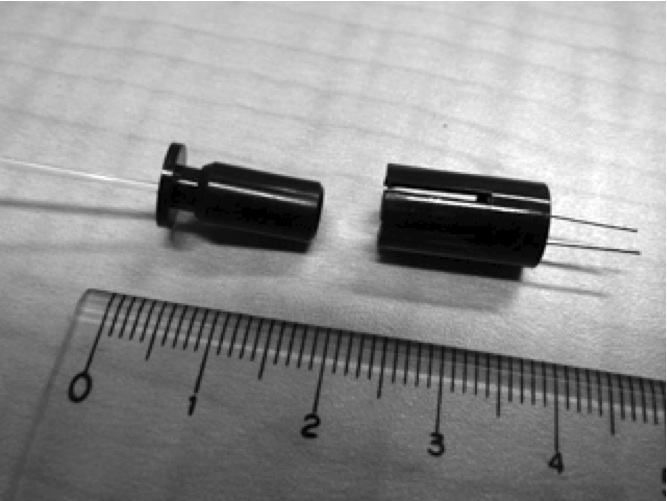} 
\caption{Coupler to connect MPPC (right) and WLS fiber (left).}
\label{fig:MPPC-coupler}
\end{figure}

The light loss due to coupling was measured with this coupler, a 1~mm
diameter WLS fiber, and a prototype MPPC with a 1 $\times$ 1~mm$^2$ active
area.  The measured light loss was 13~\%~\cite{Kawamuko:2006zz}.  The light
loss with the final MPPC is expected to be less than this because
of the enlarged active area.

\section{Target water system}  
\label{sec:watermodules}

The second FGD alternates scintillator modules with $\sim 25$~mm
thick polycarbonate panels containing water~\cite{Roberge-thesis}.
These panels, maintained under subatmospheric pressure to prevent
water leaks, serve as targets for studying neutrino interactions on
water.  The panels are constructed so that their elemental
compositions nearly exactly match those of a mixture of water and
polystyrene scintillator, allowing the total interaction rates on
water to be determined by measuring the total interaction rate in FGD2
and subtracting out the interaction rate on polystyrene as measured in
FGD1.

\subsection{Water panel description}

The water modules for the FGD were built from rigid, hollow
polycarbonate panels, originally designed for use as wall material for
greenhouses.  These corrugated panels are 25.4~mm thick, 1809~mm wide,
and are extruded in long sections that are cut to length.  The panels
were constructed from Sunlite$\textsuperscript{\textregistered}$
polycarbonate multiwall sheets produced by the Palram company.
Polycarbonate is a strong, lightweight, waterproof material ideal for
making water panels. The panels are divided into long cells by regular
internal walls that provide rigidity and strength and also help to
maintain the shape of the panel and prevent bulging even when filled
with water. Furthermore, their oxygen content is such that it permits
matching the elemental compositions of a water-filled panel with a
mixture of water and polystyrene scintillator.

Figure~\ref{fig:panel_xsec} shows a schematic view of the internal
wall structure of a panel.  The long internal cells of the panel are
oriented vertically inside the detector, and water continuously flows
in the bottom and out the top under the control of the subatmospheric
pressure system.

\begin{figure}[htb]
\centering
\includegraphics[width=\columnwidth]{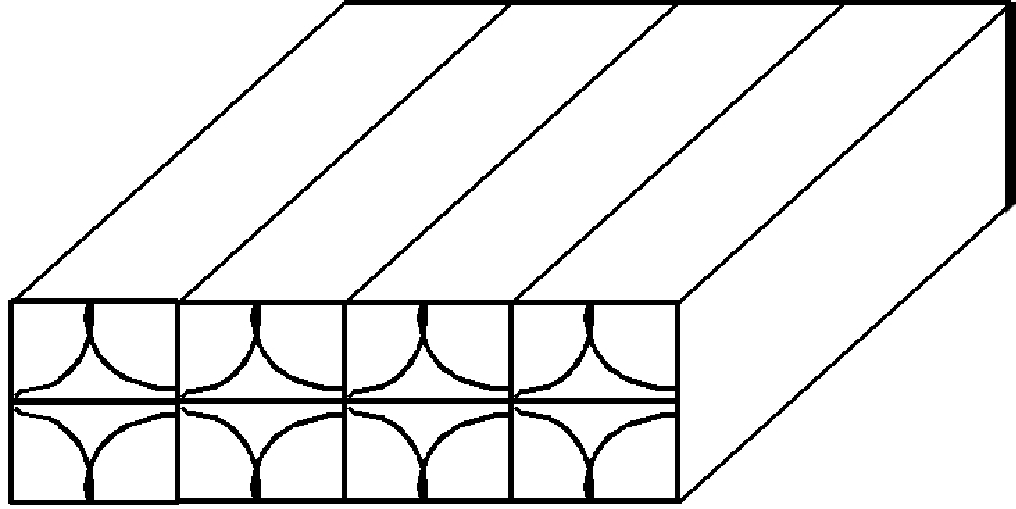}
\caption{Cross-sectional view of a section of a polycarbonate water
module panel, showing the internal wall structure. The approximate
panel thickness is 25.4~mm. Drawing is not to scale.}
\label{fig:panel_xsec}
\end{figure}

Two water panel productions were done using slightly different models
of polycarbonate panel.  The first six panels were produced using a
slightly heavier panel thickness, while another three panels were
produced using a marginally lighter panel.  Ultimately the best six
panels were deployed in the detector, including five from the first
production run and one from the second run, with the other three
panels retained as spares.

\subsubsection{Panel fabrication}

The panels as delivered from the supplier were first cut to a
specified length. For the first production of water modules, the
panels were cut to a length of 1873~mm, while for the second
production this was modified to 1863~mm.

After the panels were cut, the next step was to pierce the inner walls
of the panel at each end to create a contiguous internal volume that
could be filled and drained through a single inlet and outlet. Each of
the thin walls (round walls in Fig.~\ref{fig:panel_xsec}) had a
horizontal slot cut in it using a Dremel multi-tool and then a
vertical slash cut with a sharp pick to produce a T-shaped opening,
while the thick internal walls each had two round holes drilled in
them with the Dremel. These holes allow water to flow from one
adjacent vertical cell of the panel into its neighbors.  All of these
cuts were made 1.5\,-\,2~cm from the edge of the panel to allow room
for the sealant.

Each end of the polycarbonate panel was sealed with a polyurethane
sealant\footnote{HE\,1908, obtained from Engineering Chemicals
B.V.}. This sealant was chosen for its compatibility with
polycarbonate, its low level of absorption of water, and a low
probability of its curing process damaging the polycarbonate
substrate. Prepared sealant was poured evenly into a high-walled,
tight-fitting channel, then the end of the panel was fitted into the
channel to immerse the end in sealant. This sealing process produces
solid seals extending 1\,-\,1.5~cm into the body of the panel. Brass
fittings inserted through the polyurethane seal in the bottom right
and top left parts of the panel provide input and output ports,
respectively.

\subsubsection{Elemental composition}

Polycarbonate has a chemical formula of C$_{16}$H$_{14}$O$_3$.
Subtracting the equivalent of three water molecules from its
composition leaves a remainder of C$_{16}$H$_{8}$.  To match the C:H
ratio of the residual material to polystyrene, (CH)$_n$, thin sheets
of polypropylene, CH$_2$, were glued to the front and back surfaces of
each panel.  Two sheets of 0.8~mm thick polypropylene were attached to
each panel with thin layers of CLR\,1390/CLH\,6025 epoxy (obtained
from Crosslink Technology Inc.).  Thirty thin 25.8~mm $\times$ 51.6~mm
G10 plastic spacers were then attached to each side of the panel to
provide spatial separation between each module and its neighbor when
hanging inside the FGD. The total thickness of a panel, including
polypropylene skins and spacers, is 30.0~mm.

The areal mass densities of all components were measured and used to
compute the elemental compositions per unit area for each panel.
Table~\ref{tab:water_comp} shows the composition for Panel 0.  Also
shown is the composition of an XY module.  The final row shows the
leftover material if the scintillator module composition, scaled by
the relative amounts of carbon in a water panel and a scintillator
module, is subtracted from the water panel composition.  Essentially
the equivalent amount of ``scintillator-like'' material and H$_2$O in
the panels are determined by the carbon and oxygen contents,
respectively, with the ``remainder'' representing the total material
left over after the subtraction.  The full water panel is shown to be
equivalent to 490~mg/cm$^2$ of scintillator-like material,
2297.2~mg/cm$^2$ of water, plus a few mg/cm$^2$ of trace amounts of H,
Mg, Si, and Ti that do not cancel in the subtraction.  These results
demonstrate that the elemental composition of the panels is carefully
tuned so that each water panel is compositionally equivalent to a mass
of pure water and a mass of scintillator module material.

\ifx\review\rev
 \begin{sidewaystable}
\else
\begin{table*}[!t]
\fi
\centering
\caption{Elemental composition of Water Panel 0. All entries are in
units of mg/cm$^2$.  The first three rows give the amounts of each
element for an empty water panel, for the water inside it (for Panel
0), and the total.  The fourth row lists the elemental composition
of an XY scintillator module.  The last three rows show how the
composition of the full panel can be decomposed into the sum of a
mass of scintillator module-like material, of water, and of a
remainder.}
\vspace{2mm}
\footnotesize{
\begin{tabular}{cccccccc}
\hline
 & C & O & H & Mg & Si & Ti & Total\\\hline
Empty panel & 422.0 $\pm$ 6.9 & 92.7 $\pm$ 3.6 & 43.6 $\pm$ 0.9 & 6.8
$\pm$ 0.9 & 11.2 $\pm$ 1.4 & 0 & $577.2 \pm 12.5$ \\
Water in panel 0 & 0 & 1967.4 $\pm$ 3.3 & 248.0 $\pm$ 0.4 & 0 & 0 & 0 & 2215.4 $\
pm$ 3.7\\
Panel 0 (full) & 422.0 $\pm$ 6.9 & 2060.1 $\pm$ 4.9 & 291.6 $\pm$ 1.0 & 6.8 $\pm$
 0.9 & 11.2 $\pm$ 1.4 & 0 & 2792.6 $\pm$ 13.4\\\hline
XY Module & 1848.6 $\pm$ 9.2 & 79.4 $\pm$ 4.8 & 157.9 $\pm$ 2.1 & 0 &
21.8 $\pm$ 4.3 & $35.5 \pm 5.9$ & 2146.3 $\pm$ 14.4\\\hline
``XY''-like & $422.0 \pm 6.9$ & $18.1 \pm 1.1$  & $36.0 \pm 0.8$ & 0 &
$5.0 \pm 1.0$ &$ 8.1 \pm 1.4$ & $490.0 \pm 9.0$ \\
H$_2$O & 0 & $2042.0 \pm 5.0$ & $255.4 \pm 0.6$ & 0 & 0 & 0 & $2297.2
\pm 5.7$ \\
Remainder & 0 $\pm$ 9.8 & 0 $\pm$ 7.1 & 1.2 $\pm$ 1.4 & 6.8 $\pm$ 0.9
& 6.2 $\pm$ 1.7 & $-8.1 \pm 1.4$ & 5.4 $\pm$ 17.1\\\hline
\end{tabular}
}
\label{tab:water_comp}

\ifx\review\rev
 \end{sidewaystable}
\else 
 \end{table*}
\fi

\subsection{Subatmospheric pressure system}

A schematic of the water system is shown in
Fig.~\ref{fig:targetWaterSystem}.  The system is designed to maintain
the pressure of FGD water vessels inside the magnet below atmospheric
pressure to prevent water emission through any leaks that may arise.
Any leak results instead in only ingress of air; a substantial rate of
ingress can be tolerated by the system while still maintaining
effective operation.  The water panels are rigid enough that they can
sustain full vacuum in principle, although the system and operating
procedures are designed to limit the pressure in the panels to greater
than 200~mbarA in all conditions.  There was insufficient space in the
dark box to install level sensors in the panels or for buffer tanks
above them to contain sensors. Hence, the principle for ensuring that
the panels remain full is to maintain upward water flow at all times
at a flow rate that is somewhat less than 1~L/min.  Flow to the inlets
at the bottom of panels (about 3.8~m from the floor) starts from an
open reservoir on a 2.4~m high stage, with a maximum water level just
below the lowest water connection inside the magnet that might leak.
A typical operating water level in the reservoir is 3~m from the
floor.  Flow results from the connection of the outlets at the tops of
the panels to a vacuum vessel at floor level, with a typical half-full
operating water level of about 1~m from the floor.  The operating
pressure in the vacuum vessel is 300~mbarA, which is maintained by a
closed loop servo controller driving a proportional valve leading to a
vacuum buffer tank.  A humidity-tolerant vacuum scroll pump runs
whenever needed to keep this buffer tank at a lower pressure.  The
water level in the vacuum vessel is maintained by a closed loop servo
controller driving a water pump that returns water from this vessel to
the overhead reservoir.  This water pump is a conventional centrifugal
type, but with parameters selected conservatively to avoid cavitation
on the impeller despite the low suction pressure.  Each of the six
water panels has its own independent flow loop so that they can be
filled or drained independently.  Each loop contains a flow meter, as
well as a 3-port valve in both the supply and return legs, which are
used to fill and drain the panel without risk of positive pressure in
the panel at any time, even if there are in-leaks of air.  Draining
occurs in the reverse flow direction, and involves the slow metered
admission of air to the top of the panel to maintain subatmospheric
pressure in the panel.  The water temperature is controlled by a heat
exchanger that also carries coolant flow from the detector cooling
system. A small amount of antimicrobial and anticorrosive agents
(0.25~\% by volume of liquid Germall Plus, manufactured by Sutton
Laboratories, plus 10~ppm tolyltriazole) dissolved in the water inhibits
biological growth in the system.

\begin{figure*}[!t]
\centering
\ifx\figstyle\bw
 \includegraphics[width=\textwidth]{jpgFigures/fig17_bw.jpg}
\else
 \includegraphics[width=\textwidth]{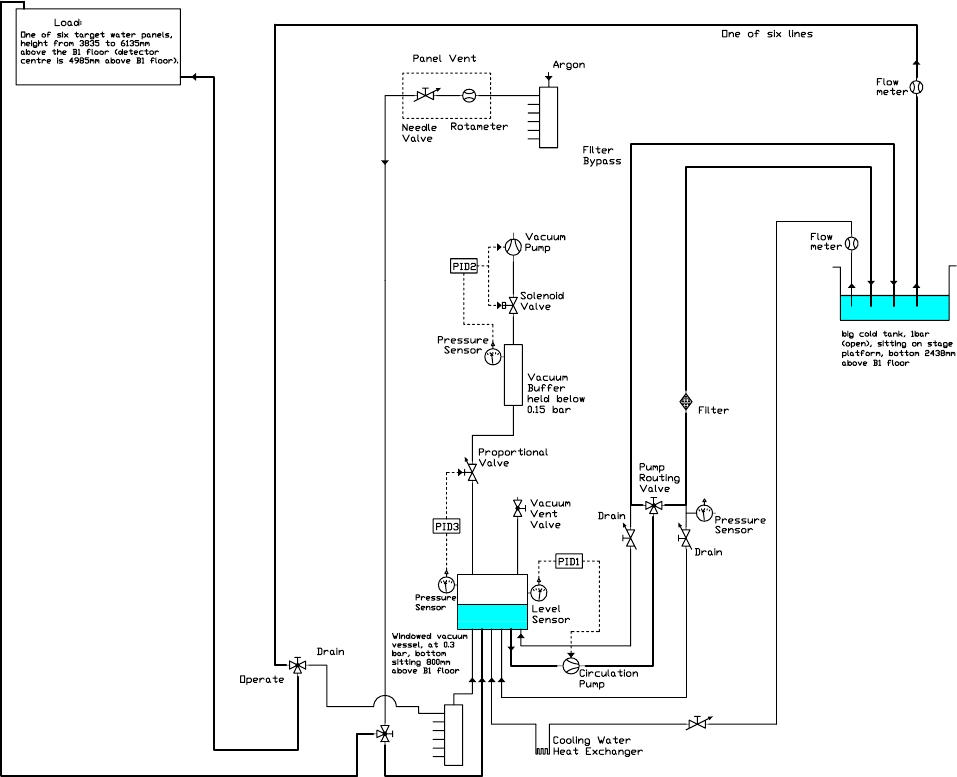}
\fi
\caption{Schematic of the subatmospheric pressure water system.}
\label{fig:targetWaterSystem}
\end{figure*}

\section{Dark Box}
\label{sec:darkbox}

The scintillator and water modules hang inside a lighttight structural
framework known as the dark box.  Each dark box has external
dimensions of 2300~mm $\times$ 2300~mm $\times$ 365~mm thick along the
beam direction, with internal dimensions of 2069~mm $\times$ 2069~mm
$\times$ 352~mm.  Inside the dark box are the scintillator modules
with their associated WLS fibers, photosensors on busboards, and the
water modules.  Front-end electronics cards are mounted around the
four sides of the dark box, outside the lighttight volume.
Twenty-four backplanes mounted along the four sides of the dark box
(six per side) provide the connection between the photosensors on
busboards inside the lighttight region and the electronics cards on
the outside.  Thin aluminum cover plates on the upstream and
downstream faces of the dark box complete the lighttight enclosure.  A
water cooling system operated at subatmospheric pressure cools the
exterior electronics.  Each dark box weighs 327 kg, not including the
electronics cards, and is made almost entirely of aluminum.

\subsection{Construction}

The main structural elements of the dark box frame are
\begin{enumerate}
\item eight long aluminum I beams that fit together to form upstream
and downstream square frames, each with outer dimensions of
2300~mm $\times$ 2300~mm
\item four side plates for the top, bottom, left and right sides of
the FGD, which bolt to the insides of the upstream and downstream
I beam frames and contain cutouts for the backplanes
\item thin square aluminum front and back covers that mount on the
upstream and downstream faces of the FGD
\item cross braces across the top face of the FGD from which the
modules inside the dark box are suspended
\end{enumerate}

A photograph of the FGD frame is shown in Fig.~\ref{fig:mech2}.  The
aluminum I beams that form the dark box frame are machined out of
104.3~mm $\times$ 50.15~mm bars of solid aluminum.  Material is
removed from either side to leave a web of thickness 4.15~mm, except
for regions near the end, thereby reducing the weight by
$\sim$~65~\%. Four I beams fit together to form a square frame
on the upstream end of the FGD, while the other four form a similar
frame on the downstream end.  Four M12\,$\times$\,150 mm stainless
steel bolts join the I beams at each corner where the beams are solid
without any material removed.  The I beam flanges on the inner side of
each square frame mount to the side plates while the outer flanges
define the outer surface of the FGD.  The distance along the beam
direction between the outer edges of the upstream and downstream cover
plates is 365 mm.  The four vertical I beam members extend 35~mm
below the bottom edge of the FGD, forming ``feet'' to which mounting
hardware is attached.

\begin{figure}[htb]
\begin{center}
\ifx\figstyle\bw
 \includegraphics[width=\columnwidth]{jpgFigures/fig18_bw.jpg}
\else
 \includegraphics[width=\columnwidth]{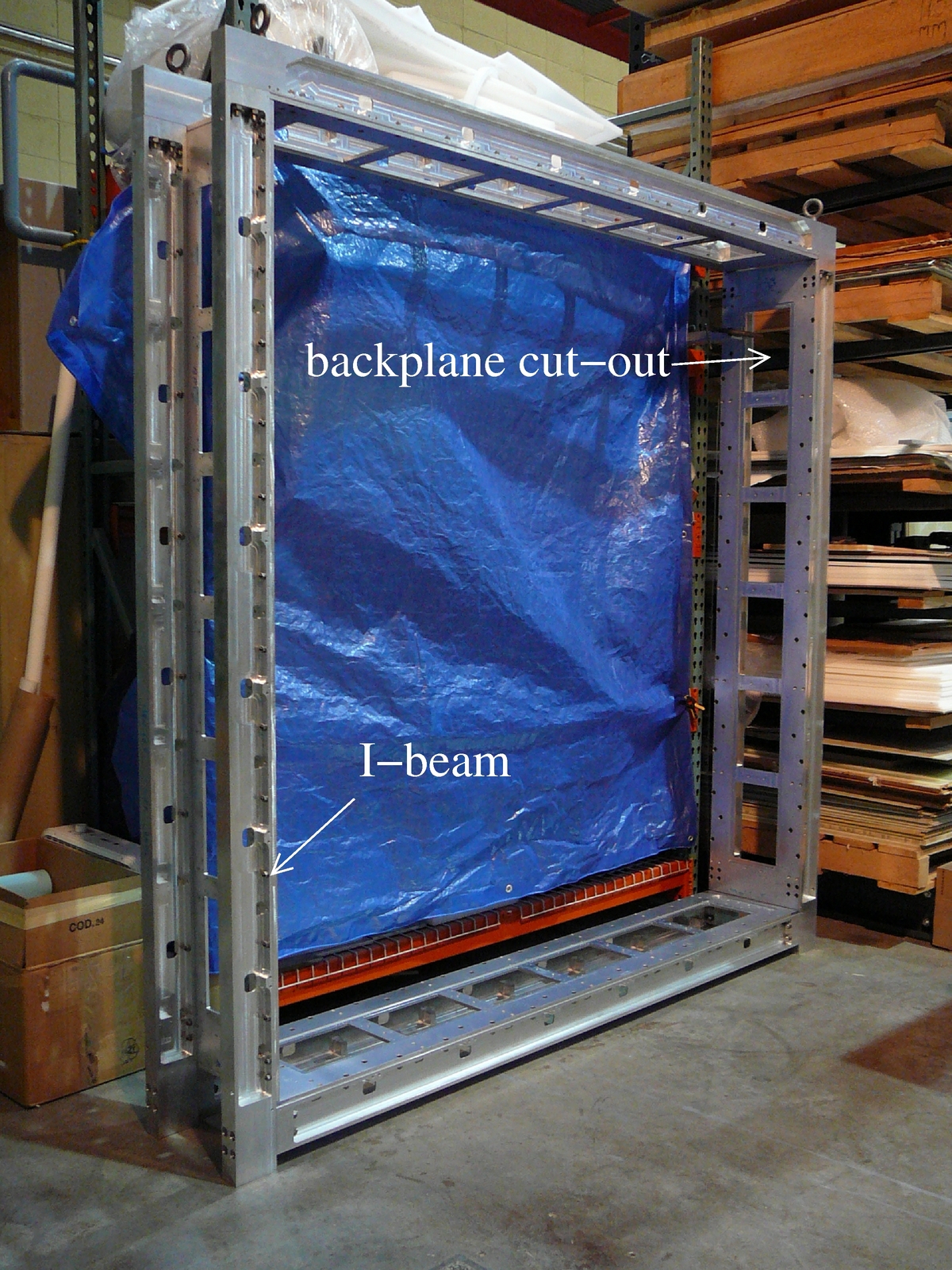}
\fi
\caption{Photograph of the dark box structure.  One of the 24 cutouts
for backplanes and one of the 8 I beams are labelled.}
\label{fig:mech2}
\end{center}
\end{figure}

The four side plates consist of 11.2~mm thick aluminum plates that are
348.2~mm long in the beam direction, which when bolted to the inner
surfaces of the I beam frames form an enclosed area with internal
dimensions of 2069~mm high by 2069~mm tall.  Each side plate is bolted
to the upstream frame along one long edge and to the downstream frame
along the other. Six rectangular cutouts (166~mm $\times$ 258.5~mm),
equally spaced along the length of each side plate, provide space for
mounting lighttight electronics backplanes. Signals from photosensors
are carried by ribbon cables from busboards inside the dark box to the inner
surface of the backplanes (see Fig.~\ref{fig:mech1}), while their
outer surfaces present connectors that front-end electronics cards
plug into.  These cards reside in the 104.3~mm tall region between the
outer edge of the I beam frame and the outer surface of the side
plate.

\begin{figure}[htb]
\begin{center}
\ifx\figstyle\bw
 \includegraphics[width=\columnwidth]{jpgFigures/fig19_bw.jpg}
\else
 \includegraphics[width=\columnwidth]{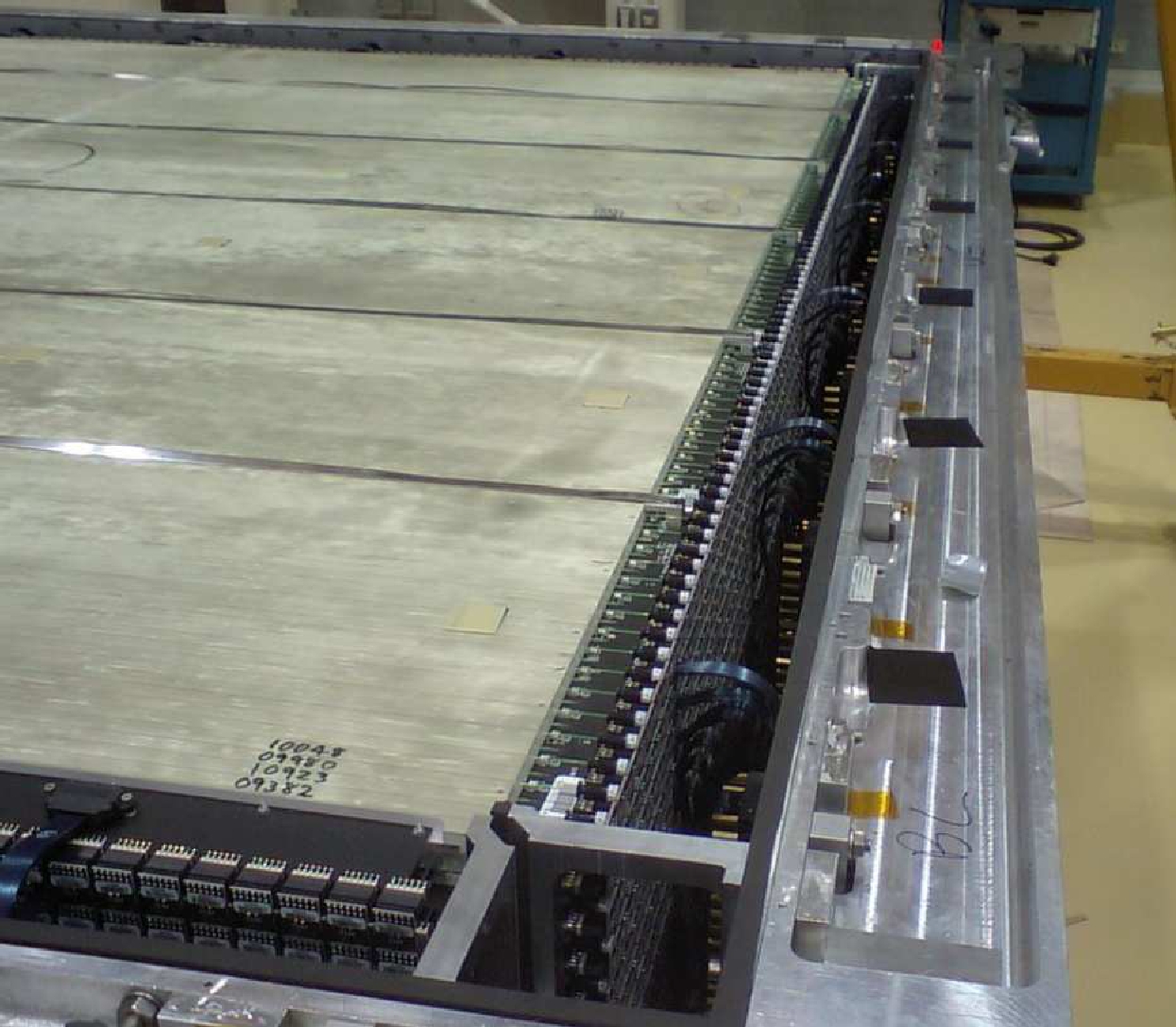}
\fi
\caption{Photograph of the XY modules inside the dark box.  Busboards
carrying photosensors are mounted along all four sides of each module.
Ribbon cables connect each busboard to the backplanes (not visible in
this photo) on the sides of the dark box. Angular support brackets in
the corners of the dark box (bottom center of photo) prevent lateral
movement of the XY modules within the dark box. The five stainless
steel straps are also visible.}
\label{fig:mech1}
\end{center}
\end{figure}

The upstream and downstream aluminum cover plates are square, with a
6.4~mm thickness around the edges and a 1.6~mm thickness over the
central 1856 $\times$ 1856~mm$^2$ area.  They are attached to the I
beam frames around their perimeter with small screws, with black
PORON$\textsuperscript{\textregistered}$ stripping along the edge to
ensure a lighttight seal.

Five steel crossbars, spaced evenly across the width of the FGD and
oriented in the beam direction, attach to the inner top side of the
dark box and provide structural support for the XY and water modules
that hang from them inside the box.  

Electrically isolated copper bars carrying low voltage power and
ground run along the inside edges of the I beams, next to the
backplanes.  Trigger and signal cables run along the outer edges of
the box, between electronics cards mounted in minicrates.  The
upstream and downstream walls of each minicrate contain slots to hold
the cards in position in the minicrates. This arrangement is shown
schematically in Fig.~\ref{fig:mech3}.

\subsection{Module support system}
\label{sec:modulesupport}

In order to minimize the inactive mass inside the FGD fiducial volume,
XY and water modules are suspended inside the dark box from thin
stainless steel straps that loop under the bottom of each module and
attach to the steel crossbars at the top of the box.  Each module is
supported by five 1~cm wide straps made of 0.025~mm thick stainless
steel, spaced evenly across the width of the module.  These straps
loop around small aluminum pads mounted at the bottom of each module
and are welded to top fixtures for attachment to the crossbars.  Each
top fixture is attached with a bolt to its crossbar, with Belleville
washers mounted on the bolt to provide cushioning and accommodate
small differences in strap tension.

Each crossbar itself has a threaded rod on its top that passes through
holes in the top of the dark box to attach to a holding crossbeam on
the outside of the box.  Nuts at the top of these threaded rods lock
the crossbars into position, and are adjusted so that the lengths of
rod, and hence the vertical positions of crossbars, are equal.

\subsection{Attachment to detector basket}

The feet of each FGD rest on a narrow steel bar that is bolted to the
inside bottom edge of the detector support basket.  Large M16 bolts
attach each foot to a bar, and thin aluminum shims are used to adjust
the level of each foot.  A support bracket attached to the top of each
side of the FGD fits into an anchor mounted on the top rail of the
basket.  The support bracket is free to move up and down within the
anchor bracket, allowing for possible sag of the basket under load,
but prevents the top of the FGD from moving laterally.

\subsection{Cooling system}
\label{sec:coolingSystem}

A water circulation system operating at subatmospheric pressure cools
the electronics in the minicrates.  Chilled water flows through hollow
aluminum extrusions mounted on the inside of each I beam.  These
extrusions are connected by hoses at the four corners to form two
cooling loops: one on the upstream side of the FGD and another on the
downstream side.  Water flows clockwise through one loop and
counterclockwise through the other. Fig.~\ref{fig:mech3} illustrates
the relative positions of the FGD electronics cards, I beam,
backplane, and cooling and power bars inside a minicrate.  Each
electronics card in a minicrate carries an aluminum cooling plate
mounted with a thermal coupling compound.  Fingers on these plates fit
into notches in a ``cooling lid'' that is bolted across the top of
each minicrate in thermal contact with the two cooling bars on either
side of it.  Wire springs on a locking mechanism push the card fingers
into contact with the cooling lids.  Thus heat is conducted from the
electronics to the cooling plates, into the cooling lid, and hence
into the cooling loops.

\begin{figure}[htb]
\begin{center}
\ifx\figstyle\bw
 \includegraphics[width=\columnwidth]{jpgFigures/fig20_bw.jpg}
\else
 \includegraphics[width=\columnwidth]{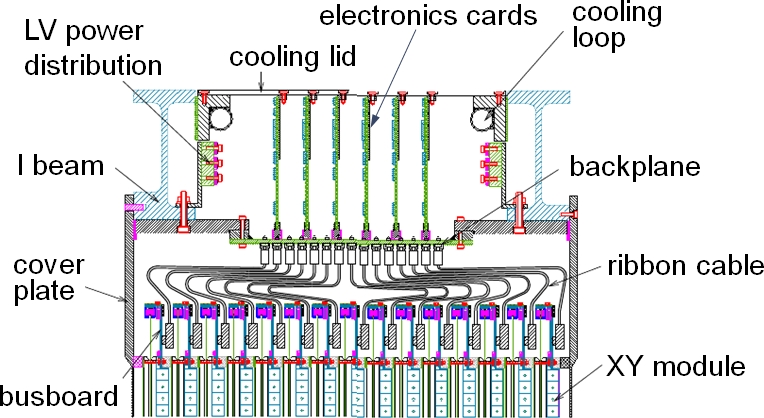}
\fi
\caption{Cross-sectional view through a minicrate, showing the
electronics cards, I beams, cooling and power bars, and backplane.}
\label{fig:mech3}
\end{center}
\end{figure}

\section{Electronics}
\label{sec:electronics}

The FGD readout electronics were designed to record a snapshot of the
detector activity before, during and after the neutrino beam spill.
The critical feature is the readout of the MPPCs by recording
waveforms using switched capacitor array chips. The electronics
provide both timing and energy (charge) measurement for all 8448
channels.

\subsection{Requirements}

The J-PARC neutrino beam structure is divided into spills of 6\,-\,8
bunches separated by $\sim$~580~ns. The electronics must be active
during each bunch.  Furthermore, it is important to keep the
electronics active between bunches and for several microseconds
after the spill in order to tag Michel electrons from the decays of
stopped pions. The total length of time recorded must therefore be
close to 10~$\mu$s.

The required charge resolution is driven by calibration demands.  In
order to calibrate the MPPCs, single pe peaks must be identified. Thus
it is desirable to keep the electronics noise to a level less than 0.1
pe.  Note that minimum ionizing particles typically produce 20\,-\,30
pe when passing through a bar. Thus the energy resolution is
$\sim$~20~\% just from Poisson fluctuations and the electronics
noise does not contribute significantly.

The overall charge dynamic range requirements are set by the
electronics noise level and by the need to fully measure electrical
signals from the largest MPPC pulses when all 667 pixels fire.

The required timing resolution is driven by two considerations. First,
a large background of neutrino interactions occur in the surrounding
magnet.  At full J-PARC beam power, each beam spill will result in
more than 50 such interactions.  Many of these interactions could
cause detector activity in the FGD. Thus the FGD must have good timing
resolution in order to reliably separate this activity from that
initiated in the FGD scintillators.  Studies showed that a timing
resolution on the order of 3~ns for an FGD \textit{vertex}, which
consists of several hits, would be satisfactory.  This is the most
critical requirement for the timing resolution.  However, better
timing resolution would make it possible to distinguish track
direction based on comparing hits in FGD1 versus FGD2.  The goal,
therefore, is to have a timing resolution of better than 3~ns for each
\textit{hit}.

Another requirement on the electronics is the time to read out the
data.  Although the interval between beam spills is 3 seconds, which
provides more than enough time, calibration triggers, such as cosmic
ray, electronic pedestal, and light pulser triggers, are taken between
beam spills. Thus the maximum allowed time to read out events is set
to 50~ms by the maximum data-taking rate, which is 20~Hz.

Finally, access to the electronics boards within the magnet is
impossible while the experiment is running. Thus it must be possible
to remotely control and monitor the condition of the
boards. Furthermore, the temperature must be recorded in order to
correct for temperature dependence of the photosensor response. In
addition, the bias voltage of each photosensor must be set
individually in order to obtain a uniform detector response. Thus, a
robust slow control system is required.

\subsection{Architecture}

A brief description of the electronics arrangement was given in
the Introduction (see section~\ref{sec:designOverview}). Further details
of its organization and operation are presented here.

\begin{figure*}[htb]
\centering
\ifx\figstyle\bw
 \includegraphics[width=\textwidth]{jpgFigures/fig21_bw.jpg}
\else
 \includegraphics[width=\textwidth]{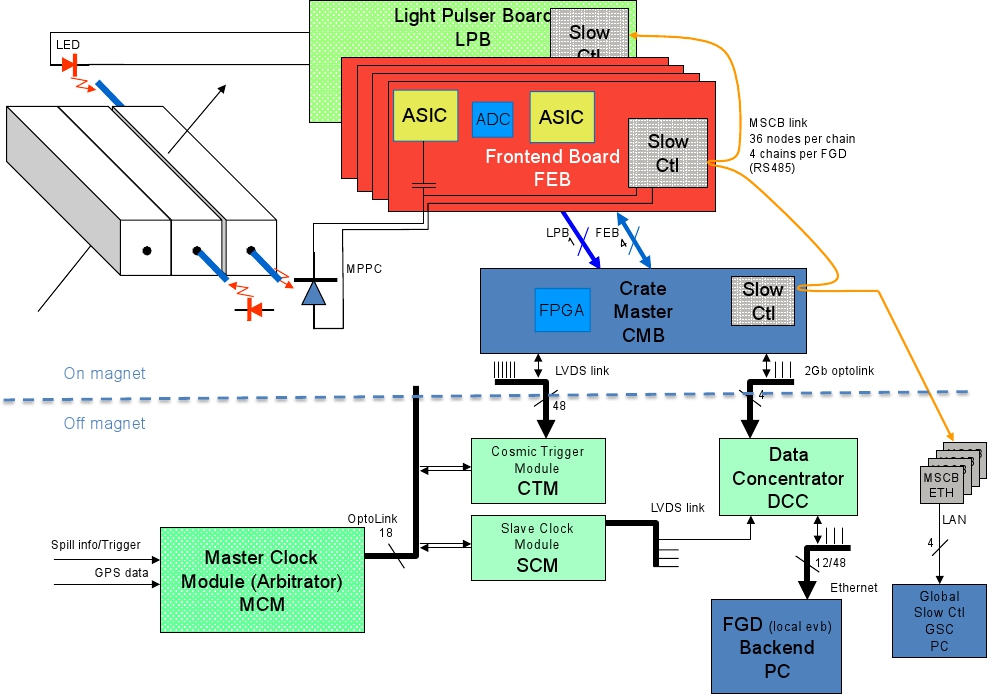}
\fi
\caption[FGD Electronics Overview] {FGD electronics overview, showing
how signals from the MPPCs are captured using the AFTER ASIC chip on
the Front-End Boards; this information is then transferred through the
CMBs and then off-magnet to the DCCs and the back-end computers.  The
upper part of this plot shows the cards that reside inside the
magnet (i.e., one crate's worth of front-end cards).  The principal
trigger for neutrino beam spills comes from the Master Clock Module
(MCM).}
\label {fig:fgd_electronics_overview}
\end{figure*}

The overall readout electronics system is shown in
Fig.~\ref{fig:fgd_electronics_overview}. As mentioned in the
Introduction, the front-end electronics sits in minicrates located
along the outside edges of the dark box.  Each minicrate in FGD1
contains four front-end boards (FEBs) (two in FGD2) that provide the
MPPCs with power and also sense and digitize the MPPC signals.  A
Crate Master Board (CMB) in each minicrate incorporates an FPGA to
control the readout procedure.  One other element in each minicrate is
a Light Pulser Board (LPB) used for calibration.  Each minicrate
services either 240 (FGD1) or 112 (FGD2) MPPCs.

The digitized MPPC data is transferred through fiber optic links to
Data Concentrator Cards (DCCs) that then transfer the data to the
back-end computers for data acquisition system (DAQ) processing.  The
DCCs are also connected to a number of common ND280 electronics cards
in order to be part of the overall ND280 triggering and clock
distribution schemes. All these elements will be described in what
follows.

\subsection{Busboards and backplane}

The MPPCs are mounted on busboards; each busboard holds 16 MPPCs (see
Fig.~\ref{fig:mech1}). Sixteen LEDs are also mounted on each busboard
to flash the WLS fibers near the mirrored end.  The chosen LED is the
LTST-T670TBKT (obtained from Lite-On Inc.) whose peak wavelength is
468~nm, well matched to the WLS fiber. It was important to ensure that
each LED on a busboard would produce similar amounts of light when
driven by an identical electrical pulse.  Because of manufacturing
differences, this uniformity could not be guaranteed without
additional sorting of the LEDs. A sophisticated robotic system to sort
them was therefore devised to automatically flash each LED and measure
the resultant light output.  Each busboard then received LEDs with
similar responses. Usage of the LEDs is described in
Section~\ref{sec:fgd_elec_lpb}.

The busboards are connected by ribbon cables to the electronics
backplane.  The backplane provides the electrical and mechanical
interface between the inside of the dark box, containing the busboards
and MPPCs, and the outside of the box containing the front-end
electronics cards (FEBs, CMBs and LPBs).  Thus besides providing
common services for all the cards in the minicrate, such as power,
slow control and readout, the backplanes also provide the electrical
connections between the busboards and the FEBs. Care must be taken to
ensure there is minimal electrical crosstalk between analog MPPC lines
and that the large digital signals do not influence the small analog
signals.  The layout and routing of the backplane was therefore a
complicated affair, with care being taken to route the analog and
digital signals on different layers of the board.

\subsection{Front-end electronics slow control}
\label{sec:fgd_slow_control_arch}

One critical aspect of the FGD electronics is the slow control system.
\textit{Slow Control} refers to the electronics circuitry and
associated firmware located on all the front-end cards, as well as the
associated software system described in
Section~\ref{sec:slowcontrol}. A separate slow control power source is
supplied to each card, which permits continuous monitoring of the
hardware without the necessity of full activation of the card.
The addition of an asynchronous slow control system is also useful for
providing a clean conceptual separation between elements of the
electronics system that must be recorded for each event (MPPC data,
trigger types, timestamps, etc.) and elements that can be monitored
less frequently (temperatures, voltages, etc.).  Therefore, the
inclusion of a separate slow control system makes the FGD electronics
architecture more robust, since it provides an independent way of
monitoring and debugging the state of the boards and simplifies the
types of communication among the FEBs, CMBs, DCCs and back-end
computers.

The slow control section or node on each board is operated by a
microcontroller. The interface and protocol used for the communication
is based on the Midas Slow Control Bus (MSCB)~\cite{Ritt:mscb}. This
bus is an RS485 multi-drop master/slave serial link. Several
microcontroller families can support the MSCB protocol. The FEBs
employ the C8051F133 microcontroller (obtained from Silicon Labs) with
64~kB of flash, running up to 100~MHz. The C8051F133 has internally
8~$\times~$10 bit ADCs and 32 I/O ports. Some of the I/O ports are
programmed to communicate via SPI, I2C, and SST buses with other
devices, enhancing the microcontroller capabilities. Each FGD has 4
MSCB chains (one per side). Each chain is composed of up to 36 MSCB
nodes (6 cards/crate $\times$ 6 crates).

Each MSCB-RS485 chain is converted to Ethernet protocol in order to
normalize the communication method with the rest of the ND280
apparatus. This is done by an MSCB-ETH board powered by a C8051F121
and Cypress CS9800A Ethernet chip.  The data communication mechanism
uses the UDP protocol.

\begin{figure}[htb]
\centering
\ifx\figstyle\bw
 \includegraphics[width=\columnwidth]{jpgFigures/fig22_bw.jpg}
\else
 \includegraphics[width=\columnwidth]{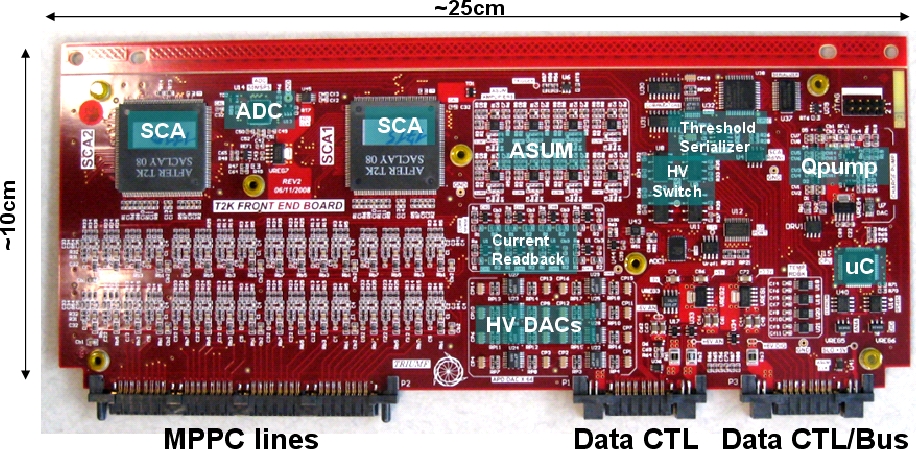}
\fi
\caption[FGD FEB] {FGD Front-End Board with critical features
indicated; its lower edge in this figure plugs into the backplane.}
\label {fig:fgd_feb_picture}
\end {figure}

As each card is described in the following subsections, the
particular slow control functionality that is implemented on that card
will also be described.

\subsection{AFTER Chip and FEB}

The key element of the FGD front-end electronics is the AFTER ASIC
that was designed for the T2K TPC~\cite{Baron:2008zza}. This ASIC
provides a preamplifier, a shaper and a switched capacitor array (SCA)
for each of 72 channels.  The purpose of these elements is as follows:

\begin{itemize}

\item The combination of preamplifier and shaper amplifies the
electrical pulse from the MPPC and also extends the pulse length.  For
the FGD, the AFTER ASIC preamp-shaper time constant is set to 30~ns,
the lowest possible value, corresponding to a 180~ns peaking time.

\item The SCA is an array of capacitors that are used to store analog
samples of the shaped MPPC pulse. When the last cell in the array is
reached, the ASIC will start overwriting the values in the first cell;
the SCA can therefore be continuously filled until the write process is
stopped by an event trigger.  The sampling frequency of the SCA is
tunable, but for the FGD it is set to 50~MHz.  Therefore the SCA is
sampled every 20~ns and the total depth of the 511-cell array is
10.2~$\mu$s.

\end{itemize}

A critical feature of the AFTER ASIC is that the preamp-shaper
combination allows sampling the waveform at a much lower frequency
than would be possible if the MPPC signal was digitized directly. This
strategy reduces the cost significantly while still allowing the
timing resolution requirements to be met.  The contents of the SCAs
are digitized by a 12 bit flash ADC running at 20~MHz and the
resulting data are transferred to the CMB. This process is described in
section~\ref{sec:fgd_elec_cmb}.

Fig.~\ref{fig:fgd_feb_picture} shows the component face of a FEB. Two
AFTER ASICs can be seen. Each ASIC reads out 32 MPPCs; hence one FEB
reads out 64 MPPCs.  Coupling an MPPC to the AFTER ASIC is not
trivial, however. The dynamic range requirement implies that two AFTER
channels must be used to record both low-gain and high-gain signals
from each MPPC.  The signal is divided as shown in
Fig.~\ref{fig:charge_divider}.  Small values are used for the
capacitors in the low and high attenuation branches and in the trigger
branch in order to minimize the input capacitance, hence minimizing
the electronics noise.  Recharging the MPPC is achieved by the 100~nF
capacitors.  The 10~k$\Omega$ resistor in series with the 100~nF
capacitor in the recovery branch means that the MPPC recovery time
constant is 667C(R/667+10\,k$\Omega$)~=~614~ns to first order.  Here
C~=~90~fF (R~=~150~k$\Omega$) is the capacitance (resistance) of a
single pixel, implying that 667C is negligible compared with
100~nF. This time constant is only visible for large pulses; otherwise
the pixels that do not avalanche quickly recharge the pixels that do.
Fig.~\ref{fig:charge_divider} also shows a third branch that is summed
with others on the board to generate a trigger signal.

\begin{figure}[htb]
\centering
\ifx\figstyle\bw
 \includegraphics[width=\columnwidth]{jpgFigures/fig23_bw.jpg}
\else
 \includegraphics[width=\columnwidth]{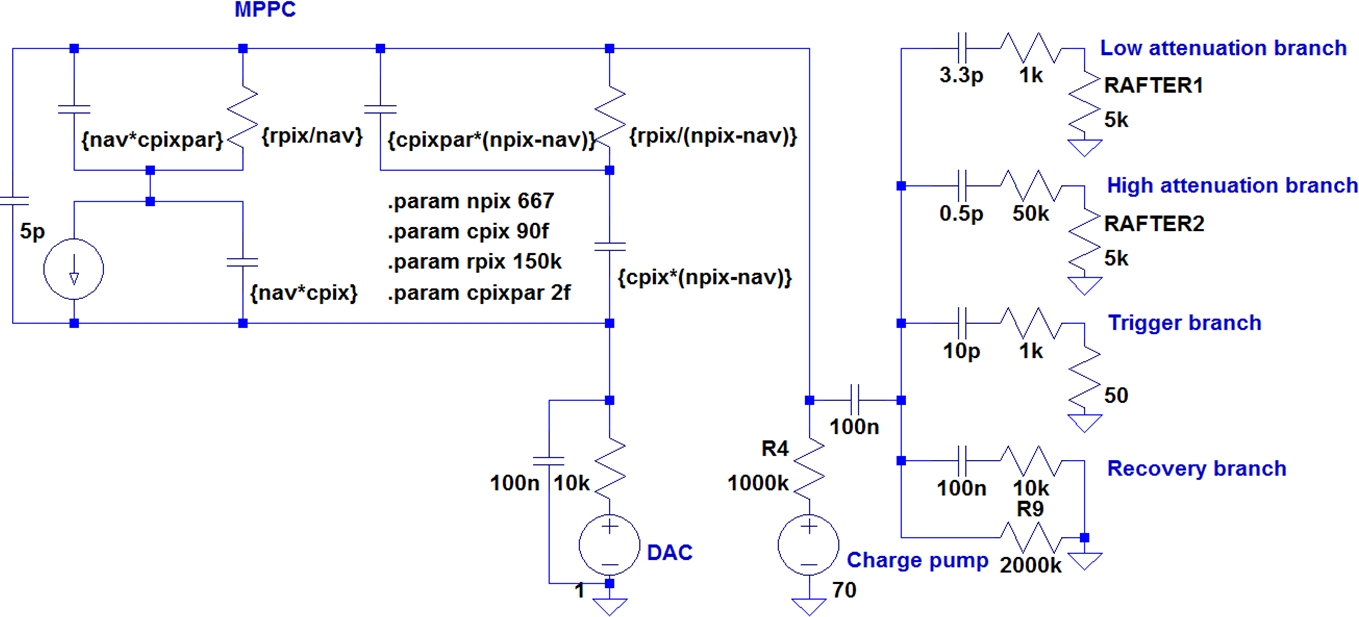}
\fi
\caption[FEB Charge Divider] {SPICE model of the MPPC and of the Front
End Board / MPPC-AFTER coupling circuit.}
\label {fig:charge_divider}
\end{figure}

Fig.~\ref{fig:fgd_example_waveform} shows an example for the high- and
low-gain waveforms for a single MPPC, for a particular event.  These
data are from beam tests where a pion stops in one bar and the
subsequent Michel $e^+$ is detected. The pion pulse almost saturates
the high-gain channel but it is clearly resolved in the low-gain
channel. The undershoot following the high-gain pulses is a
consequence of the capacitive coupling on the AFTER ASIC inputs.  With
the MPPC operating at 0.8~V overvoltage, the high-gain channel
saturates at $\sim$~80 pixels (4096 ADC counts), while a minimum
ionizing particle typically produces light activating 25\,-\,30
pixels. Thus the low-gain channel does not need to be used for minimum
ionizing particles.

The rest of the FEB is used for monitoring and slow control.  The slow
control infrastructure is handled by the MSCB protocol that is is
described in section~\ref{sec:fgd_slow_control_arch}.  The MPPC bias
voltage is generated on-board by a charge pump circuit. Using a charge
pump is necessary because only 6~V is provided to the FEB but the MPPC
operating voltage is $\sim$~70~V. The charge pump voltage is common to
all MPPCs. The individual MPPC voltages are adjusted by trimming the
MPPC signal return voltage using 64 Digital to Analog Converters.
Both the charge pump voltage and the trim voltage can be remotely set
through MSCB.  Because the MPPC gain is sensitive to the temperature
(5\,-\,10~\% change in pulse height per degree Celsius), each FEB also
reads out eight temperature sensors mounted on its busboards close to
the MPPC themselves.  If the temperature variations are large, the FEB
slow control system allows for adjustment of the MPPC bias voltage in
order to maintain equalized MPPC gains (this feature is not presently
used).

\begin{figure}[htb]
\centering
\includegraphics[width=\columnwidth]{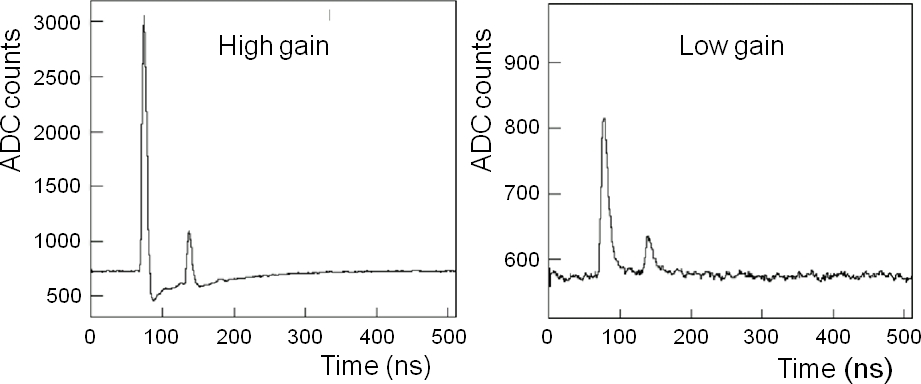}
\caption[FGD FEB Example Waveform] {FGD FEB example waveform from the
TRIUMF beam test showing results for the high- and low-gain
channels for a single MPPC. Time is measured in 20 ns wide bins. This
event can be interpreted as the particle stopping in the bar and a
Michel positron subsequently emerging. }
\label{fig:fgd_example_waveform}
\end{figure}

\subsection{Crate Master Board}
\label{sec:fgd_elec_cmb}

The Crate Master Board is the next element of the electronics; an
image of a CMB is shown in Fig.~\ref{fig:fgd_cmb_picture}.  The core
of the CMB is a Xilinx Virtex~2 Pro FPGA. The FPGA controls the record
and readout phases of the AFTER ASIC, using a set of parallel LVDS
clock and signal lines dedicated to each FEB.  The waveform samples
from each AFTER SCA are serially read out, digitized by the ADC and
stored locally on the CMB.  The local storage uses a pair of Zero Bus
Turnaround (ZBT) synchronous SRAMs.  Stopping of data acquisition and
digitization of the stored waveform happens automatically upon
receipt of a trigger from the DCC.  The whole process is done in
$\sim$~2~ms by simultaneously reading out the SCAs from all ASICs that
are attached to the CMB (up to eight in total). This parallel readout
is critical, because the analog information in the SCA cells degrades
if the readout takes longer.  As an added benefit, digitization is
sufficiently fast that it does not significantly contribute to the
overall FGD readout time.

The locally stored data can then be transferred to the Data
Concentrator Cards.  The connection to the DCCs is done through a
full duplex RocketIO optical port running at 2~Gb/s. The DCC
to CMB transfer is initiated by a packet request from the DCC
(rather than by the original DCC trigger).  This CMB to DCC transfer
procedure will be described further in section~\ref{sec:cmb_data_comp}.

In addition to the core AFTER data handling, the CMB FPGA also
provides the following functionality:

\begin{itemize}

\item configuration of the AFTER ASIC parameters (such as the
amplifier gain and the shaping time) using a serial protocol.
\item processing the trigger signals from the FEBs according to
programmed logic conditions.
\item triggering of the LPB.
\end{itemize}

For configuration purposes, the CMB has an flash memory with
a CPLD\footnote{Complex Programmable Logic Device};
this allows new FPGA code to be uploaded through the optical
link.  The CMB FPGA firmware can therefore be upgraded and improved
even after the ND280 magnet is closed.  This procedure has been
successfully used a number of times.  As an additional safety feature,
the flash memory has two banks, one of which is kept loaded with a
stable version of the firmware; this ensures that the CMB can recover
from upload of nonfunctional firmware.

\begin {figure}[htb]
\centering
\ifx\figstyle\bw
 \includegraphics[width=\columnwidth]{jpgFigures/fig25_bw.jpg}
\else
 \includegraphics[width=\columnwidth]{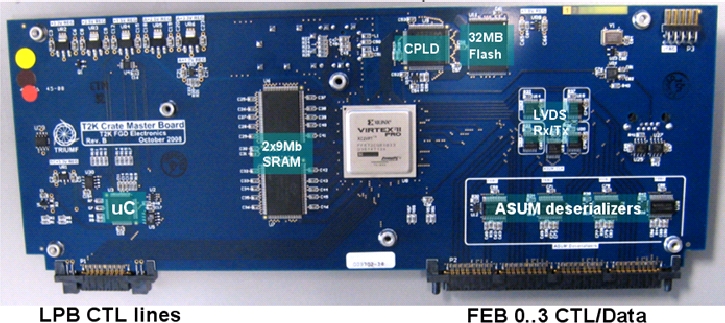} 
\fi
\caption[FGD CMB] {Crate Master Board with critical features
indicated; its lower edge in this figure plugs into the backplane.
The core of the CMB is the central Virtex II FPGA. The SFP (Small
form-factor pluggable transceiver) fiber connector for connection to
the DCC is mounted on the backside of the CMB and cannot be seen in
this image.}
\label {fig:fgd_cmb_picture}
\end {figure}

In addition to the link to the DCC, the CMB also has a second,
electrical LVDS link to the Cosmic Trigger Module (CTM); the purposes
of this link will be described in more detail in
section~\ref{sec:link_nd280_electronics}.

In order to reduce the volume of data recorded, data compression
capability is programmed in the CMB. Part of this compression scheme
makes use of a pulse finding/pulse height algorithm.  Both of these
processes are fully described in Section~\ref{sec:cmb_data_comp}.

Finally, the CMB slow control microcontroller can be used to turn
on/off the FPGA and to monitor the main operating voltages, currents
and temperatures of the board. A separate serial link between the
slow control microcontroller and the FGPA is available for transmission of
particular Status/Control bits such as the geographical address of the
board and the optical link state, as well as allowing for the reset of
the FPGA.

\subsection{Light Pulser Board}
\label{sec:fgd_elec_lpb}

The Light Pulser Board is the last of the cards in each minicrate.
The LPB controls the pulsing of the LEDs mounted on the busboards. The
purpose of the LED system is twofold:

\begin{itemize}

\item to test the integrity of the signal path starting from the fiber
to the digitization for every MPPC channel.  Light is injected near
the end of the WLS fiber opposite from
the MPPC and therefore can detect possible mechanical issues such as
microcracks in the fiber or problems with the fiber/MPPC coupling.

\item To provide a coarse gain calibration of the MPPCs.  This
calibration can either be performed continuously between the beam
spills or can be done in dedicated LPB runs.

\end{itemize}

There are up to 240 LEDs being driven by each LPB.  Driving each of
those LEDs individually is impractical.  Therefore, sets of four LEDs
on each busboard are ganged together and driven by a single LPB
pulser.  Hence there are 60 independent pulsers on the LPB.

Triggering the LED pulses is controlled by the CMB. Specifically, if
the CMB receives a `LED-type' trigger then it will send a signal to
the LPB to fire the LEDs.  The CMB will then wait for a couple of
microseconds before initiating readout of the AFTER ASICs.  This
will ensure that the LED pulses actually occur near the middle of the
AFTER waveforms.

Like the FEB and CMB, the LPB has a slow control microcontroller
to monitor board functions.  In addition to monitoring voltages and
temperature, the LPB slow control also allows the user to specify the
intensity of the LED pulse by setting a DAC level for each group of
four pulsers (i.e.\ a separate DAC is used for each busboard).

\subsection{Infrastructure}

The front-end electronics require a number of different voltage levels
for the various cards.  Specifically, +6~V digital, +6~V analog,
+4~V and -6~V voltages are supplied to the cards using a
pair of Wiener PL508 power supplies.  The slow control on the
front-end cards is powered separately using a +5~V level from a
Xantrex power supply.  The Wiener power supplies are 
hardware-interlocked to the Xantrex power supply; if the Xantrex supply drops
below 4.5~V the Wieners will automatically shut off.  This ensures
that the cards will not be fully powered without slow control
monitoring.

The total power consumed by the front-end electronics is approximately
1500~W; this heat is extracted from inside the magnet by the cooling
system described in section~\ref{sec:coolingSystem}.
              
\begin {figure}[htb]
\centering
\ifx\figstyle\bw
 \includegraphics[width=\columnwidth]{jpgFigures/fig26_bw.jpg}
\else
 \includegraphics[width=\columnwidth]{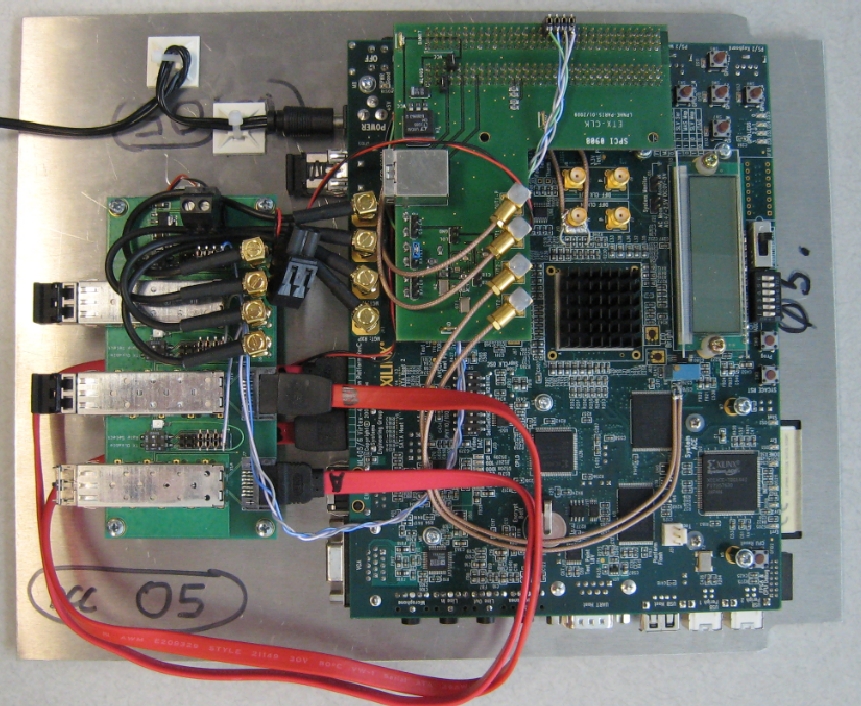}
\fi
\caption[FGD DCC] {FGD Data Concentrator Card assembly; ML405 with
extension cards mounted on support.}
\label {fig:fgd_dcc_picture}
\end {figure}

\subsection{Data Concentrator Card}

Each of the 48 CMBs has an optical link connected to a Data
Concentrator Card. The DCCs start and stop the data acquisition process
on the front-end electronics; also, after the front-end acquisition has
finished, the DCCs request channel data packets from the CMBs and format
the data packet before sending it to the DAQ over ethernet.  Each DCC
is presently composed of an ML405 evaluation board and 2 custom cards.
Each DCC possesses four RocketIO optical links and can read out data
from four CMBs.  Twelve DCCs are therefore needed to read out the full
FGD.

Each ML405 board is based on a Xilinx-IV FPGA with a PPC405/300MHz
microprocessor and Gigabit Ethernet port.  The DCC functions are
implemented as FPGA firmware and microprocessor tasks.  The FPGA part
of the DCC has custom VHDL firmware that interfaces with the RocketIO
ports and provides the synchronous trigger distribution.  The DCC
microprocessor has been configured to run a simple Linux kernel
(2.6.23-rc2) using the BusyBox utility package.  This Linux kernel
hosts a reduced Midas Data Acquisition front end directly on the DCC;
more details can be found in section~\ref{sec:midas_daq}.

The DCC assembly is shown in Figure~\ref{fig:fgd_dcc_picture}.  There
are two custom extension boards added to the ML405. The first board
provides a 4-to-1 SMA/RocketIO Optical link converter and 3
SATA/RocketIO to 2 Optical link converters. The second board provides
the interface to the trigger module and also provides the global
200~MHz clock for RocketIO operation.

\subsection{Interface to ND280 electronics}
\label{sec:link_nd280_electronics}

There are a number of elements of the FGD electronics that must
interface with the global ND280 electronics system. Many of these
elements can be seen schematically in
Fig.~\ref{fig:fgd_electronics_overview}.  They are:

\begin{itemize}

\item MCM/SCM Trigger Reception: events in ND280 are triggered by
  either the Master Clock Module (MCM) if in global mode or by the FGD
  Slave Clock Module (SCM) if in local mode.  In either case, the
  triggers are passed through the FGD SCM which distributes an LVDS
  signal over Cat.-6e cables to the twelve DCCs.  These triggers
  are then passed on to the CMBs over the RocketIO optical link and
  initiate the AFTER ASIC readout procedure that was described in the
  section~\ref{sec:fgd_elec_cmb}.

\item Global Clock: the FGD SCM also provides a 100~MHz clock to the
DCCs, which is in turn passed on to the CMBs.  The FGD SCM is
phase-locked to the MCM 100~MHz clock; the FGD electronics is therefore
phase-locked to the other ND280 subdetectors, thereby simplifying
cross-detector time calibration.

\item FGD Cosmic Trigger: the FEBs have dedicated Analog Sum (ASUM)
circuitry capable of generating a local trigger.  Each FEB sends 8
separate ASUM triggers to the CMB, which then makes a decision whether
or not the crate has satisfied its trigger logic conditions.  If so, a
trigger is sent over a Cat.-6e LVDS link to the FGD Cosmic Trigger
Module.  If the triggers from the CMBs satisfy the
overall CTM trigger condition (which usually requires that at least
two CMBs in each FGD have fired), then a global ND280 trigger can be
generated.  This trigger can select cosmic rays that pass through both
FGDs, which is a useful tool for calibration.

\item FGD Timing Marker: the CTM provides to each CMB an LVDS timing
marker pulse, which is injected asynchronously into spare channels on
the AFTER ASICs.  Since the timing marker pulses are injected into all
FEBs at the same time (modulo cable length differences), the recorded
timing markers can be used to correct for clock phase differences and
digital jitter.

\end{itemize}

\section{Data acquisition and compression}
\label{sec:daq}

The Data Acquisition System (DAQ) is responsible for transferring the
digitized MPPC waveform from the CMB RAM to the DCC and from the DCC
to the back-end computers.  The DAQ therefore consists of a
combination of FPGA firmware on the CMBs and DCCs, as well as software
on the DCCs and back-end computers.  One of the major challenges for
the DAQ is to reduce the volume of data written to disk.  If the full
uncompressed FGD data for every event were saved, the maximum recorded
event rate would be $\sim$~2.5~Hz, with a data rate of
$\sim$~60~MByte/s (limited by the CMB to DCC bandwidth).  However, the
DAQ is required to record events at a trigger rate greater than 20~Hz
with a data rate of a MByte/s for the FGD.  Clearly the DAQ must
therefore provide a significant amount of data compression while
maintaining a high trigger rate.

A typical particle interaction will deposit charge in only
approximately 60 of the 8448 scintillator bars; this means that all
relevant information about a particular event can be collected from
only a small fraction of the detector.  In addition, MPPC dark noise
pulses must also be saved for calibration purposes; however, not all
the waveform information for noise hits must be saved.  These two
facts suggest that considerable data reduction is possible.

\subsection{CMB online data compression}
\label{sec:cmb_data_comp}

The choice was made to implement data compression using VHDL firmware
on the CMB FPGA~\cite{Kirby-thesis}. The advantage of implementing the
data compression at the CMB level is that it allows the process to be
highly parallelized.  It would have probably been impossible to
provide the same degree of compression if it had been done at the DCC
or back-end computer level.  The following explains how this CMB
compression is implemented.

The custom DCC to CMB data transfer protocol starts when the DCC
requests from the CMB a data packet from a list of channels that the DCC
specifies. The CMB processes the list of channels until it finds a
pulse to process and transmit in a data packet, or reaches the end of
the list.  The small size of the compressed data packets allows the
DCC to read out all four CMBs in parallel.  This transfer protocol is
efficient only when a large fraction of the FGD channels are
completely zero-suppressed because the number of DCC data requests and
CMB transaction overhead is reduced; the efficiency of the transfer
protocol is therefore dependent on effective data compression.

\begin {figure}[htb]
\centering
\includegraphics[width=\columnwidth]{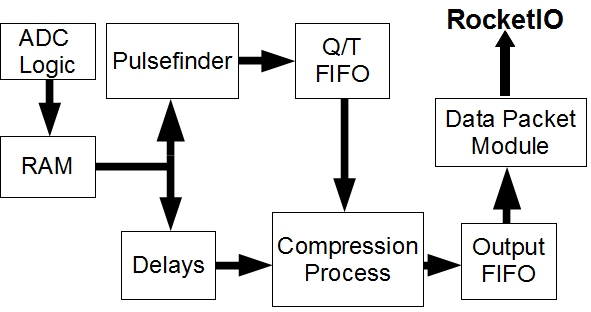}
\caption[FGD Data Compression Scheme] {FGD channel readout and data
compression processes implemented in CMB firmware.}
\label {fig:fgd_data_compression_scheme}
\end {figure}

There are two elements to the data compression: the ability to quickly
and efficiently identify pulses (pulse finding) and the compression of
the information from that pulse. The procedure is shown schematically
in Fig.~\ref{fig:fgd_data_compression_scheme}. The pulse finder
algorithm is a local peak detector that compares the sum of amplitudes
during consecutive rising and falling edges to a threshold to identify
a pulse.  The current system uses a threshold of 30 ADC counts,
roughly equivalent to peak pulse heights of 15 ADC counts. This
provides greater than 98~\% pulse finding efficiency for pulse heights
above 20 ADC counts (assuming nonoverlapping pulses).  This threshold
is well below the single-pixel pulse height distribution (with mean
pulse height measured at $\sim$~40 ADC counts and $\sigma$ $\sim$~6
ADC counts), but is large enough to suppress most baseline
fluctuations (with mean $\sigma$ of 3 ADC counts).  The pulse-finding
efficiency increases with larger pulse height with an efficiency of
effectively 100~\% for large pulses from minimum ionizing particles.

Pulse height is measured as the difference between the pulse peak
height and a sample measured a fixed time before the peak to estimate
the channel baseline.  The approximately constant gain across all
MPPCs established by overvoltage tuning allows the same threshold to
be used for the entire system.  ADC samples are read from the RAM and
processed by the firmware-based pulse finder.  When a pulse is
identified, the pulse finder writes the pulse height and time
information into the Q/T FIFO (Q/T = charge/time). The Q/T FIFO feeds
a separate compression process to zero-suppress a delayed sequence of
ADC samples. The amount of delay depends on the desired number of
samples in the data packet prior to the pulse rising edge, as well as
the time required for the pulse finder to identify a pulse and
calculate its height.

Additional data reduction decisions are made by the compression
process based on the identified pulse height.  Specifically, a
decision must be made whether to save only the pulse height and time
for a pulse or whether to also save the ADC samples around the peak
position.  Saving the ADC samples is critical for calculating a
precise time for the pulse during offline data processing.  But having
a precise time is only important for pulses associated with physics
hits; it is not important for dark noise pulses.  The pulse height
spectrum obtained from cosmic-ray trigger data (see
Fig.~\ref{fig:pulse_height_compression}) shows that the dark noise
pulses dominate the spectrum for pulse heights below 200 ADC counts,
while larger pulses are primarily produced by cosmic muons interacting
in the scintillator bars.  Therefore, ADC samples for a pulse are
saved only if the pulse height is greater than the dark noise
threshold, which is 200 ADC counts.

\begin {figure}[htb]
\centering
\includegraphics[width=\columnwidth]{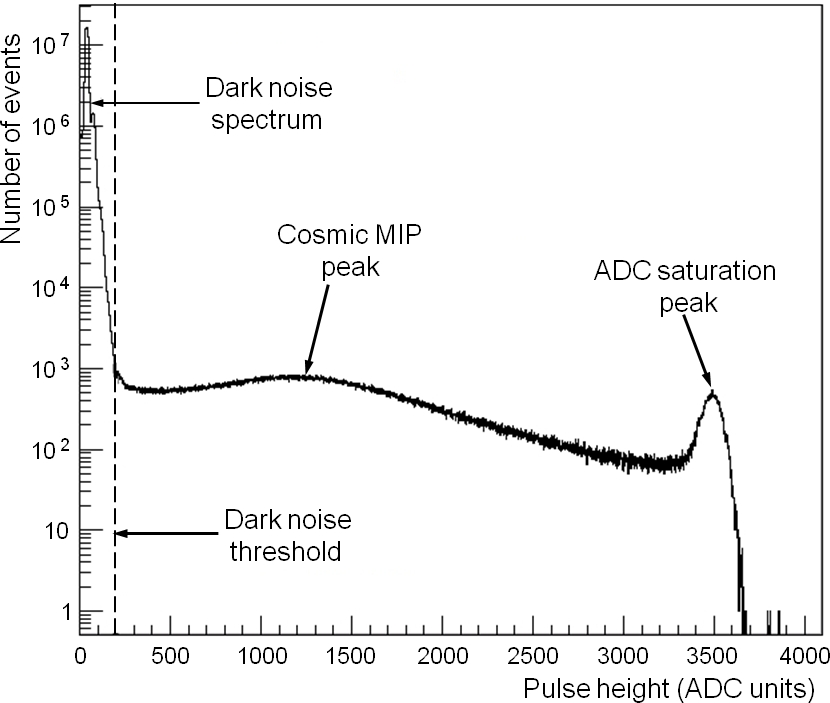}
\caption[MPPC Pulse Height Distribution] {MPPC pulse height
distribution for cosmic trigger events, with dark noise threshold
displayed.}
\label {fig:pulse_height_compression}
\end {figure}

In fact, even discarding ADC samples for dark noise pulses does not
provide a sufficient degree of data compression.  The total average
data size for dark noise pulse information processed using this scheme
is $\sim$~280~kB per event, compared with $\sim$~20~kB for pulses due
to particle interactions.  Dark noise information can be discarded or
retained on an event-by-event basis since it is necessary only for
long-term calibration; therefore the total data rate can be adjusted
by discarding dark noise pulses for some fraction of the events.  In
the current implementation, the dark noise data are typically saved only
10~\% of the time.\footnote{An alternative method is to always retain
the dark noise data in the CMB but discard it in a back-end process
after updating an online charge histogram.  This scheme will be
implemented in the future.}

After all the samples for each channel are read from RAM, the contents
of the output FIFO are moved into the RocketIO clock domain of the CMB
and used to form the data packet that is sent over the optical link to
the DCC.

\begin {figure}[htb]
\centering
\ifx\figstyle\bw
 \includegraphics[width=\columnwidth]{jpgFigures/fig29_bw.jpg}
\else
 \includegraphics[width=\columnwidth]{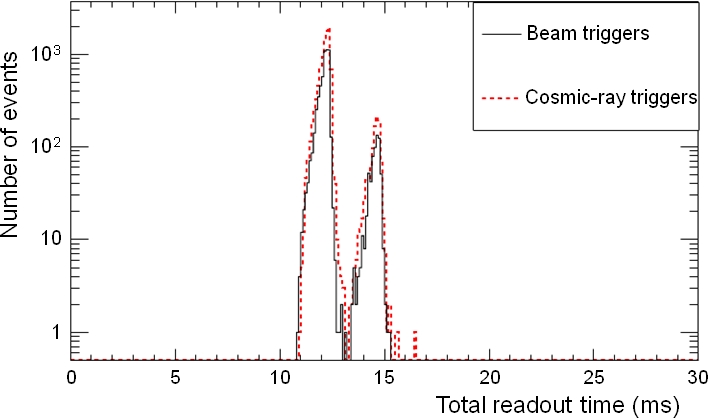}
\fi
\caption[FGD Total Readout Time] {Total time to read out FGD, for both
beam and cosmic-ray events. These measurements are provided by the FGD
SCM.  The two distinct peaks correspond to events where we read out
all dark noise pulses (right peak) and where we completely suppress
the dark noise pulses (left peak).}
\label {fig:readout_time}
\end {figure}

\subsection{Midas Data Acquisition System}
\label{sec:midas_daq}

The FGD (like the rest of ND280) uses the MIDAS
system~\cite{Ritt:mdap} as the basis for the DAQ.  The core of the FGD
DAQ system is a set of Midas front-end programs that run on each of
twelve DCCs; as mentioned above, running Midas programs directly on
the DCCs is possible because of the Linux kernel running on the DCCs.
This DCC program initiates the CMB to DCC transfer that was described
in the previous section; the programs also handle the transfer of data
from the DCC to the back-end FGD computer.

One complication of the FGD DAQ is that it is necessary to be able to
run in both a local FGD-only mode, as well as in a global mode with
all the rest of the ND280 detectors.  A cascade system has been
designed to provide this functionality.  The first level of the
cascade is the FGD DAQ that runs on the FGD back-end computer.  In
FGD-only local mode the data is stored only on a disk on the FGD DAQ.
In the global ND280 mode~\cite{Thorpe:daq}, the FGD data is
transferred to a different global DAQ computer where it is merged with
data from other subdetectors before being saved to disk.

All of the processing and transfers by the FGD DAQ must be completed
within a time period that is compatible with the overall ND280 DAQ.
Fig.~\ref{fig:readout_time} shows FGD SCM measurements of the total
time between the reception of the trigger to the time at which the
busy signal was disabled by the FGD DAQ software.  As can be seen, the
time to read out the FGD is always less than 20~ms.  Thus in normal
experimental conditions the FGD can run at a 50~Hz trigger rate,
comfortably exceeding the 20~Hz requirement.

\section{Slow Controls and monitoring of FGD components}
\label{sec:slowcontrol}

As described in section~\ref{sec:electronics}, several of the FGD
components communicate with a PC for control functions and
monitoring. All of these functions are performed by front-end tasks
connected to one instance of the MIDAS framework
system~\cite{Ritt:mscb}. This system is similar to the DAQ system and
is common to all ND280 detectors.  The MIDAS framework provides
templates for front-end tasks as well as drivers for the various means
of connecting to the hardware.

The slow control front-end tasks for the FGDs are: 
\begin{itemize}
\item front-end electronic boards (FEB, CMB, LPB) control and monitoring:
connected via MSCB
\item Wiener power supplies (3) control and monitoring: connected via
Ethernet
\item Xantrex power supplies (2) monitoring: connected via MSCB providing
power to slow control section of front-end electronic boards and DCC
\item DCC power control module: connected via MSCB
\item target water system monitoring: connected via MODBUS
\item target water temperature sensor monitoring: connected via MSCB
\item FGD SCM and CTM temperature sensor monitoring: connected via MSCB 
\end{itemize}

Each of the front-end tasks connects to a common MIDAS online database
(ODB) where all settings and sensed values are stored. Users
communicate with the ODB via web pages.  The MIDAS framework provides
generic web pages for display of settings and sensed variables but
custom web pages have been written to facilitate user interaction for
control of power supplies and for the FGD electronics front-end
boards.

The most extensive front-end program controls and reads back the bias
voltage for each MPPC as well as temperatures on the busboards and on
each electronic board, and a few humidity sensors. Power to the main
section of the boards is also controlled through this program.  Eight
instances run in parallel connecting to the eight MSCB communication
chains (see section~\ref{sec:fgd_slow_control_arch})

The MIDAS task MLOGGER logs all sensed values stored in the ODB at
fixed intervals defined by the user for each type of variable. The
usual frequency is 5 minutes but faster frequencies were used when
debugging or watching a critical front-end component. For example a
frequency of 10 seconds was used to troubleshoot problems with the FGD
water target water system. The FGD electronics front end is a special
case where the sensed values are stored at a fixed interval of 5
minutes or when the user modifies electronics settings such as the
bias voltages.

These variables are logged in a MYSQL ``Slow Control Database'' by
MLOGGER, which creates tables automatically, based on the labels of
the variables in the ODB. There are 420 tables populated with FGD slow
control variables.  The MIDAS web server provides a versatile
mechanism to define history plots of any of these variables, which the
user can consult for any given period or see the plots grow in real time.

\section{Calibration}
\label{sec:calibration}

The calibration involves two almost independent tasks: charge
(Sec.~\ref{sec:qcalib}) and time (Sec.~\ref{sec:timecalib})
calibration. The procedures described in this paper are those
currently used to produce physics results. (Further improvements are
expected to be implemented in the future.)

\subsection{Charge calibration}
\label{sec:qcalib}

The purpose of the charge calibration chain is to convert a raw pulse
height ($PH$), measured in ADC channels from a digitized MPPC
waveform, to a normalized value representing the energy deposited in a
scintillator bar.  This normalized value should be insensitive to any
changes in operating conditions encountered during the course of a
data-taking run.

\subsubsection{Electronic high-/low-gain calibration}

The low-gain channel extends the dynamic range to include energy
depositions yielding more than about 90 avalanches, which saturate the
12-bit ADC on the high-gain channel.  The signal-to-noise ratio in the
low-gain channel is too small to allow efficient detection of
single-pixel avalanches, but becomes good above 10-pixel avalanches.
If the high-gain pulse corresponds to more than about 65 avalanches,
then the waveform from the low-gain channel is used.  The relationship
between the responses of corresponding high and low-gain channels is
linear, characterized by the constant
\begin{equation}
\label{eqn:cali1}
C_{HL} \equiv \frac{PH_{H}}{PH_L}.
\end{equation}
Here $PH_{H[L]}$ denotes the pulse height measured in the
high-(low-)gain channel.  The value of $C_{HL}$ is measured for each
MPPC using cosmic rays that deposit energy in the bars creating
avalanches in more than 10 pixels while encompassing the full range of
energies measurable by the high-gain channel. The value of $C_{HL}$
averaged over all the MPPCs is 8.81, with an RMS spread of 0.27.

\subsubsection{Normalization by temperature-dependent single-avalanche gain}

The next step in the calibration chain is to calculate the number
$N_{av}$ of pixels avalanching by normalizing the measured pulse
height $PH$ by the average pulse height $\langle PH_1 \rangle$
corresponding to a single-pixel avalanche:
\begin{equation}
\label{eqn:cali2}
N_{av} = PH/\langle PH_1 \rangle
\end{equation}
The distribution of dark noise pulse heights is used to measure
$\langle PH_1 \rangle$ for each MPPC.  The first peak of the
distribution is fitted by a truncated Gaussian, $\langle PH_1 \rangle$
being the location of this peak. A typical example is shown in
Fig.~\ref{cali:fig1}.  Dark noise spectra are accumulated in dedicated
runs taken with periodic triggers.  Such spectra are also obtained
during physics data-taking, since dark noise pulses are present in the
10~$\mu$s time windows during which MPPC waveforms are digitized.

\begin{figure}[htb]
\centering
\includegraphics[width=\columnwidth]{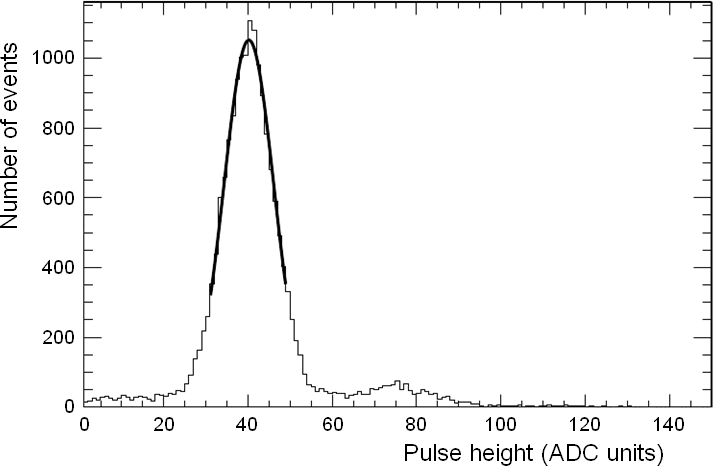}
\caption{Typical dark noise pulse height distribution, showing
the truncated Gaussian fit.}
\label {cali:fig1}
\end {figure}

The single-avalanche pulse height $\langle PH_1 \rangle$ scales with
the MPPC gain, which is proportional to the overvoltage, the
difference between the operating and breakdown voltages.  The
breakdown voltage depends on temperature~\cite{MPPC:2007mp}, which can
change during running.  The operating voltage is kept constant in
normal operation, but the temperature varies, by at most $\pm 2 ^{\circ}C$.
The change in gain is expected to be less than 10~\%, which does
not justify developing a feedback system to adjust the operating
voltage while the experiment is running.  Instead it is compensated in
the analysis.

Rather than analyzing dark-noise spectra to extract $\langle PH_1
\rangle$ often enough to track possible temperature changes, it is
more convenient and adequately accurate to correct for the temperature
changes by applying a correction based on the temperature measured in
the vicinity of each MPPC.  At a fixed temperature, $\langle PH_1
\rangle$ is expected to vary linearly with operating voltage, if the
pixel capacitance were independent of voltage.  However, it was
empirically determined that a quadratic dependence provided a better
description of the measured variation (see Fig.~\ref{cali:fig2}), in
agreement with ~\cite{Vacheret:2011zza}.  For each MPPC, $\langle PH_1
\rangle$ is calculated using the relation
\begin{equation}
\label{eqn:cali3}
\hspace{-8mm}V - C_T(T-T_0) = V_{bd,0} + (1/G) \langle PH_1 \rangle + C_G \langle PH_1 \rangle^2
\end{equation}
where $V$ and $T$ are the operating voltage and temperature at the
time of measurement, $V_{bd,0}$ is an estimate of the breakdown
voltage at $T_0$, $G$ is an empirically determined parameter in units
of ADC counts/V, which can be interpreted as the product of the
electronic gain with the capacitance of the MPPC in parallel with that
of its busboard trace, and $C_G$ an empirical parameter accounting for
the quadratic dependence.  This quadratic term receives contributions
from both the slight nonlinearity of the intrinsic dependence of the
MPPC gain on overvoltage as well as the effect of a large resistor in
series with the MPPC bias voltage source in combination with the
dependence of the MPPC dark noise rate on overvoltage.  The parameter
$C_T$ is a constant with an empirically determined value of $57\pm
3$~mV/deg, in agreement with expectations~\cite{MPPC:2007mp} from lab
bench tests.  The parameters $V_{bd,0}$, $G$ and $C_G$ are determined
separately for each MPPC from a series of dark-noise spectra acquired
at various voltage settings.  Each such ``voltage scan'' is done in a
time sufficiently short to ensure that the temperature did not vary
significantly from $T_0$.  During the subsequent analysis of data,
periodic extraction of $\langle PH_1 \rangle$ from a dark noise
spectrum at the temperature of that time is used to confirm that
Eq.~\ref{eqn:cali3} is producing an appropriate estimate. It is found
that values of $\langle PH_1 \rangle$ extracted from dark noise
spectra and those estimated using Eq.~\ref{eqn:cali3} agreed to better
than 2~\% during the entire first year of running.
\begin{figure}[htb]
\centering
\ifx\figstyle\bw
 \includegraphics[width=\columnwidth]{jpgFigures/fig31_bw.jpg}
\else
 \includegraphics[width=\columnwidth]{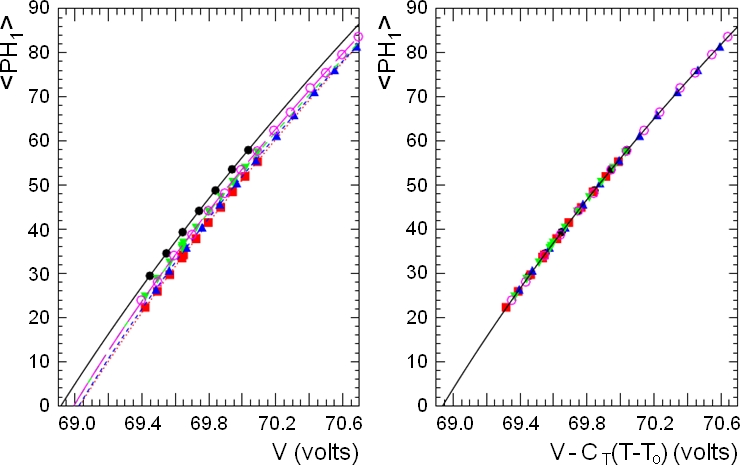}
\fi
\caption{Pulse height $\langle PH_1 \rangle$, measured in ADC counts,
for single-pixel avalanches vs applied voltage for a typical MPPC,
measured at various temperatures. The plot on the left is without
temperature correction; the points on the right plot have been
temperature corrected.}
\label {cali:fig2}
\end {figure}

Each MPPC temperature is estimated as the average value from eight
sensors located on the MPPC busboards in that region of the FGD.  The
temperature and operating voltage values are recorded every few
minutes and logged in the slow control database.  The constants
$V_{bd,0}$, $G$, and $C_G$ are stored for each MPPC in the calibration
database.

Values of breakdown voltage determined from fits to voltage scan
measurements match well with the appropriately scaled recommended
operating voltages for individual MPPCs as provided by the
manufacturer.

\subsubsection{Effects on $N_{av}$ of variations in overvoltage}
\label{sec:tempcor}

The number $N_{av}$ of pixels avalanching for a given number of
photons hitting the MPPC depends on the photodetection efficiency,
and the cross talk and after-pulsing probabilities.  Temperature
variations change the MPPC breakdown voltage and hence overvoltage,
the difference between the operating and breakdown voltages.  This
affects not only the MPPC gain but also the above three attributes, as
well as the dark noise rate~\cite{Vacheret:2011zza}.  The variation of
$N_{av}$ with temperature for a constant light pulse intensity must be
compensated.

The variation of $N_{av}$ with overvoltage was studied using an
electron beam during the M11 beam tests at TRIUMF (see
Sec.~\ref{sec:m11tests}), and with self-triggered high-energy cosmic
rays {\em in situ} at J-PARC. The operating voltage was intentionally
varied while keeping the beam or cosmic trigger condition constant,
and the data were taken while the temperature was not changing
significantly.  Events were selected to define a reproducible narrow
distribution in energy deposition in each bar, from either passing
electrons or muons.  At each operating voltage setting, the mean pulse
height $\langle PH \rangle$ from this distribution was measured, as
well as $\langle PH_1 \rangle$ from dark noise.  Over a limited range
of overvoltage, the data can be fit by this empirical relation:
\begin{equation}
\label{eqn:cali4}
\langle N_{av} \rangle  \equiv \frac{\langle PH \rangle}{\langle PH_1 \rangle} 
\propto C_o + C_{\epsilon}  \langle PH_1 \rangle
\end{equation}
Such a relationship would be expected for such a fixed (small)
light-pulse intensity if both the photodetection efficiency
$\epsilon$ and MPPC gain increased linearly with overvoltage.  Lab
bench tests demonstrated such a linear behavior of the gain, and an
approximate such behavior of $\epsilon$ over a limited
range~\cite{Vacheret:2011zza}.  The values of the constants $C_{o}$
and $C_{\epsilon}$ were found to be $-0.0885 \pm 0.0071$ and $0.0338
\pm 0.0007$ ADC counts$^{-1}$, respectively, where their normalization
was chosen so that the factor $C_o + C_{\epsilon} \langle PH_1
\rangle$ is unity for the average value of $\langle PH_1 \rangle$
across all the MPPCs.  It was found that only one set of values was
required for all MPPCs.

Over a wider range of overvoltage $\Delta V$, departures from
Eq.~\ref{eqn:cali4} can be expected.  The lab bench tests
in~\cite{Vacheret:2011zza} measured not only the dependencies of the
gain and photodetection efficiency $\epsilon$, but also the number
$n_{sec}$ of secondary avalanches (cross talk plus after-pulses) per
primary avalanche.  The behavior of $\epsilon$ is consistent with the
expected shape~\cite{Orme:2009mppc} for saturation of its only
overvoltage dependent factor, the Geiger efficiency
$\epsilon_{Geiger}$ (the probability for a photoelectron to cause an
avalanche), while $n_{sec}$ was found to vary quadratically with
$\Delta V$ for $\Delta V < 1.5$V.
\footnote{The lab bench test data of Ref.~\cite{Vacheret:2011zza}
indicate that these two sources of nonlinearity cause opposite
curvatures in the plot of $\langle PH \rangle$ versus $\langle
PH_1\rangle$, with that of secondary avalanches expected to dominate.
On the other hand, the data from the M11 test beam studies suggest a
curvature opposite to that expected from this dominance.  This
inconsistency is presently not understood.}  Hence a more general
functional form might be employed for the relevant product $\epsilon
(1+n_{sec})$ expressed directly in terms of $\Delta V = V -
C_T(T-T_0)$ (see Eq.~\ref{eqn:cali3}).  Nevertheless,
Eq.~\ref{eqn:cali4} is used
in the calibration chain because it is a simple parameterization that
matches the data well enough over the temperature range experienced.
Thus a temperature-corrected value of $N_{av}$ is computed:
\begin{equation}
\label{eqn:calib5.5}
N_{av,cor} = \frac{N_{av}}{C_o + C_{\epsilon}  \langle PH_{1,proxy} \rangle}.
\end{equation}
Here $\langle PH_{1,proxy} \rangle$ serves as a proxy for $\Delta V$
through the intrinsic properties of the MPPC, so the value used here
for each MPPC is its value for $\langle PH_1 \rangle$ corrected for
the effects of the capacitance of its busboard trace, the length of
which varies among those on each busboard.

\subsubsection{Saturation of the MPPC pixel population}
\label{sec:satcorr}

Since each MPPC has a finite number of pixels, it is expected that
measured pulse heights will saturate with increasing light levels, as
each pixel may be hit by multiple photons and the fraction of pixels
avalanching begins to become of order unity.  (This is why the
discussion in the preceding subsections was limited to small
light-pulse intensities.)  For the case of a uniform distribution of
the photons hitting the entire MPPC and assuming infinite recovery
time (i.e., a pixel cannot avalanche multiple times within the
electronic pulse shaping time), an analytic formula for the saturation
can be derived:
\begin{equation}
\label{eqn:cali6}
\hspace{-8mm}
\langle N_{av,cor}\rangle = N_{pix} \left(1-e^{-\epsilon (1+n_{sec}) \langle N_{ph}\rangle / N_{pix}}\right).
\end{equation}
Here $\langle N_{ph}\rangle$ is the mean number of photons hitting the
MPPC, $\epsilon$ is its photodetection efficiency, $n_{sec}$ is the
number of cross talk or after-pulse pixels fired per primary avalanche,
and $N_{pix}$ is the number of pixels in the device.  Bench tests
using a nonuniform photon distribution from a wavelength-shifting
fiber have shown that this formula adequately models the FGD MPPCs'
response when the number of avalanches is less than 300, provided that
$N_{pix}$ is replaced by a smaller effective value $N_{pix,eff}$.  In
these studies~\cite{simon-thesis}, a 405 nm laser was used to excite
the Y11 fiber.  The green light reemitted by the Y11 fiber was
detected by an MPPC using the FGD electronics system. No blue light
from the laser could reach the MPPC directly.  At a low light level
where there was a substantial probability $p_0$ of the MPPC producing
no response from the light pulse, this probability was measured and
Poisson statistics was assumed to calculate $\epsilon \langle
N_{ph,1}\rangle = -\ln(p_0)$.  (Note that this normalization point
does not involve secondary avalanches.)  Scaling from this value of
$N_{ph,1}$ at low light level, the amount of light $N_{ph}$ hitting
the MPPC was tuned using digital variable attenuators and monitored
using a PIN diode.  The parameters that best fit Eq.~\ref{eqn:cali6}
to the dependence of $N_{av}$ on $\epsilon \langle N_{ph}\rangle$
measured at the nominal operating voltage are $N_{pix,eff} = 476$ and
$n_{sec} = 0.03$~\cite{simon-thesis}.  This fit is over the range
$0<N_{av}<450$, and there agrees with the data within 10~\%, as shown
in Fig.~\ref{cali:figsimon}.  The fit diverges from the data for
larger light levels, which are rarely encountered in normal FGD
operation.  The parameter $N_{pix,eff}$ is smaller than $N_{pix}=667$,
the total number of physical pixels in the MPPCs, because the fiber
does not in fact uniformly illuminate the MPPC's active area.  For the
largest light levels, the data exceeds the number of physical pixels.
This is possible because a pixel is able to quickly recharge by
drawing charge from other pixels in the same device, and refire within
the light decay time constant of the fiber, which is about 7~ns.

\begin{figure}[htb]
\centering
\ifx\figstyle\bw
 \includegraphics[width=\columnwidth]{jpgFigures/fig32_bw.jpg}
\else
 \includegraphics[width=\columnwidth]{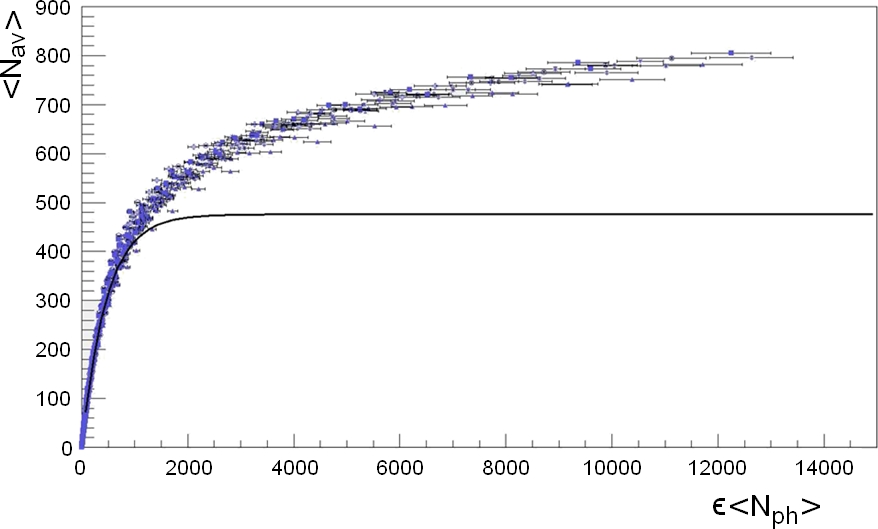}
\fi
\caption{Measured response of an MPPC in terms of the number $N_{av}$
of avalanching pixels, when illuminated by light from several
different wavelength-shifting fibers, as a function of $\epsilon
\langle N_{ph}\rangle$.  The experimental conditions and fitted curve
are described in the text.}
\label {cali:figsimon}
\end {figure}

In the present analysis of FGD data, the number $N_{DPE} \equiv
\epsilon (1+n_{sec}) N_{ph}$ of ``detectable\footnote{The quantity
$N_{DPE}$ differs from the number of photoelectrons by the factor
$\epsilon_{Geiger}$.}  photoelectrons'' (primary plus secondary) is
calculated from $N_{av,cor}$ of Eq.~\ref{eqn:calib5.5} by inverting
Eq.~\ref{eqn:cali6}, using the values $N_{pix,eff}= 396$ for all
MPPCs.  The light injection system is expected to eventually provide
individual values of these constants for each MPPC.

\subsubsection{Correction for bar-to-bar variations}
\label{sec:Cbar}

Given enough data, it becomes possible to check whether the number of
photons hitting each MPPC not only remains constant for a constant
energy deposit in that FGD scintillator bar, but also is the same for
all bars.  There are a number of reasons why small differences might
be expected. These include minor variations in the fiber/MPPC
coupling, variations in scintillator material, variations in fiber
mirroring, variations in the diameter of the hole in the bars and the
exact position of the fiber within the hole, etc.  It was determined
that such variations could be accounted for by introducing an
additional correction constant $C_{bar}$ for each bar, representing
the factor by which the efficiency for conversion of energy deposition
in the bar to $N_{ph}$ photons hitting the MPPC differs from its value
averaged over the whole FGD.

The values of $C_{bar}$ were determined from cosmic ray data by using
tracking information to compute the track length $\Delta l$ through
each bar, and then by comparing the distributions of $N_{DPE}/\Delta
l$ for all bars.  Consistent results were obtained from two separate,
similar, but independent, data sets.  The spread of values of
$C_{bar}$ is about 7~\%, as illustrated in Fig.~\ref{cali:fig3}.  The
values are stored in the Calibration Database for use in analysis.

\begin{figure}[htb]
\centering
\includegraphics[width=\columnwidth]{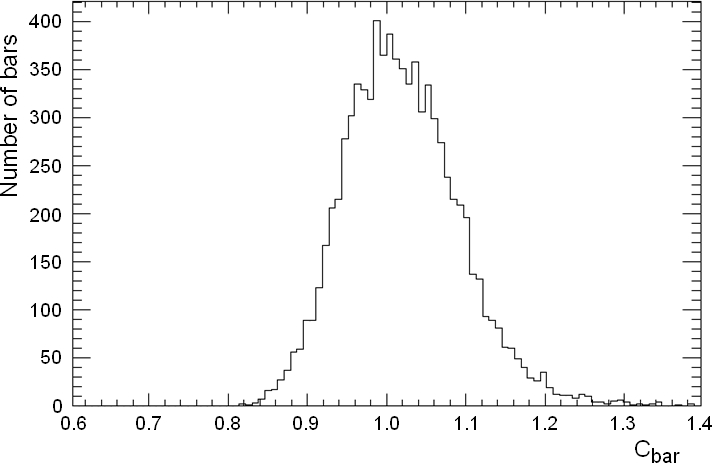}
\caption[Fig. 3]{Distribution of the bar-to-bar correction constant
$C_{bar}$ for all FGD MPPC's for several ``voltage scans''.}
\label {cali:fig3}
\end {figure}

\subsubsection{Correction for light loss along the bar}
\label{sec:attcorr}

The light attenuation in each wavelength-shifting fiber was measured
on a test bed, as reported in Sec.~\ref{sec:fiber}.  However, the
correction of the FGD data for the dependence of the MPPC response on
the position along the scintillator bar of the passing particle track
is based on {\em in situ} measurements of cosmic rays using tracking
information to identify this intersection point.

\begin{figure}[htb]
\begin{center}
\includegraphics[width=\columnwidth]{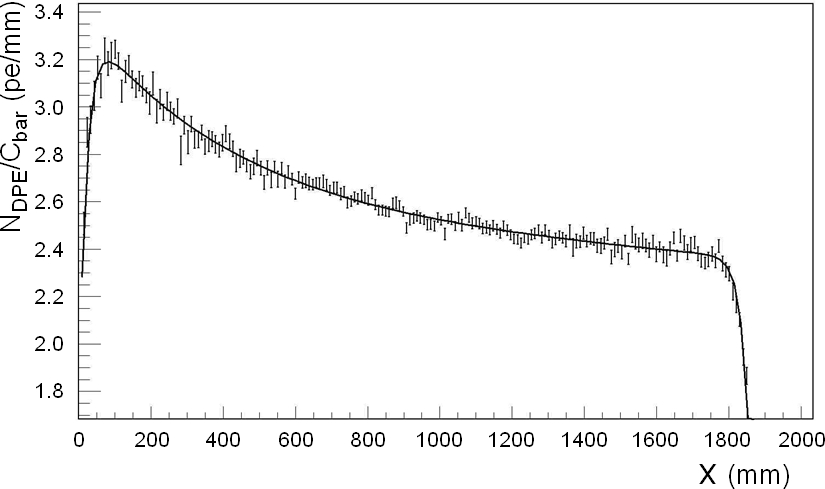}
\caption{Number $N_{DPE}/C_{bar}$ of intercalibrated detectable
photoelectrons for cosmic ray events, as a function of hit position
along the length of the bar. The data are compared to the empirical
fit given in Eq.~\ref{eq:hit_vs_q}.
\label{fig:q_vs_x}
}
\end{center}
\end{figure}

Fig.~\ref{fig:q_vs_x} shows the measured yield $N_{DPE}$ of detectable
photoelectrons from cosmic ray events as a function of the distance of
the hit from the MPPC.  The light yield is highest nearest the MPPC,
dropping by $\sim 25~\%$ at the far end.  The overall shape of the
attenuation curve is consistent with that measured during the fiber
testing (see Section~\ref{sec:fiber}), except near either end of the
bar.  For hits occurring within $\sim 5$~cm of the end of the bar
there is a deficit in light yield that is due to light leaking out the
uncovered end of the bar before it is all absorbed in the wavelength
shifting fiber.  Empirically the light yield as a function of hit
distance from the MPPC is well-described by:
\begin{eqnarray}
\hspace{-8mm} I(x) & = & I_0 \left(1 - \frac{1}{2}e^{-x/M} - \frac{1}{2}e^{-(D-x)/M}\right) \nonumber \\
& & \times \left(e^{-(x+A)/S} + B e^{-(x+A)/L}\right)
\label{eq:hit_vs_q}
\end{eqnarray}
where $I_0$ is a normalization constant, $x$ is the distance of the
hit from the end of the bar closest to the MPPC, $D = 1864.3$~mm is
the length of the bar, $A=41.0$~mm is the extra length of fiber
extending beyond the end of the bar between the MPPC and the bar's
end, $S=410 \pm60$~mm and $L=23600\pm 2900$~mm are the short and long
attenuation lengths, and $B=0.739\pm0.005$ is the relative
normalization between the long and short components of the fiber's
attenuation curve.  The factor in the first set of parentheses
represents an exponential decrease in light yield very close to the
ends of the bars with length scale $M=21.55\pm 0.28$~mm, while the
second factor is the overall attenuation from the fiber itself.

Eq.~\ref{eq:hit_vs_q} is used together with the factor $C_{bar}$
defined in Sec.~\ref{sec:Cbar} to correct the quantity $N_{DPE}$
defined in Sec.~\ref{sec:satcorr}, which is proportional to the number
$N_{ph}$ of photons incident on the MPPC, to yield a quantity that is
assumed to be proportional to the number $N_{scint}$ of photons
produced in the scintillator.

\subsubsection{Scintillation photons to energy}
\label{sec:scint-energy}

The calibration steps described in the previous subsections correct
the MPPC response for its temperature dependence, differences among
bars, and the location of the energy deposition along the bar.  The
resulting corrected response is assumed to be proportional to the
number $N_{scint}$ of scintillation photons.  The conversion of this
response to energy deposition in the scintillator requires not only an
empirically determined normalization constant (about 21 detectable
photoelectrons per MeV), but also consideration of nonlinearity in the
scintillator response.  It is well known that when charged particles
lose energy by ionization in a scintillator (especially a plastic
scintillator), the light yield at high ionization density is quenched
below the linear extrapolation from low ionization density.  The
scintillator response is well represented by Birks'
formula~\cite{Birks:1964jb}:
\begin{equation}
\frac{dN_{scint}/dx}{C_{scint}\, dE/dx} = \frac{1}{1+C_B\, dE/dx}.
\end{equation}
Birk's constant $C_B$ was measured by the SCIBAR
Collaboration~\cite{Hasegawa:2006am} for protons incident on extruded
scintillator composed of the same material as the FGD bars. For this
analysis, we adopt their value ($C_B =
0.0208\pm0.0003$(stat)$\pm0.0023$(sys)~cm/MeV).

The final normalization factor required to convert the resulting
corrected response to energy deposition incorporates the product
$\epsilon (1+n_{sec}) C_{scint}$ as well as the efficiencies for
optical coupling of the scintillator to the fiber and thence to the
MPPC.  This factor is empirically determined as that required to
produce a measured distribution of energy depositions by cosmic rays
that matches that predicted by a detailed simulation accounting for
the kinematic distribution of the cosmic ray flux, as well as the
dependence of the FGD trigger efficiency on the cosmic ray track
properties.  Great care was taken to ensure that the simulated spatial
and angular distributions of the cosmic ray tracks reproduce the data.
Section~\ref{sec:lightyield2} provides more information about this
simulation.  Note that since the deposited energy depends on the path
length through the scintillator, measured energy depositions can be
compared to expectations only for hits belonging to accurately
reconstructed tracks that define the path length for each hit.

Figures~\ref{fig:pid1} and \ref{fig:pid2} show energy deposition by
particles stopping in an FGD, reconstructed from experimental data
recorded with the neutrino beam and a cosmic ray trigger,
respectively.  The agreement of the distributions with the curves
representing simulated energy deposition for the same particle ranges
demonstrates that the calibration chain operates correctly.

\subsection{FGD timing calibration}
\label{sec:timecalib}

The FGD time calibration involves the following elements:

\begin{itemize}
\item Pulse fitting.
\item FGD internal time calibration using timing markers.
\item FGD internal time calibration using FEB-to-FEB corrections.
\item Corrections to align FGD times with other subdetectors (not
described in this paper).
\item Calculation of average time for tracks in each FGD using timing
for each hit, including light travel time correction.
\end{itemize}

\subsubsection{Pulse fitting}

Recorded waveforms are fitted in order to determine the hit time as
accurately as possible. In order to save computation time, only
waveforms with pulse heights larger than 150 ADC counts ($\sim$~4
avalanches) are fit. The functional form of the fitted pulse is:
\begin{eqnarray}
\label{eqn:cali8}
\hspace{-8mm}y(t > t_0) & = & B + A \left(\frac{t-t_{0}}{\tau}\right)^{4}\left(1-\frac{t-t_{0}}{\lambda\tau}\right) \nonumber \\
 & & \times \hspace{2mm} e^{-\frac{t-t_{0}}{\tau}}\,,
\end{eqnarray}
with $B$ the waveform baseline, $A$ the pulse amplitude, $t_0$ the start
time of the pulse, $\tau$ the amplifier-shaper time constant and
$\lambda$ the time constant of the undershoot following the
pulse. Only the parameters $A$, $B$ and $t_{0}$ are allowed to float in
the fit; the parameters $\tau$ and $\lambda$ are fixed to 43.4~ns and
6.93, respectively.  Only the leading edge of the FGD pulse is fit;
specifically, the fit is over the region from 240~ns before the actual
peak ADC sample up to the peak. Fitting the leading edge was found to
provide optimum timing resolution by minimizing the sensitivity of the
measure to after-pulsing and late photons (typically reflected from
the mirrored end).  The fitted time $t_{0}$ is stored and processed by
the subsequent time calibration steps.  If the high-gain pulse
corresponds to more than about 65 avalanches, then the fitted time
from the low gain channel is used.

\subsubsection{Corrections based on CMB timing markers}

The fitted time for each hit is corrected for the variation among the
phases of the various clock domains and for the differences in fiber
cable lengths.  The correction for these effects is made with the FGD
timing markers.  A diagram of the FGD timing marker distribution is
shown in Fig.~\ref{fig_fgd_timing_marker_diagram}.

\begin{figure}[htb]
\centering
\ifx\figstyle\bw
 \includegraphics[width=\columnwidth]{jpgFigures/fig35_bw.jpg}
\else
 \includegraphics[width=\columnwidth]{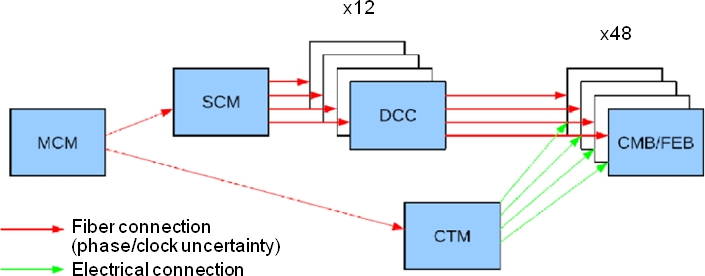}
\fi
\caption[FGD Example Timing Marker] {Diagram showing how the FGD
timing marker pulses are distributed through the FGD-CTM.  As can be
seen, the timing marker generation is largely independent of the
various optical fiber clock shifts associated with the FGD data
readout chain.}
\label {fig_fgd_timing_marker_diagram}
\end {figure}

The FGD Cosmic Trigger Module (CTM) has electrical LVDS connections to
each of the 48 CMBs.  When the MCM issues a trigger, the CTM sends a
set of approximately simultaneous LVDS pulses to each CMB.  Each CMB
fans out these signals asynchronously and injects them into a spare
channel on the FEB ASICs.  Because the signals do not propagate
through the digital SCM$\rightarrow$DCC$\rightarrow$CMB connections,
they are not influenced by the various digital clock domains.  The
times of arrival of the timing markers at the FEBs are fixed relative
to one another.  Hence the differences in the times recorded for the
timing marker pulses provide a measurement of the effects of these
clock domain jitters and hence a way of correcting for the jitter.  In
practice the correction works as follows: for a particular hit, the
difference between the timing marker for the ASIC of that hit relative
to the timing marker for the first ASIC in the FGD is calculated; this
difference is subtracted from the particular hit time.

\begin{figure}[htb]
\centering
\includegraphics[width=\columnwidth]{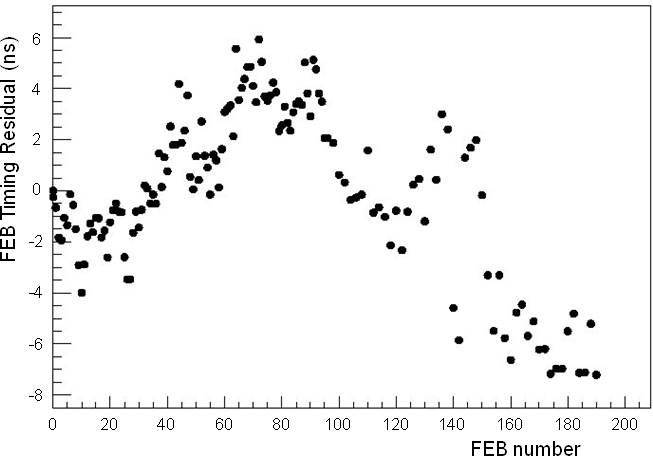}
\caption[FEB-to-FEB timing corrections]
{Measured FEB-to-FEB timing corrections.} 
\label {fig_feb_to_feb_corrections}
\end {figure}

\subsubsection{FEB-to-FEB time corrections}

A series of studies using cosmic rays has shown that while the timing
markers account for clock jitters that vary from run to run and event
to event, there remain residual FEB-to-FEB time differences.  These
residual differences are not unexpected, since the CTM-CMB cable
lengths are not identical, and the FGD CTM does not actually produce
the timing markers at exactly the same time; the design of the CTM
hardware and firmware makes this unavoidable.

However, the differences in the timing marker production times are
found to be constant from event to event.  We can therefore make an
empirical correction for this residual difference using offsets
calculated using cosmic tracks. For each track, the difference between
the individual hit time and the track time averaged over all the hits
is calculated. The residuals used in the correction are the average
values of the difference calculated for a large number of cosmic
rays. The set of measured corrections is shown in
Fig.~\ref{fig_feb_to_feb_corrections}.

\subsubsection{Calculation of final FGD time}

The main utility of the FGD hit times is to calculate a time of
passage through the FGD of a reconstructed track as an average time
for the hits in the track.  The hit times are first corrected for the
light travel time down the WLS fiber, using the track information
about the hit location. The dependence of this propagation time on hit
location is calibrated by averaging over a large number of cosmic ray
tracks while using the average track time as reference. The
relationship between the measured time and the distance between the
track crossing point along the bar and the MPPC is found to be not
strictly linear. This phenomena is well reproduced by simulations
accounting for the 17.2 cm/ns propagation speed, the photon
reflections on the mirror and MPPC, and convoluting the photon arrival
timing distribution with the electronics response function. The
measured curve is used to correct for the effective propagation delay.

\begin{figure}[htb]
\centering
\includegraphics[width=\columnwidth]{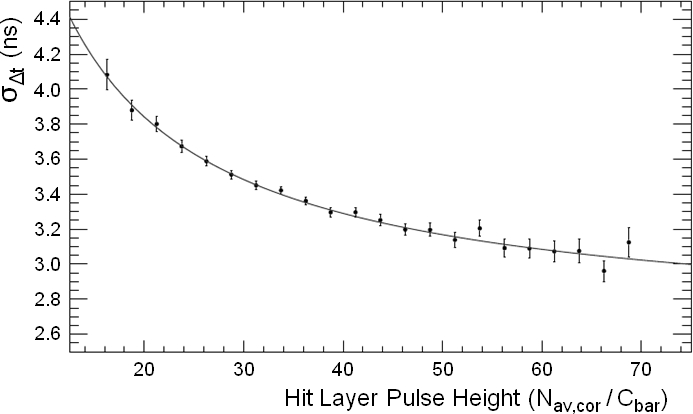}
\caption[FGD Timing Resolution vs Hit Charge] {Spread of timing
residuals $\Delta$t between the hit layer and reference layer as a
function of the hit layer charge.}
\label {fgd_timing_reso_vs_hit_charge}
\end {figure}

The single-hit timing resolution was measured by constructing
histograms of the time difference between a given hit and a reference
hit in the first FGD layer (with a correction for particle travel
time) as a function of both the hit charge $N_{av}$ and the reference
hit charge.  The widths of these curves for a particular reference hit
charge are shown in Fig.~\ref{fgd_timing_reso_vs_hit_charge} as a
function of the ``test'' hit charge.  They are well fitted by a
constant contribution accounting for the average resolution for the
reference hits added in quadrature with a contribution depending on
the charge of the ``test'' hit:
\begin{equation}
\sigma_t(N_{av}) = \frac{12.5 ns}{\sqrt{N_{av}}}\,.
\label{eqn:tresn}
\end{equation}
Equation~\ref{eqn:tresn} could be interpreted as the single-hit time
resolution if the statistical complications arising from, e.g.,
secondary avalanches and light reflection from the mirrored fiber ends
could be neglected.  The value of $12.5\pm 0.6$~ns might be
interpreted as the single-avalanche timing resolution, which is
closely related to the light decay time constants of scintillator and
wavelength-shifting fiber, but smeared by light reflection from the
mirrored fiber ends.

\begin{figure}[htb]
\centering
\includegraphics[width=\columnwidth]{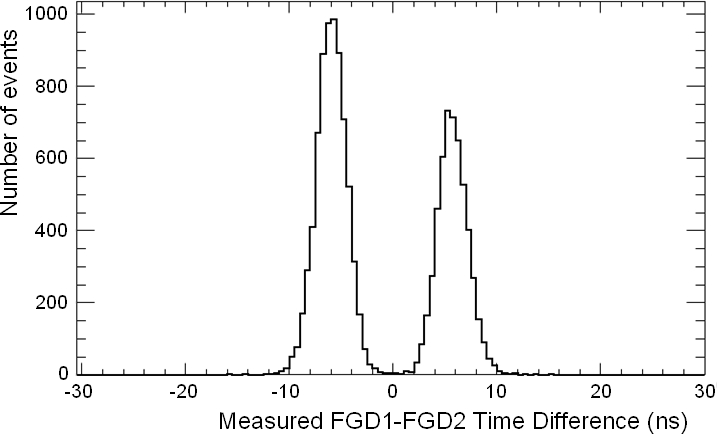}\\
~~\\
\ifx\figstyle\bw
 \includegraphics[width=\columnwidth]{jpgFigures/fig38b_bw.jpg}
\else
 \includegraphics[width=\columnwidth]{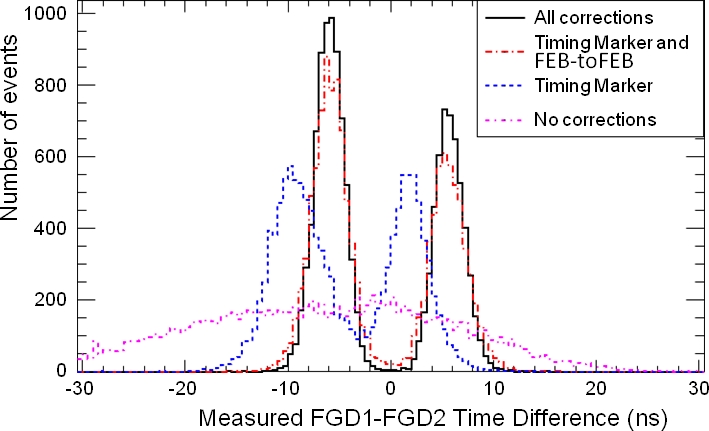} 
\fi
\caption[Difference between FGD1 and FGD2 times] {FGD1 - FGD2 time
difference for FGD-triggered cosmic rays.  The two peak structure is
from cosmic rays that hit either FGD1 or FGD2 first.  The upper
plot shows the FGD1-FGD2 time differences with all corrections.  The
lower plot shows the FGD1-FGD2 time differences as the various
corrections are applied.}
\label {fgd_fgd1_fgd2_time_diff}
\end {figure}

\subsubsection{Tests of FGD timing resolution}

A weighted average time for each FGD is calculated for reconstructed
tracks, where the inverse weight is the expected timing resolution for
a single hit associated with the track.  The principal test of the FGD
timing calibration is to look at the time difference between the two
FGDs for FGD-triggered cosmic-ray events that hit both FGDs. In
approximately half the events the muon hits FGD1 first; in the other
half FGD2 is hit first.  We use the fully calibrated time for each FGD
track, calculated in the FGD reconstruction described in the previous
sections.

The distribution of the difference between FGD1 and FGD2 track times
is shown in the upper plot of Fig.~\ref{fgd_fgd1_fgd2_time_diff}.
This plot shows the expected two peak structure, which comes from the
fact that the cosmic muon hits either FGD1 or FGD2 first.  The width
of each FGD1-FGD2 peak for the fully corrected data curves is 1.47 ns.
The clear separation of the two peaks shows that the FGD1-FGD2 time
difference can be used to distinguish track directions between the two
FGDs.  The lower plot shows the effect of the various calibration
stages.  As expected, each calibration stage makes the distributions
narrower.

\section{M11 tests}
\label{sec:m11tests}

Prior to shipment to Japan, assembled FGDs were installed in the M11
secondary beam line at TRIUMF for testing. The goal of these tests was
to calibrate and assess the performance of the FGD with beams of
electrons, muons, pions and protons of known momenta. The tests were
conducted during two running periods: in the fall of 2008 and spring
of 2009.  The earlier tests employed FGD1, with only a limited number
of instrumented channels.  Subsequently, one of the ND280 TPCs was
installed upstream of the FGD.  For later tests, FGD1 was fully
instrumented.  Finally, FGD1 was replaced by FGD2.

The location of the end of the M11 beamline is fixed.  However, for
the tests, the FGD was mounted on a rail system and could thus be
moved horizontally to expose its entire width to the beam.  It could
also be rotated by up to 45$^{\circ}$.  Its vertical position could
not be changed.

\subsection{Beamline and M11 area}
\label{sec:m11area}

The TRIUMF cyclotron provides a 500 MeV proton beam composed of
3\,-\,4~ns wide bunches every 43~ns.  Secondary particles are created
via interactions of the primary beam in a beryllium production target
(T1).  The M11 channel views T1 at an angle of 7$^{\circ}$.  It
consists of two bending magnets and six quadrupole magnets.  Its
length from T1 to the center of the final quadrupole is 13 m.  The channel
provides secondary charged particles: $e$'s, $\mu$'s, and $\pi$'s, of
either polarity, with momenta up to 400 MeV/c.  It also provides $p$'s
up to 400 MeV/c, but these penetrate the FGD to a depth of only 6~cm.

For the FGD tests, a {\em front} plastic scintillator trigger counter
was installed at the end of the beamline just downstream of the last
magnet.  A second, much larger, {\em back} plastic scintillator
trigger counter was installed downstream of the FGD.  Another possible
trigger was provided by a movable plastic scintillator hodoscope
located at an intermediate dispersed focus near the center of the
channel.

At lower beam momentum, particle identification was based on
time-of-flight through the channel using the relative times of the
various trigger counters as well as the time provided by a capacitive
pickup probe just upstream of the T1 production target.  This method
ceased to be useful above $\sim$~250~MeV/c because the times of
flight become too similar to be distinguished.  Nevertheless, it was
possible to identify $p$'s by their large energy deposit in the front
trigger counter.  Below $\sim$~200 MeV/c, demanding a coincidence
between front and back trigger counters would select purely $e$'s,
since heavier particles would range out in the FGD material.

The size of the beam spot at the FGD was $\sim$~15 cm FWHM in the
horizontal direction and $\sim$~8~cm FWHM in the vertical.  The
angular spread of the beam was $\sim$~3$^{\circ}$ FWHM.  The beam
flux varied with momentum and could be limited using sets of vertical
and horizontal slits and jaws in the channel.  Typical rates of
$\sim$~10~Hz were used for testing in order to match the capabilities
of the developing data acquisition system.

\subsection{Performance}

Some measurements could be compared to Monte Carlo simulations, which
were carried out using the same GEANT code as for the full ND280
detector~\cite{Abe:2011ks}, but with the addition of the M11 front
trigger counter.

The simplest such comparison, not requiring calibration beyond a
simple pulse height threshold, involved histogramming the number of
hits in individual FGD layers for pions and muons of various incident
momenta (electrons passed right through the FGD and protons stopped in
the first few centimeters). The range can be varied through the
thickness of the FGD by varying the incident momentum.  One expects to
see a uniform distribution of hits versus layer number in layers
upstream of the layer corresponding to the particle's range at the
given momentum, at which point most particles stop, so there is a
sharp drop and few hits in layers downstream.  The results of
measurements at 190 MeV/c are shown in Fig.~\ref{fig:rangePlot}.
Excellent agreement can be seen between data and Monte Carlo for this
comparison.  The distributions for pions are not as clean as for muons
because of pion decays in flight upstream of the FGD.  Qualitatively
similar results were obtained for other momenta.

\begin{figure}[htb] 
\centering
\ifx\figstyle\bw
 \includegraphics[width=\columnwidth]{jpgFigures/fig39a_bw.jpg}\\
~~\\
 \includegraphics[width=\columnwidth]{jpgFigures/fig39b_bw.jpg}
\else
 \includegraphics[width=\columnwidth]{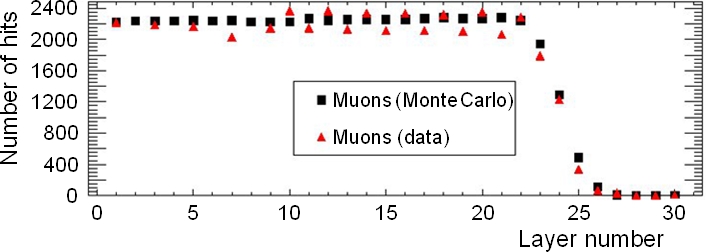}\\
~~\\
 \includegraphics[width=\columnwidth]{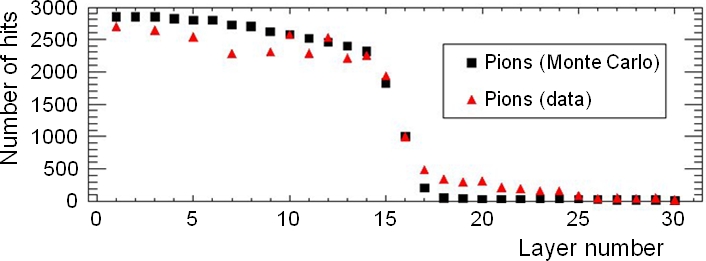}
\fi
\caption{Comparison of data and Monte Carlo for the ranges of pions
and muons in FGD1. The M11 channel was tuned to 190 MeV/c. Some
channels were not instrumented at the time these measurements were
made.}
 
\label{fig:rangePlot}
\end{figure}

Having identified the FGD bar in which a muon has stopped, one can
look for a second hit in the same bar at a later time, arising from
muon decay.  One expects the time distribution of such late hits to
decrease exponentially, corresponding to the muon lifetime.  Such
measurements were made and found to be in agreement with expectations.

Electrons and positrons at M11 momenta are expected to pass through
the FGD, depositing the same amount of energy in every layer.  This
feature proved useful in developing the FGD calibration chain
described in detail in Section~\ref{sec:calibration}.  Various effects
that required correction were first identified in the M11 data
analysis, and possible correction algorithms were tested and optimized
using the $e$ beam data.  Temperature-dependent effects were
particularly evident in the earliest runs since the cooling system
described in Section~\ref{sec:coolingSystem} was not yet in place.

After calibration, several additional analyses and data/Monte Carlo
comparisons could be performed:

\begin{itemize}
\item
The most basic of these was a comparison of the pulse height
distributions of $e$'s, $\mu$'s, and $\pi$'s in various FGD layers.
Simulations agreed with the relative distributions observed and the
data were used to adjust Monte Carlo parameters that had not yet been
optimized.  For $\mu$'s and $\pi$'s stopping in the FGD, dE/dx
increases as the particles slow down and the corresponding increase in
pulse height was seen and matched by the simulations.

\item
The layer with the largest pulse height was taken to be a measure of
muon range and the range was observed to vary with incident muon momentum in
the way expected.

\item
Using an $e$ beam, the hit information from vertical FGD bars was used
to determine the hit location along the length of the horizontal bars.
By translating the FGD horizontally, the dependence of pulse height on
distance from the MPPC readout was measured to determine the light
attenuation in the WLS fibers.  The light attenuation curve measured
in M11 agrees well with those determined from the fiber test-bed
measurements described in Section~\ref{sec:fiber}, except at the very
ends, where a deficit was observed in the test-beam results as
described in Sec.~\ref{sec:attcorr}.  This feature was subsequently
confirmed using cosmic rays {\em in situ} (see
Section~\ref{sec:results}), as well as in bench tests.
\end{itemize}

\subsection{MPPC saturation and Birks' effect}

Figure~\ref{fig:birks} shows the observed FGD response (M) for
$\mu$'s, $\pi$'s, and $p$'s, measured at various incident momenta in
the first two layers of FGD1, normalized by the response measured for
350~MeV/c $\pi$'s, which we take as the value for minimum ionizing
particles (mips).  The results are plotted versus the similarly
normalized calculated energy loss, also in units of mips.  The left
panel of the figure shows results corrected by a combination of
elements of the calibration chain yielding $N_{av,cor}/C_{bar}$ (see
Eq.~\ref{eqn:calib5.5} in Sec.~\ref{sec:tempcor}, and also
Sec.~\ref{sec:Cbar}).  The middle panel shows the results with the
MPPC saturation correction also applied, as described in
Sec.~\ref{sec:satcorr}, while the right panel shows the results with
the Birks' correction described in Sec.~\ref{sec:scint-energy} applied
in addition.  As can be seen, each correction improves the fit.

\begin{figure}[htb]
\centering
\ifx\figstyle\bw
 \includegraphics[width=\columnwidth]{jpgFigures/fig40_bw.jpg}
\else
 \includegraphics[width=\columnwidth]{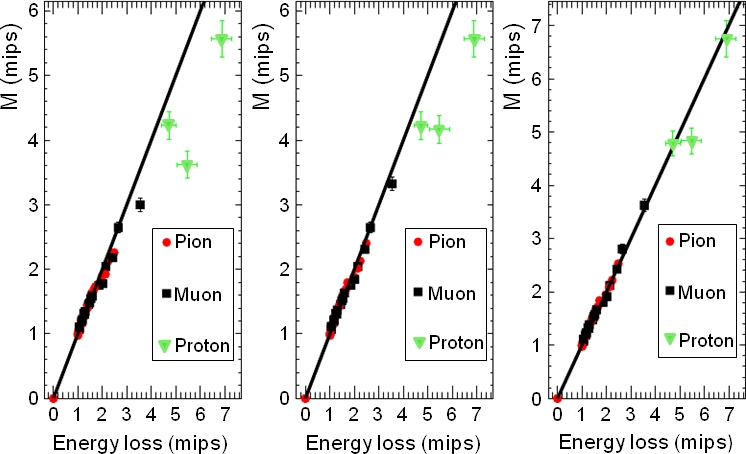}
\fi
\caption{FGD responses for muons, pions and protons from various
stages of the calibration chain, versus calculated energy loss.  Both
axes are normalized by the corresponding values for 350~MeV/c pions,
which is taken to be the value for minimum ionizing particles. Left:
calibration chain corrections applied; middle: saturation corrections
also applied; right: correction for Birks' effect additionally
applied.  See text for details.}
\label{fig:birks}
\end{figure}

\section{{\em In situ} Performance Measurements}
\label{sec:results}

The FGDs were shipped to the J-PARC laboratory in summer 2009, where
they were inspected, re-assembled, and tested in the assembly building
adjacent to the ND280 pit.  In October 2009 both detectors were
installed into the ND280 basket by overhead crane and services were
connected.  Commissioning and integration with the global DAQ
continued through the remainder of 2009, and in February 2010 the
ND280 detector as a whole began operations in the neutrino beam.  The
FGDs operated successfully during normal neutrino running
(February\,-\,June 2010, December 2010\,-\,March 2011), until
operations were interrupted by the large March 11, 2011 earthquake in
eastern Japan.  Subsequent tests of the detectors with cosmic-ray
triggers taken in the summer of 2011 showed no ill-effects resulting
from the earthquake.

This section will highlight the operational performance of the
detectors after installation into ND280.

\subsection{Reliability}

The FGD hardware has operated reliably with minimal downtime since its
installation.  Regular studies of dark noise data and voltage scan
calibrations have been used to identify channels with abnormal
signals.  During the first period of T2K running (February\,-\,June 2010),
approximately 30 FGD channels (out of 8448) appeared to be
nonfunctional, typically showing an absence of dark noise pulses.
These channels were scattered randomly across both detectors.  In June
2010 one front-end board malfunctioned, turning off an additional 64
channels for the remainder of the month.  This is the only example to
date of a board-wide failure.

During the summer of 2010 the failed front-end board was replaced and
an effort was mounted to reduce the number of isolated malfunctioning
channels.  It was found that about half the dead channels were
caused by problems on the front-end board itself, most often a poor
solder connection between the signal trace and the connector pin that
plugs into the minicrate backplane.  By replacing or repairing such
cards the number of dead channels was reduced to 17 (0.2~\% of the
detectors).  These remaining channels are believed to be due to
isolated faults on the backplanes themselves or else MPPC failures.
Their number is low enough to represent a negligible problem for
data-taking.

Because the MPPCs themselves are quite sensitive to temperature, the
slow control system constantly monitors the environment inside the
dark box.  Fig.~\ref{fig:temp} shows the typical temperature variation
over several days during normal beam running with the ND280 magnet
closed and powered.  The temperature variations are small enough that
there is no need to adjust the operating voltages of the MPPCs to
compensate, but instead these effects are compensated by the
calibration chain (see Section~\ref{sec:calibration}).

\begin{figure}[htb]
\centering
\ifx\figstyle\bw
 \includegraphics[width=\columnwidth]{jpgFigures/fig41_bw.jpg}
\else
 \includegraphics[width=\columnwidth]{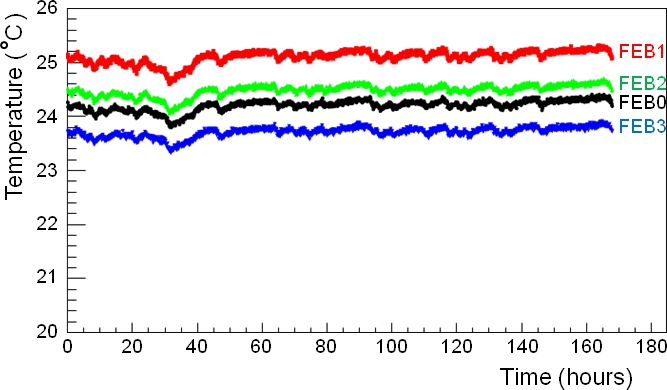}
\fi
\caption{Temperature measured on four front-end boards inside the FGD1
 dark box during neutrino running.}
\label{fig:temp}
\end{figure}

\subsection{Hit efficiency}

\begin{figure}[ht]
\centering
\includegraphics[width=\columnwidth]{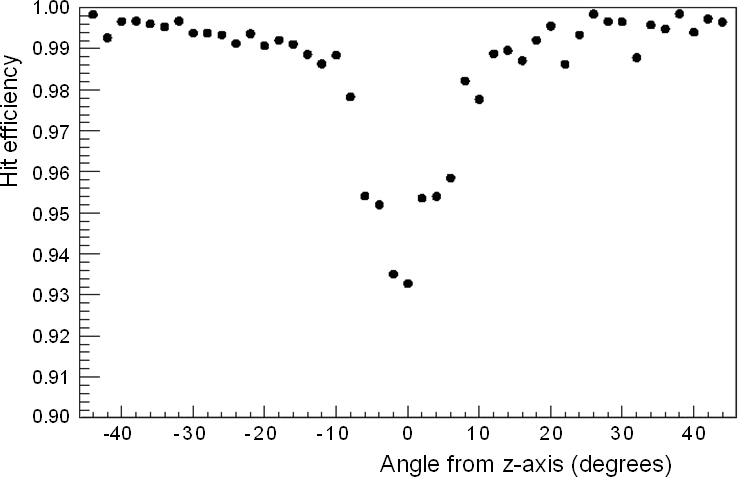}\\
~~\\
\includegraphics[width=\columnwidth]{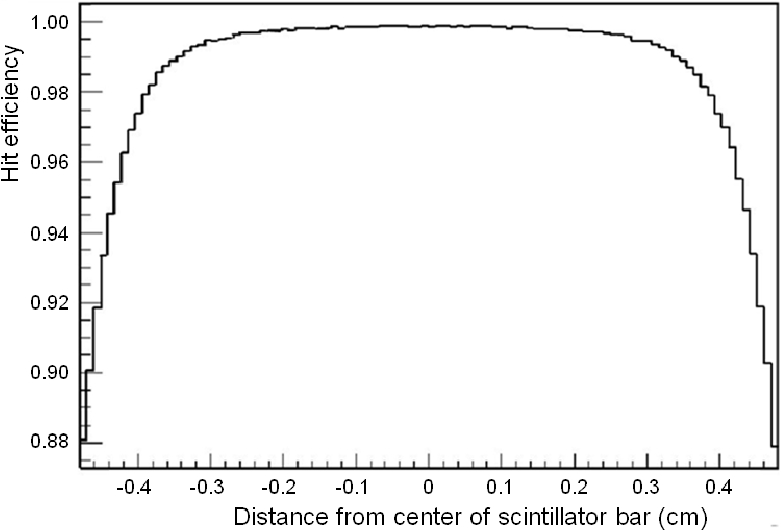}
\caption{Measured hit efficiency as a function of track angle relative
to bar axis (top) and position relative to bar width (bottom).}
\label{fig:hiteff}
\end{figure}

The hit efficiency of a scintillator bar may be estimated by analyzing
cosmic-ray events that pass through an FGD.  For simple cosmic-ray
events with isolated tracks, hit efficiency may be estimated by
looking at tracks that pass through several XY modules.  If the track
is reconstructed as passing through the $n^{th}$ and $n+2^{th}$ layer,
then a hit should have been recorded in layer $n+1$.  The fraction of
the tracks where the middle layer is missing a hit gives the hit
inefficiency.

Because the number of photoelectrons produced by a typical minimum
ionizing particle ($\sim 30$~pe) is much larger than the effective
threshold of the pulse finder ($\sim 0.5$~pe) or any analysis
threshold ($\sim 5$~pe), the hit inefficiency is dominated by tracks
that pass through the inactive coating of the scintillator bar.  This
is illustrated in Fig.~\ref{fig:hiteff}, which shows the hit
efficiency for cosmic ray events as a function of angle or position
across the bar.  As a function of angle the inefficiency occurs
primarily for tracks that move parallel to the coating of the bar
itself.  The size of the dip at zero angle corresponds approximately
to the thickness of the coating divided by the bar width.  In other
words, the missed tracks are those that skim along the coating at the
edge of a bar rather than passing through active scintillator.  Viewed
as a function of position on the bar, the efficiency is very high for
tracks passing through the center of the bar, but drops off for tracks
that hit the edges of the bar, which depending on their angle may
again pass through the inactive coating and miss the active volume.

\subsection{Light yield}
\label{sec:lightyield2}

\begin{figure}[htb]
\centering
\includegraphics[width=\columnwidth]{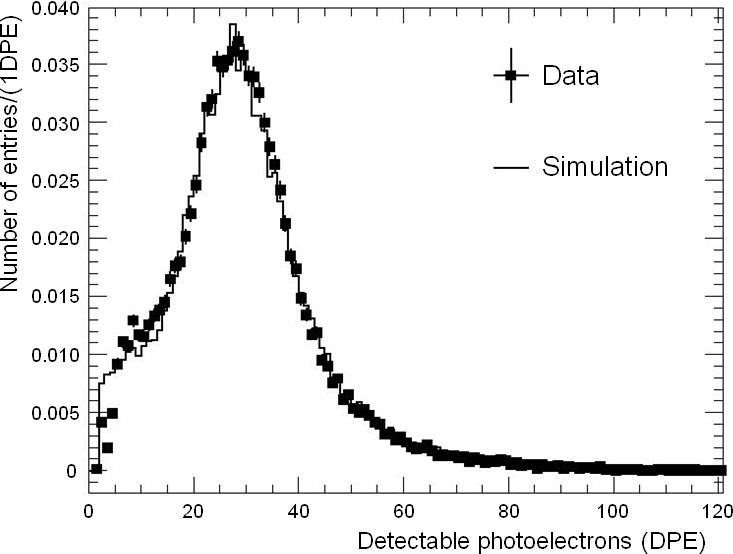}
\caption{Measured spectrum of pulse heights processed to represent a
quantity proportional to the number $N_{scint}$ of scintillation
photons (see Sec.~\ref{sec:attcorr}), for cosmic ray events triggered
by the FGDs.  The data are compared to the corresponding result
extracted from simulated events.  The horizontal scale of the
simulated spectrum is normalized to the data.}
\label{fig:pulseheight}
\end{figure}

The ND280 software is used to do a full Monte Carlo simulation of
cosmic rays passing through the detector.  The simulation includes a
detailed GEANT4 model of the detectors.  The processes of photon
production and propagation in scintillator and fiber are not treated
in a Monte Carlo technique, but rather using empirical analytic models
for light attenuation in the fibers (as in the calibration chain) and
the spatial distribution of the light incident on the MPPC pixel array
that is derived from measurements with
fibers~\cite{Bryant:2005mb}.  Then the simulation returns to a
Monte Carlo technique for the behavior of the MPPC.  Optical photons
are generated with this spatial distribution and with a time distribution
given by the decay constants of light production in the scintillator
and fiber together with the superposition of direct and reflected
paths in the fiber.  The MPPC behavior is simulated by a model that
may generate a primary and possibly a secondary (possibly delayed)
avalanche from each incident photon~\cite{Vacheret:2011zza}.  Each
avalanche produces a charge impulse, the series of which is then
processed by a model of the electronic analog and digital system.  The
simulation produces digitized MPPC waveforms, as well as logic signals
from a model of the FGD's self-trigger system.

Figure~\ref{fig:pulseheight} shows the spectrum, for FGD-triggered
cosmic ray events, of a measured quantity assumed to be proportional
to the number of scintillation photons (see last paragraph of
Sec.~\ref{sec:attcorr}), superimposed upon a histogram of the same
quantity extracted from reconstructed events from this
simulation\footnote{ The version of the simulation used for this
figure employed a uniform distribution of photons incident on the
MPPCs from the fibers.}.

\begin{figure}[htb]
\centering
\ifx\figstyle\bw
 \includegraphics[width=\columnwidth]{jpgFigures/fig44_bw.jpg}
\else
 \includegraphics[width=\columnwidth]{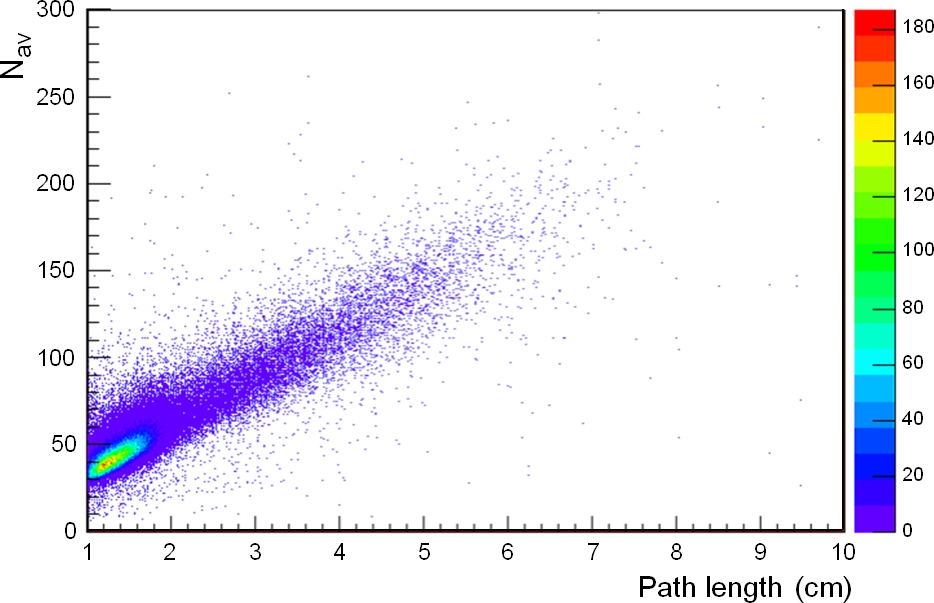}
\fi
\caption{Correlation of the number $N_{av}$ of MPPC avalanching pixels 
with path length through the scintillator bar.}
\label{fig:dedx}
\end{figure}

The total number of photons detected in an MPPC should be proportional
to the total deposited energy by a minimum ionizing particle in the
active part of the scintillator bar, and hence to the path length of
the track through the bar.  Fig.~\ref{fig:dedx} illustrates the
measured MPPC avalanche yield from cosmic rays travelling through an
FGD versus the calculated path length of the track through the bar.
Although each bar is only 9.61~mm wide, much longer path lengths are
possible for tracks with a direction component parallel to the bar's
axis.

\subsection{Particle identification}

\begin{figure}[htb]
\centering
\ifx\figstyle\bw
 \includegraphics[width=\columnwidth]{jpgFigures/fig45_bw.jpg}
\else
 \includegraphics[width=\columnwidth]{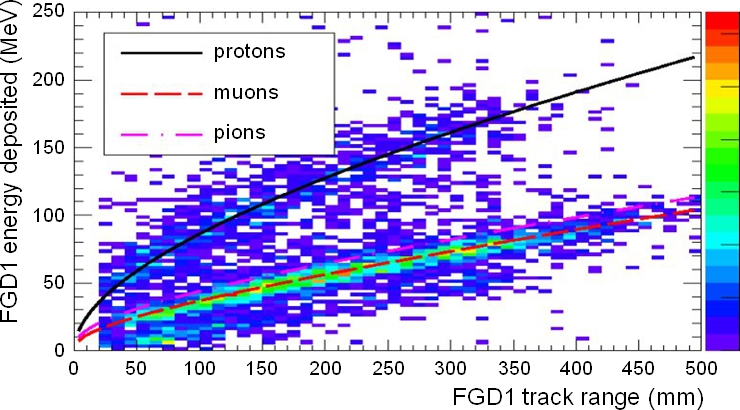}
\fi
\caption{Deposited energy vs range for particles stopping in FGD1.
The scatterplot shows stopping particles in neutrino beam data, while
the curves show the MC expectations for protons, muons, and pions.}
\label{fig:pid1}
\end{figure}

\begin{figure}[htb]
\centering
\ifx\figstyle\bw
 \includegraphics[width=\columnwidth]{jpgFigures/fig46_bw.jpg}
\else
 \includegraphics[width=\columnwidth]{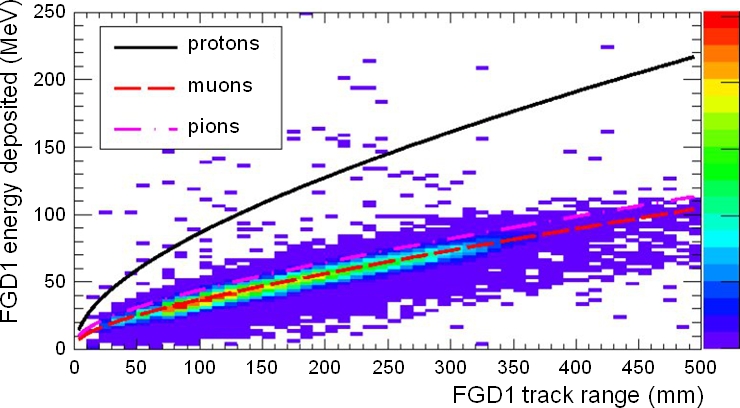}
\fi
\caption{Deposited energy vs range for particles stopping in FGD1.
The scatterplot shows stopping particles from cosmic ray triggers,
while the curves show the MC expectations for protons, muons, and
pions.}
\label{fig:pid2}
\end{figure}

Because slower moving particles deposit more energy per path length,
the energy loss summed over bars can be used to identify the type of
particle that produced a track stopped in the
FGD~\cite{Licciardi:thesis}.  The measured light yield of a hit along
a track is first converted into an equivalent energy deposit in MeV by
the calibration chain described in Sec.~\ref{sec:calibration}, which
includes corrections for MPPC gain and saturation, Birks' constant
effects, and the attenuation curve of the bar/fiber combination.  By
comparing the measured total energy deposit for a given particle range
in the FGD to the theoretically expected energy deposit for particles
with that range, protons can be distinguished from muons and pions.
Fig.~\ref{fig:pid1} shows a scatterplot of deposited energy vs. range
for particles produced by neutrino interactions and stopping in FGD1.
The solid, dashed and dot-dashed lines show the expected locations of
protons, muons, and pions, respectively, on this plot.  A distinct
population of protons can be discerned.  In Fig.~\ref{fig:pid2},
similar distributions are shown for cosmic rays that stop in the FGD.
In this case the muon line is well populated, as expected, but protons
are of course absent.

\subsection{Michel electron studies}

\begin{figure}[ht] 
\centering 
\includegraphics[width=\columnwidth]{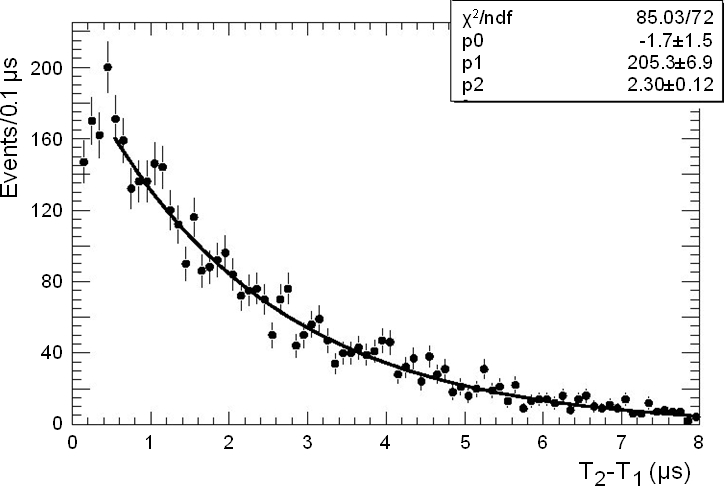}\\
~~\\ 
\includegraphics[width=\columnwidth]{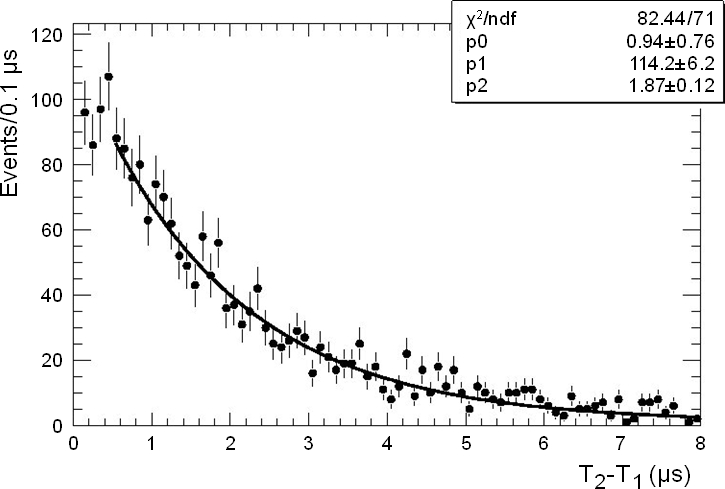} 
\caption{Measured lifetime distributions for positive muons (top) and
negative muons (bottom) that stop inside an FGD. The data are fitted
with an exponential plus a flat background. See text for details.}
\label{fig:michel}
\end{figure}

Michel electrons produced by muons that stop in an FGD can be
identified by looking for a delayed cluster of hits following the
initial neutrino interaction.  The FGD electronics reads out a
10~$\mu$s waveform for each beam spill, while the beam bunches span an
interval of just over 4~$\mu$s.  The electronics is therefore
sensitive to Michel electrons for over four times the muon lifetime
for muons produced early in the beam spill, dropping to about two and
a half muon lifetimes for neutrino interactions late in the spill.

As a demonstration of the FGD's ability to identify Michel electrons,
we have looked for delayed clusters of hits following cosmic ray muon
events.  Stopping cosmics are selected by identifying muons that pass
entirely through one FGD and enter the other FGD, without leaving hits
in the far side of the second FGD. A ``delayed cluster'' is defined as
a group of at least two hits that occurs at least 100~ns after any
previous hits.

Fig.~\ref{fig:michel} shows the distribution of the time interval
between the primary muon and the delayed cluster of hits for tracks
stopping in the FGDs.  The sign of the initiating muon can be
determined from the track curvature in the magnetic field.  Each
fitted curve is the sum of an exponential decay with normalization
$p1$ and lifetime $p2$ on top of a flat background $p0$.  The fitted
backgrounds are consistent with zero, as expected. For positive muons
the fitted lifetime of $2.23 \pm 0.09$~$\mu$s is consistent with the
muon lifetime, while for negative muons, the fitted lifetime ($1.87
\pm 0.12$~$\mu$s) is shorter, since negative muons can become trapped
in atomic states and then absorbed.  The measured lifetime for
negative muons is consistent with previous measurements of the $\mu^-$
capture lifetime on $^{12}$C~\cite{Suzuki:1987jf}.

\subsection{Neutrino beam monitoring and stability}

\begin{figure}[htb]
\centering
\ifx\figstyle\bw
 \includegraphics[width=\columnwidth]{jpgFigures/fig48_bw.jpg}
\else
 \includegraphics[width=\columnwidth]{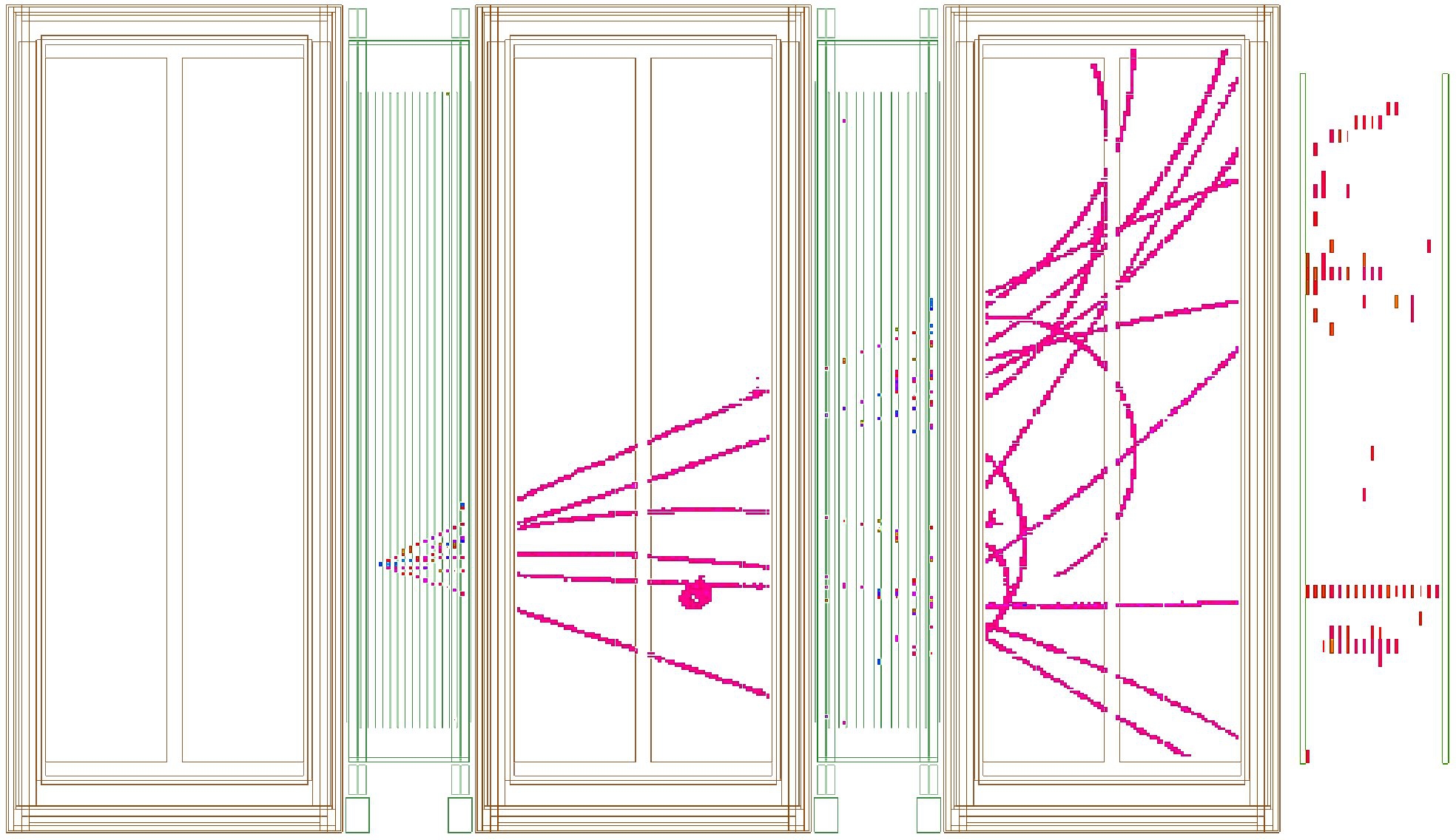}
\fi
\caption{Sample ND280 event display for a neutrino interaction in the
tracker.  This appears to be a deep inelastic scattering interaction
in FGD1.}
\label{fig:eventdisplay}
\end{figure}

Fig.~\ref{fig:eventdisplay} shows an event display of an actual
neutrino event in the ND280 tracker.  Most likely a neutrino has
undergone deep inelastic scattering inside the first FGD, resulting in
several particles that leave tracks in both FGDs and in the TPCs.  The
TPC track information is a key element of the FGD reconstruction and
is used to associate hits to tracks for those particles that penetrate
the TPCs.

\begin{figure}[htb]
\centering
\includegraphics[width=\columnwidth]{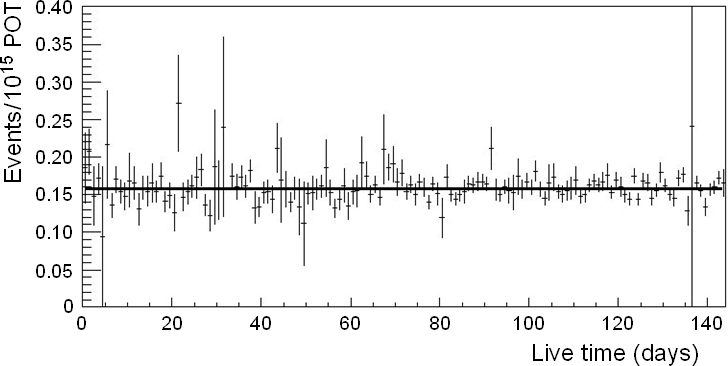}
\caption{Measured neutrino interaction rate in the FGDs per POT as a
function of day of beam running.}
\label{fig:ratevspot}
\end{figure}

Using a set of low-level selection criteria, one can search for events
that occur inside an FGD.  Hits that occur in time with the expected
neutrino beam bunch timing are selected and grouped in time if they
occur within 10~ns of each other.  Noise hits are rejected by a hit
threshold of 2.5~pe on individual hits and a requirement that there be
at least two hits that total at least 10~photoelectrons.  Finally,
interactions that occurred in the FGD are selected by vetoing events
with hits in the most upstream layers of the FGDs (to remove entering
events from upstream), and at least 3 spatially contiguous hits are
required in order to select track-like events.  (These selection
requirements are not the same as those used for T2K's normal analysis,
but rather provide a low-level cross-check that uses only information
from the FGDs but not other detectors.)

Fig.~\ref{fig:ratevspot} shows the number of candidate interactions
satisfying these criteria per proton delivered on target (POT), as a
function of calendar time.  The rate is constant, as expected.  As the
beam power gradually increased over the running period, one can see
that the statistical uncertainties on each day's rate measurement
generally decreased.

\begin{figure}[ht] 
\centering 
\includegraphics[width=\columnwidth]{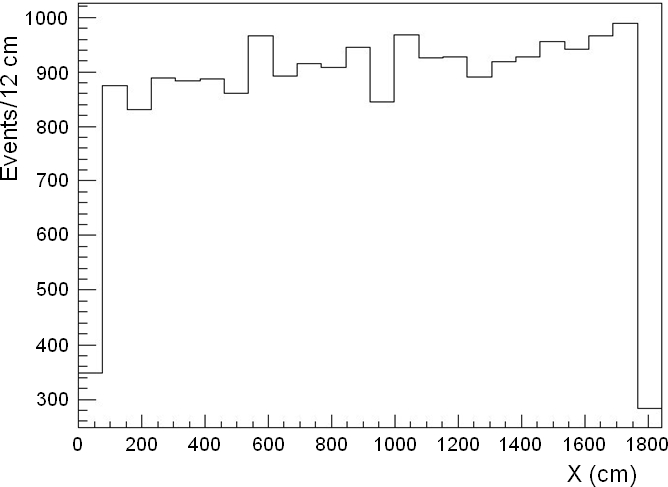}\\
~~\\ 
\includegraphics[width=\columnwidth]{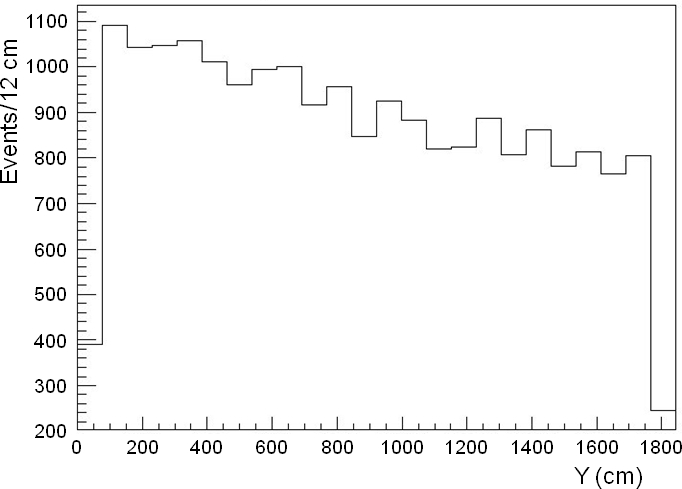} 
\caption{Vertex distribution in $x$ (top) and $y$ (bottom) inside the
FGDs.  The slopes indicate the change in flux with off-axis angle.}
\label{fig:profile}
\end{figure}

Fig.~\ref{fig:profile} shows the spatial distribution of hits in the
$x$ (horizontal) and $y$ (vertical) coordinates of the FGD.  Because
T2K runs with an off-axis beam, we expected to see a decrease in the
neutrino flux with increasing off-axis angle.  This is observed, as
the beam center is directed at X=+3.222 m, Y=-8.146 m, in the ND280
coordinate system.

Overall the results demonstrate that the FGDs are well-suited for
measuring neutrino interactions from T2K's beam.

\section{Acknowledgements}

We would like to thank the support of the following agencies that made
the T2K FGD project possible: Canada Foundation for Innovation, the
Natural Sciences and Engineering Research Council (NSERC) of Canada,
TRIUMF, and the National Research Council, Canada; MEXT and JSPS with
Grant-in-Aid for Scientific Research on Priority Areas 18071007, Young
Scientists S 20674004, JSPS Fellows, the Global COE Program ‘‘The Next
Generation of Physics, Spun from Universality and Emergence’’, Japan;
and by the UK Science and Technology Facilities Council STFC, UK.

We would like to acknowledge D.Kolybaba, K.Wolbaum, C.Carbno,
C.Ingram, B.Freitag, S.Schneider, B.Harack, A.Urichuk, A.Lim,
T.Tolhurst and N.Hogan for their contributions to testing the WLS
fibers, M. Goyette, W. Faszer, S. Sooriyakumaran, A. Starr and
K. Hamano for their help with gluing the XY modules, and P. Lu for
assistance with assembly, machining, and testing.

Thanks are owed to the entire T2K ND280 group for their support of
this project.  We would also like to thank the staff of the J-PARC
center for their support and hospitality, the J-PARC Accelerator Group
for providing the proton beam, and the T2K Beam Group for providing
the neutrino beam. Support provided by KEK is also gratefully
acknowledged.





\bibliographystyle{elsarticle-num}
\begin{flushleft}
\rm{\small{
\bibliography{manuscript}
}}
\end{flushleft}


\end{document}